\documentclass[aps,superscriptaddress,nofootinbib,showpacs,eqsecnum,preprint,tightenlines]{revtex4}
\usepackage{hyperref}
\usepackage{epsfig,rotating}
\usepackage{amsmath,amssymb,graphicx}
\usepackage{dsfont}
\usepackage{bbm}
\numberwithin{equation}{section}

\newcommand{\be}{\begin{equation}}
\newcommand{\ee}{\end{equation}}
\newcommand{\bea}{\begin{eqnarray}}
\newcommand{\eea}{\end{eqnarray}}

\newcommand{\vx}{\vec{x}}

\newcommand{\vp}{\vec{p}}

\newcommand{\vq}{\vec{q}}

\newcommand{\vk}{\vec{k}}

\newcommand{\antinu}{\bar{\nu}_{l}}

\newcommand{\produ}{\mathcal{P}}

\begin{document}

\title{Searching for sterile neutrinos from $\pi$ and $K$ decays.}

\author{Louis Lello}
\email{lal81@pitt.edu}

\author{Daniel Boyanovsky}
\email{boyan@pitt.edu} \affiliation{Department of Physics and
Astronomy, University of Pittsburgh, Pittsburgh, PA 15260}

\date{\today}

\begin{abstract}
The production  of heavy sterile neutrinos from $\pi^-,K^-$ decay at rest yields charged leptons with \emph{negative} helicity (\emph{positive} for $\pi^+,K^+$). We obtain the branching ratio for this process and argue that a Stern-Gerlach filter with a magnetic field gradient leads to spatially separated domains of  both helicity components with abundances determined by the branching ratio. Complemented with  a search of the monochromatic peak, this setup can yield both the mass and mixing angles for sterile neutrinos with masses in the range $3\, \mathrm{MeV} \lesssim m_s \lesssim 414 \,\mathrm{MeV}$ in next generation high intensity experiments. We also study    oscillations of light Dirac and Majorana sterile neutrinos   with $m_s \simeq \,\mathrm{eV}$  produced  in meson decays including decoherence aspects arising from lifetime effects of the decaying mesons and the stopping distance of the   charged lepton in short baseline experiments.  We obtain the transition probability from production to detection via charged current interactions including these decoherence effects for $3+1$ and $3+2$ scenarios, also studying  $|\Delta L|=2$ transitions from $\overline{\nu} \leftrightarrow \nu$ oscillations for Majorana neutrinos and the impact of these effects on the determination of CP-violating amplitudes.    We argue that   decoherence effects are important in current short baseline accelerator experiments, leading to an \emph{underestimate} of masses, mixing and CP-violating angles. At MiniBooNE/SciBooNE we estimate that these effects lead to an $\sim 15 \%$ underestimate for sterile neutrino masses $m_s \gtrsim 3 \,\mathrm{eV}$. We argue that reactor and current short baseline accelerator experiments are fundamentally different and suggest that in future high intensity experiments with neutrinos produced from $\pi,K$ decay at rest, stopping the charged leptons on distances much smaller than the decay length of the parent meson suppresses considerably these decoherence effects.

\end{abstract}

\pacs{14.60.Pq;13.15.+g;14.60.St}

\maketitle

\section{Introduction}

Neutrino masses, mixing and oscillations are the clearest evidence yet of physics beyond the standard model \cite{book1,revbilenky,book2,book3}. They provide an explanation of the solar neutrino problem \cite{msw,book4,haxtonsolar} and have   important phenomenological \cite{book1,book2,book3,grimuslec,kayserlec,mohapatra,degouvea,bilenky}, astrophysical \cite{book4,book5,haxton} and cosmological \cite{dolgovcosmo} consequences. A remarkable series of experiments have confirmed  mixing and  oscillations among three ``active'' neutrinos with $\delta m^2 = 10^{-4}-10^{-3}\,\mathrm{eV}^2$ for atmospheric and solar oscillations respectively. The current bounds on these specifically are $\Delta m^2_{21} =7.62 \times 10^{-5} eV^2$ (best fit) with a $1\sigma$ range ($7.43-7.81 \times 10^{-5} eV^2$)  and $\Delta m^2_{31} =2.55 \times 10^{-3} eV^2$ (best fit) with a $1\sigma$ range ($2.46-2.61 \times 10^{-3} eV^2$)  respectively \cite{forero}, for a complementary global analysis see\cite{global}.

However, several experimental hints have been accumulating that cannot be interpreted within the ``standard paradigm'' of mixing and oscillations among three ``active'' neutrinos with $\delta m^2 \simeq 10^{-4}-10^{-3}$. Early results from
 the LSND experiment\cite{lsnd} have recently been confirmed by   MiniBooNE running in antineutrino mode\cite{miniboone} both suggesting the possibility of new ``sterile'' neutrinos with $\delta m^2 \sim \mathrm{eV}^2$. The latest report from the MiniBooNE collaboration\cite{minilast} on the combined $\nu_\mu\rightarrow \nu_e$ and $\overline{\nu}_\mu\rightarrow \overline{\nu}_e$  \emph{appearance}  data is consistent with neutrino oscillations with $0.01 < \Delta m^2 < 1.0\,\mathrm{eV}^2$. This is consistent with the evidence from LSND antineutrino oscillations\cite{lsnd}, which bolsters the case for the existence of sterile neutrinos; however, combined MiniBooNE/SciBooNE analysis\cite{minisci} of the $\overline{\nu}_\mu$ \emph{disappearance} data are consistent with \emph{no short baseline disappearance} of $\overline{\nu}_\mu$. Recently, a re-examination of the antineutrino flux\cite{flux1} in anticipation of the Double Chooz reactor experiment resulted in a small increase in the flux of about $3.5\%$  for reactor experiments leading to a larger deficit of $5.7\%$ suggesting a \emph{reactor anomaly}\cite{reactor}. If this deficit is the result of neutrino mixing and oscillation  with baselines $L\lesssim 10-100\,\mathrm{m}$, it requires the existence of at least one sterile neutrino with $\delta m^2 \gtrsim 1.5 \,\mathrm{eV}^2$ and mixing amplitude $\sin^2(2\theta) \simeq 0.115$\cite{reactor}. Taken together these results may be explained by models that incorporate one or more sterile neutrinos that mix with the active ones\cite{sorel,kara,kara2,giunti1,sterile1,sterile3,sterile2,giuntishort} including perhaps non-standard interactions\cite{akshwet}; although, there is some tension in the sterile neutrino interpretation of short-baseline anomalies\cite{tension}. These tensions present themselves in the "goodness of fit" parameter, which is obtained by comparing the fit of LSND with MiniBooNE antineutrino data and all other data, which is presently too low. A  comprehensive review of short baseline oscillation experiments summarizes their interpretation in terms of one or more generations of sterile neutrinos\cite{conrad,euro}.

 Hints for the existence of sterile neutrinos also emerge  from cosmology. The   analysis of the cosmic microwave background anisotropies by WMAP\cite{wmap} suggests that the effective number of   neutrino species is $N_{eff} = 3.84\pm 0.40$ and $\sum(m_\nu)<0.44\,eV$, suggesting  the case for sterile neutrino(s) with $m \lesssim  \mathrm{eV}$, however the recent results from (SPT), (ACT)\cite{sptact}  and PLANCK\cite{planck} weaken the bounds considerably. These bounds are obtained assuming 3 active neutrinos, 2 sterile neutrinos and incorporate CMB data, matter power spectrum information and a prior on the Hubble constant \cite{giuntibounds}. More recently stronger bounds on active-sterile neutrino mixing including Planck data has been reported\cite{neff2}. Complementary cosmological data suggests that $N_{eff} >3$ at the $95\%$ confidence level\cite{cosmodata}; although, accommodating an $\mathrm{eV}$ sterile neutrino requires a reassessment of other cosmological parameters\cite{raffelt}. For recent reviews on ``light'' sterile neutrinos see ref.\cite{abarev}. Furthermore, sterile neutrinos with masses in the $\sim \,\mathrm{keV}$ range \emph{may} also be suitable warm dark matter candidates\cite{dodelson,aba,kuse,shapo,hector,dani} compatible with the $\Lambda CDM$ model and may potentially solve the small scale problem. An experimental confirmation of sterile neutrinos would obviously bolster the argument for a cosmologically relevant warm dark matter candidate.

 When taken together, these emerging hints motivate several experimental proposals to search for sterile neutrinos (see the reviews in ref.\cite{abarev}). Various experimental searches have been proposed, such as Higgs decay and matter interactions of relic sterile neutrinos\cite{kuseexpt}, the end point of $\beta$-decay in $^{187}\mathrm{Re}$ with a value of $Q= 2.5 \,\mathrm{keV}$\cite{mare,hectormare} (although the statistics will be hindered by the long lifetime of the source $\simeq 4.3\times 10^{10} \,\mathrm{years}$), and electron capture decays of $^{163}Ho \rightarrow  ^{163}Dy$\cite{holmium} with a $Q$-value $\simeq 2.2 \,\mathrm{keV}-2.8\,\mathrm{keV}$. More recently, the focus has turned on the possible new facilities at the ``intensity frontier,'' one such proposal being project $X$ at Fermilab\cite{projectX} which would deliver high-power proton beams of energies ranging from 2.5-120 GeV and offers flexibility in the timing structure of beams. Another proposal involves using alternative high intensity sources\cite{abarev,intensity} such as mono-energetic electron neutrinos from an $Ar^{37}$ source and detecting the nuclear recoil. There are also recent proposals to study sterile-active oscillations with pion and kaon decay at rest (DAR)\cite{anderson,spitz} where a cyclotron-based proton beam can be used to produce a low energy pion and muon decay-at-rest neutrino source as well as proposals that employee the use of muons from a storage ring\cite{storm}. In addition, the possibility of discrimination between heavy Dirac and Majorana sterile neutrinos\cite{shrockboris} via $|\Delta L|=2$ processes in high luminosity experiments\cite{dibmajo} has been proposed, this is summarized in    recent reviews\cite{conrad,euro}.

 \vspace{3mm}

 \textbf{Goals:} Our goals are the following:
  \begin{itemize}

  \item \textbf{a:)} Motivated by the possibility of high intensity sources, we assess the signals of \emph{heavy} sterile neutrinos from meson (DAR) (both $\pi^-;K^-$)   by focusing on searching for charged leptons of \emph{negative helicity} (or positive helicity for their antiparticles in $\pi^+;K^+$ in (DAR)) in a setup akin to the Stern-Gerlach type experiment where opposite helicity components are spatially separated by a magnetic field gradient. Meson (DAR) produces a monochromatic beam of charged leptons back-to-back with (anti) neutrinos. Massive neutrinos yield a \emph{negative helicity} component for the charged lepton which, in a collimated beam, may be separated from the (larger) positive helicity component by a magnetic field gradient. We study the branching ratio for the \emph{negative helicity} component as a function of the sterile neutrino mass, as a complement to the search for monochromatic lines. We find that for pion (DAR) the electron channel is the most efficient for $3\,\mathrm{MeV} \lesssim m_s \lesssim 135 \,\mathrm{MeV}$ whereas for K-(DAR) both muon and electron channels are similar in the mass range allowed by the kinematics. We obtain an estimate for the upper bound on the branching ratio from previous experiments with typical upper bounds $Br \lesssim 10^{-8}-10^{-6}$ perhaps accessible in the next generation of high intensity experiments.

     \item \textbf{b:)} We  assess decoherence effects of sterile-active neutrino oscillations in \emph{short baseline experiments} as a consequence of i) the decay width of the meson, and ii) the  stopping distance of the charged lepton. As previously found in refs.\cite{boya,hernandez} the decay width of the meson leads to decoherence of oscillations quantified by the dimensionless ratio

         $$ \mathcal{R} = \frac{\delta m^2}{2 E\Gamma_M}$$

         where $E$ is the neutrino energy and $\Gamma_M$ is the meson decay width. For example, a Pion (DAR) with $E \simeq 30 \,\mathrm{MeV}$ and $\Gamma_\pi \sim 2.5 \times 10^{-8}\,\mathrm{eV}$ leads to $\mathcal{R} \simeq \big( \delta m^2/\mathrm{eV}^2 \big)$ and there could be considerable suppression of the appearance and disappearance probability in experiments with baseline $L \simeq 30-100\, \mathrm{mts}$\cite{boya,hernandez}. Another source of decoherence is the distance at which the charged lepton is stopped $L_c$:  if the charged lepton is correlated with the emitted mass eigenstate over a long time scale, the quantum state is projected onto an energy eigenstate and oscillations are suppressed\cite{boya,glashow}. Both effects, meson lifetime and charged lepton stopping scale, are sources of decoherence in sterile-active oscillations that are more prominent in short-baseline experiments and mass scales $\delta m^2 \simeq \mathrm{eV}^2$, as discussed in refs.\cite{boya,hernandez}. These effects can potentially impact the assessment of   the sterile neutrino mass, mixing angle and CP-violation phases.  We study both Dirac and Majorana neutrinos and show that these processes also affect CP-violating transitions. For Majorana neutrinos we study both $\Delta L=0$ oscillations  and $|\Delta L=2|$ (L is lepton number) transitions from $\overline{\nu} \leftrightarrow \nu$ oscillations. We focus in detail on $3+2$ and $3+1$ schemes with   new generations of sterile neutrinos and obtain the general CP-even and CP-odd expressions for the transition probabilities including $|\Delta L|=2$ processes with Majorana neutrinos.

         \item \textbf{c:)} \emph{If} sterile neutrinos are massive Majorana particles there are neutrino-antineutrino oscillations, these are lepton number violating transitions with $|\Delta L|=2$. In short baseline oscillation experiments, massive Majorana neutrinos yield two oscillation channels: the usual one with $\Delta L=0$  and another with $|\Delta L|=2$. While this latter channel is suppressed by the ratio  $m/E$, we seek to study these lepton number violating oscillations in detail as potential discriminators between Dirac and Majorana neutrinos in future high luminosity experiments.  Furthermore, neutrino-antineutrino oscillations can distinctly yield information about CP-violating Majorana phases\cite{moha} and one  of our goals is to assess the impact of the above mentioned decoherence effects on the potential measurement of these transitions for new generations of sterile neutrinos.

     \end{itemize}

 \vspace{3mm}

 \vspace{3mm}

Several appendices provide the technical details.

\section{Heavy sterile neutrinos in rare $\pi^{\pm},K^{\pm}$ decays at rest:}\label{sec:heavy}
In this work, our overarching goals are to assess the impact of sterile neutrinos in experimentally relevant situations. We begin this endeavor with the study of $\pi/K$ decay at rest experiments and focus on  helicity effects as potential experimental signals. The possibility of the existence of \emph{heavy} ``sterile'' neutrino states had received early attention both theoretically\cite{shrock} and experimentally\cite{heavyexpts1,heavyexpts2,heavyexpts3,heavyexpts4,heavyexpts5,heavyexpts6,heavyexpts7,heavyexpts8}; a review of the experimental bounds is presented in ref.\cite{kuseexpt2}.
In this section we analyze \emph{possible} observational signatures of \emph{heavy} sterile neutrinos in $\pi^{-}, K^{-} \rightarrow l^{-}_\alpha \, \overline{\nu}_\alpha$  decay at rest (DAR) but focus on \emph{negative helicity charged leptons} (or positive helicity for $\pi^{+}, K^{+} $ decay). If the neutrino is massless, the charged lepton emerges from $\pi,K$ (DAR)  with right handed helicity (in the rest frame of the meson, which for (DAR) is the laboratory frame). However; if the neutrino is massive,   a fraction of the charged lepton  yield has left handed helicity. If the charged leptons are collimated along an axis $z$ and there is a magnetic field that features a gradient along this direction, the situation is akin to the Stern-Gerlach experiment: the magnetic field gradient leads to a force $\vec{F} \propto -\vec{\nabla} (\vec{\mu}\cdot \vec{B})$ where $\vec{\mu}$ is the charged lepton magnetic moment. This force spatially separates the charged leptons with spins polarized parallel and antiparallel to the magnetic field gradient, just as in a Stern-Gerlach filter. The ratio of the helicity population is determined by the branching ratio of the production process into \emph{negative helicity charged lepton states}.  Our goal is to obtain this branching ratio, which measures the relative intensity of the negative helicity states and \emph{could} serve as a complement to the searches of monochromatic lines.

 While there has been a substantial experimental effort\cite{heavyexpts1,heavyexpts2,heavyexpts3,heavyexpts4,heavyexpts5,heavyexpts6,heavyexpts7,heavyexpts8} searching for monochromatic lines associated with heavy sterile neutrinos from $\pi,K$ decays, we are not aware of experimental efforts searching for \emph{wrong} helicity charged lepton signals in mesons (DAR). The  bounds obtained from the various experiments\cite{heavyexpts1}-\cite{heavyexpts8} are summarized as exclusion regions in ref. \cite{kuseexpt2}, which imply mixing angles (rather elements of the active-sterile mixing matrix) $\lesssim 10^{-6}$,  making the branching  ratios for these processes very small. However, high intensity beams as envisaged in the proposals\cite{anderson,spitz,projectX,abarev,intensity} \emph{may} provide the experimental setting to search for these signals complementing searches for monochromatic lines.

For a $\pi$ or K meson, M, the interaction   Hamiltonian for a $M \rightarrow l \,\antinu $ decay is given by

\be
H_i =F_{M} \sum_{\alpha=e,\mu} \int d^3x \left[  \,\overline{\Psi}_{l_\alpha} (\vx,t)\, \gamma^{\mu} \mathbbm{L}\Psi_{ {\nu_\alpha}} (\vx,t) J^{M}_{\mu} (\vx,t) \right] ~~;~~\mathbbm{L} =\frac{1}{2}(1-\gamma^5) \label{Hint}
\ee
where the label $\alpha $ refers to the charged leptons,  $J^{M}_{\mu} (\vx,t) = i   \partial_{\mu} M (\vx,t)$ and M is a complex (interpolating)  field that describes the charged pseudoscalar mesons $M=\pi^-,K^-$. For a $\pi^-$ meson, we have that $F_{\pi} = \sqrt{2}\, G_F\, V_{ud} \, f_{\pi}$   and for the $K^\pm$ meson, we have that $F_{K} =  \sqrt{2}\, G_F V_{us}\, f_{K}$,  where $f_{\pi,K}$ are the decay constants. The flavor neutrino fields and the fields that create/annihilate neutrino mass eigenstates are related  by
 \be \Psi_{ {\nu_\alpha}}=  \sum_{j}\,U_{\alpha j}\Psi_{ {\nu_j}}\,.\label{massnus}\ee

 For $n$ generations of Dirac neutrinos the matrix $U$ is $n\times n$, unitary and features $(n-1)(n-2)/2$ CP-violating Dirac phases. For Majorana neutrinos
 \be U \rightarrow \widetilde{U} = U\,D~~;~~ D = \mathrm{diag}\big[e^{i{\theta_1}/2},e^{i{\theta_2}/2},\cdots, e^{i{\theta_n}/2} \big] \label{mayomat}\ee where $U$ is the mixing matrix for Dirac neutrinos and  we have allowed an inconsequential overall phase. It follows that
 \be \widetilde{U}_{\alpha j} = U_{\alpha j}\,e^{i{\theta_j}/2}\, .  \label{mayomatels}\ee The Majorana CP-violating phases, $\theta_i - \theta_j$, \emph{only} contribute to $\nu \leftrightarrow \overline{\nu}$ oscillations and $|\Delta L|=2$ processes\cite{moha} which will be studied in detail in section (\ref{sec:majorana}).

 The details of the quantization of the different fields are provided in   appendix (\ref{app:quant}). From these results, it follows that, after integration over the spatial variables, the relevant Hamiltonian to obtain the production amplitudes is given by (see appendix (\ref{app:quant}) for notation)
 \be
H_i = \frac{F_{M}}{\sqrt{V}} \sum_{\vq,\vp} \sum_{h,h'}\sum_{\alpha, j} U_{\alpha j} \frac{\left[ \overline{\psi}_{l_\alpha} (\vk,h)\gamma^{\mu} \mathbbm{L} \psi_{\nu_j}(\vq,h') p_{\mu} (M^{+}_{\vp} - M^{-}_{\vp}) \right]}{\sqrt{8E_M(p)E_\alpha(k)E_j(q)}}~~;~~ \vk=\vp+\vq \label{hamint}
\ee where the Fermi quantum fields, $\psi_{\nu_j}$, are expanded as in (\ref{app:quant}) either for Dirac or Majorana fermions.

We identify the  production matrix element $M^-(\vec{p}) \rightarrow l_\alpha (\vec{k})\,\overline{\nu}_\alpha(\vec{q}) $  as
 \be \mathcal{M}^{\produ}_{\alpha,\alpha}(\vk,\vq,h,h') = \sum_{j} U_{\alpha, j}~ \mathcal{M}^{\produ}_{ \alpha j}(\vk,\vq,h,h')  \label{Mprodu} \ee where
 \be \mathcal{M}^{\produ}_{ \alpha, j}(\vk,\vq,h,h')= F_M \, \overline{\mathcal{U}}_{\alpha,h}(\vk)\,\gamma^\mu \mathbbm{L} \mathcal{V}_{j,h'}(\vq)\, p_\mu ~~;~~ \vk=\vp+\vq \label{mfiprod} \ee is the transition matrix element for meson decay into a charged lepton, $\alpha$, and an antineutrino mass eigenstate, $j$. For Dirac neutrinos,   the spinors $\mathcal{V}_{j,h'}(\vq)$ in (\ref{mfiprod}) are given by (\ref{Vspi}), whereas for Majorana neutrinos
 \be \mathcal{V}_{j,h'}(\vq) \rightarrow \mathcal{U}^{\,c}_{j,h'}(-\vq) \ee given by (\ref{ucs}) and the mixing matrix $U \rightarrow \widetilde{U}$ given by  (\ref{mayomat},\ref{mayomatels}).
 The separation of helicity contributions is frame dependent and the most clear identification of processes that reveals a massive neutrino is provided by the decay of the pseudoscalar meson at rest ($\vp=0$)  so that the laboratory coincides with the rest frame of the meson and helicity states are unambiguously recognized.
 The contributions to the production amplitude from the different helicity states in (DAR) are given by

\be
\begin{split}
& \overline{\mathcal{U}}_{\alpha,+}(\vq)\,\gamma^\mu \mathbbm{L} \mathcal{V}_{j,+}(\vq)\, p_\mu = - m_{M} \varepsilon_{l} N_l N_{\bar{\nu}}\\
& \overline{\mathcal{U}}_{\alpha,+}(\vq)\,\gamma^\mu \mathbbm{L} \mathcal{V}_{j,-}(\vq)\, p_\mu = 0 \\
&  \overline{\mathcal{U}}_{\alpha,-}(\vq)\,\gamma^\mu \mathbbm{L} \mathcal{V}_{j,+}(\vq)\, p_\mu  =0 \\
& \overline{\mathcal{U}}_{\alpha,-}(\vq)\,\gamma^\mu \mathbbm{L} \mathcal{V}_{j,-}(\vq)\, p_\mu = m_{M} \varepsilon_{\bar{\nu}} N_l N_{\bar{\nu}}
\end{split} \label{amphels}
\ee where (see notation in appendix (\ref{app:quant}))
\be \varepsilon_a(q) = \frac{m_a}{E_a(q)+ q}~~;~~N_a=\sqrt{E_a(q)+q}~~;~~E_a(q)=\sqrt{q^2+m^2_a}~~;~~ a=l,\nu \label{facs}\ee
Gathering these results, we obtain the helicity contributions to the $\pi,K$ decay widths either for Dirac or Majorana neutrinos:
\bea
\Gamma^{++}_{\pi/K \rightarrow l\bar{\nu_s}}  & = &    \frac{G_F^2}{4\pi} |U_{ls}|^2 \, |V_{ud/us}|^2 \, f^2_{\pi/K} ~  q^*\,m^2_l \Bigg[ \frac{  E_{\nu_s}(q^*)+q^*}{E_{l}(q^*)+q^* }  \Bigg] \label{decayspp} \\
\Gamma^{--}_{\pi/K \rightarrow l\bar{\nu_s}}  & = &    \frac{G_F^2}{4\pi} |U_{ls}|^2 \, |V_{ud/us}|^2 \, f^2_{\pi/K} ~  q^*\,m^2_{\nu_s} \Bigg[ \frac{E_{l}(q^*)+q^* }{  E_{\nu_s}(q^*)+q^*}  \Bigg]  \label{decaysmm}
\eea where
\be q^* = \frac{1}{2m_M}\Bigg[\big(m^2_M-(m_l+m_s)^2\big)\big(m^2_M-(m_l-m_s)^2\big) \Bigg]^\frac{1}{2} ~~;~~m_s \leq m_M-m_l \label{qstar}\ee and
here we refer to the heavy sterile \emph{mass eigenstate} as $\underline{s}$ rather than identifying it with a fourth or fifth generation.

 In the limit $m_s \rightarrow 0$, the usual result for $\pi,K $ decay at rest, where the antineutrino and the lepton are both right handed polarized, is obtained. We are particularly interested in the branching ratio for the process in which both the antineutrino and the charged lepton feature left handed helicity, given by (\ref{decaysmm}). The branching ratio for this process is obtained by normalizing to the total meson width and since these are rare processes, we can instead normalize to the proxy to the total width
\be \Gamma^{tot}_{\pi/K} \equiv \frac{\Gamma_{\pi/K \rightarrow \mu \bar{\nu}}}{Br(\pi/K \rightarrow \mu \bar{\nu})} \label{gammatot} \ee where $Br(\pi/K \rightarrow \mu \bar{\nu}) = 0.999, 0.635$ is the branching ratio for the purely leptonic decay into muons and \emph{massless} neutrinos  for $\pi,K$ decay respectively. Specifically, we have
\be Br^{--}_{M \rightarrow l\bar{\nu_s}}  \equiv  \frac{\Gamma^{--}_{M \rightarrow l\bar{\nu_s}} }{\Gamma^{tot}_{M}} =   |U_{ls}|^2~ \frac{ 2\, Br(M \rightarrow \mu \bar{\nu}) ~ q^*\,m^2_{\nu_s}}{m^2_\mu m_M \,\Big(1-\frac{m^2_\mu}{m^2_M}\Big)^2} \Bigg[ \frac{E_{l}(q^*)+q^* }{  E_{\nu_s}(q^*)+q^*}  \Bigg]  \label{BRmm}\ee

\begin{figure}
\begin{center}
\includegraphics[keepaspectratio=true,width=3.2in,height=3.2in]{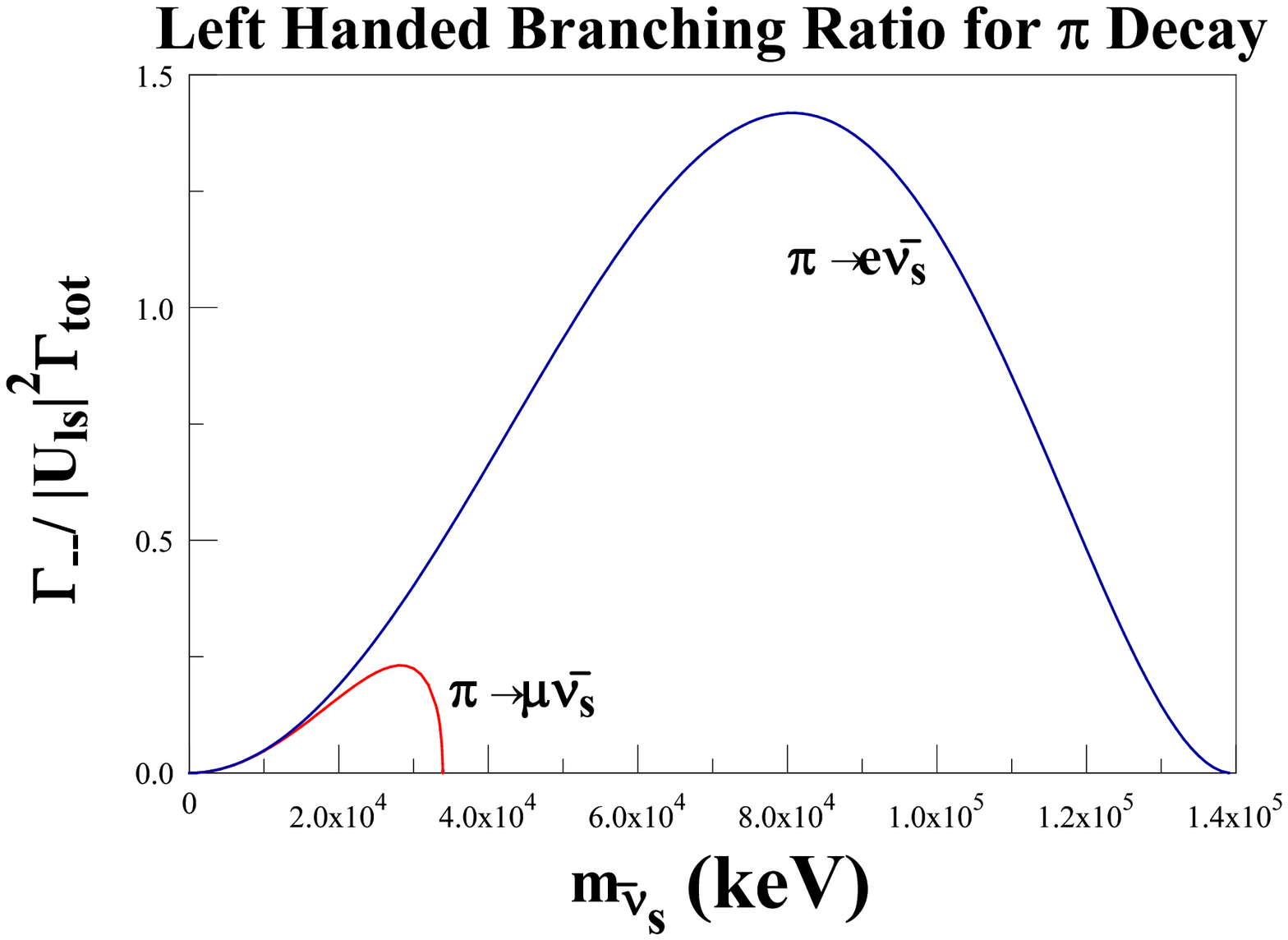}
\includegraphics[keepaspectratio=true,width=3.2in,height=3.2in]{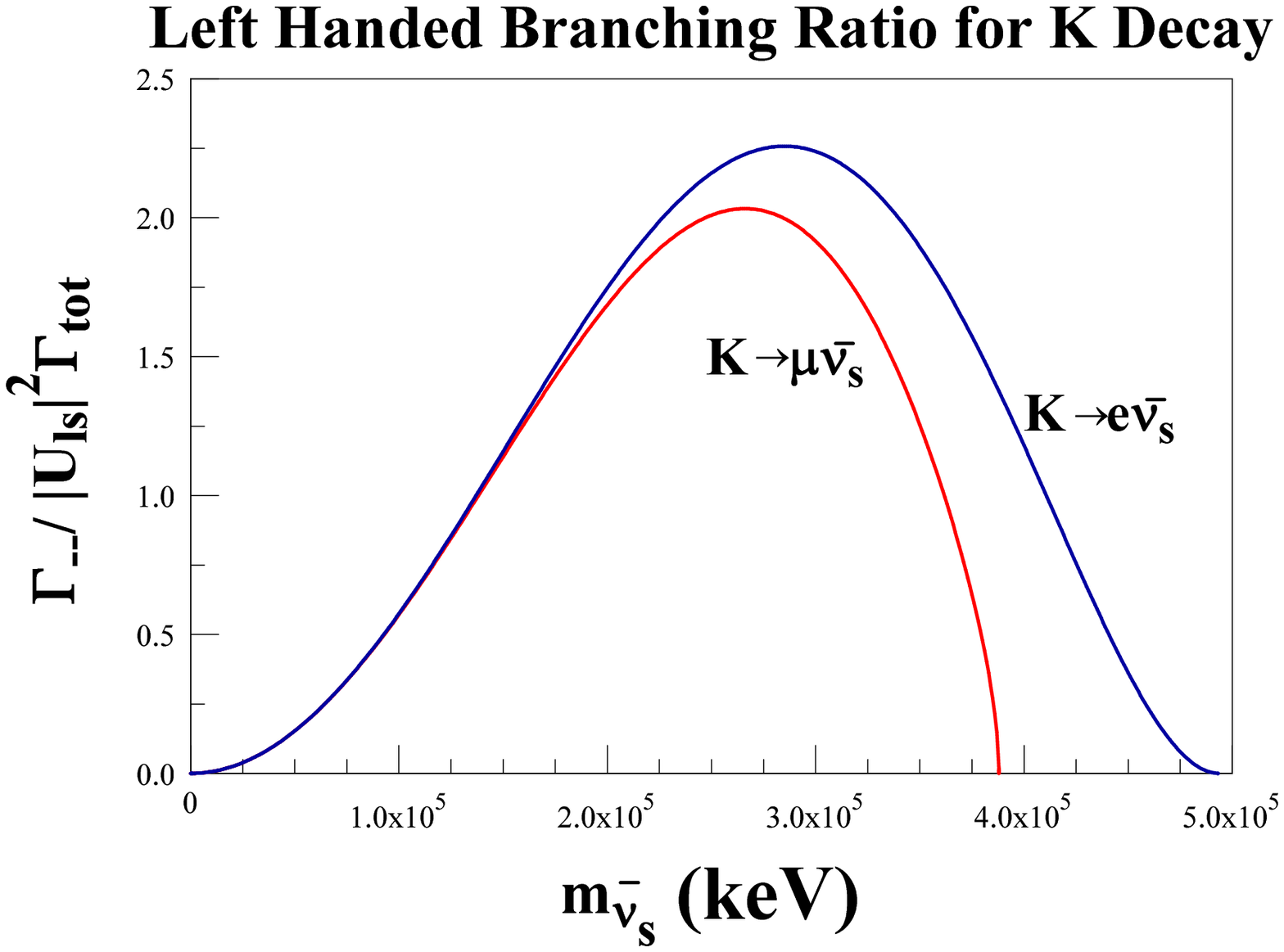}
\caption{Left panel:$Br^{--}_{\pi \rightarrow \mu, e\bar{\nu_s}}/|U_{ls}|^2$ , right panel:$Br^{--}_{K \rightarrow \mu, e\bar{\nu_s}}/|U_{ls}|^2$  vs. $m_s$ for $l=\mu,e$.  }
\label{fig:brs}
\end{center}
\end{figure}

Fig.(\ref{fig:brs}) show the branching ratios (\ref{BRmm}) for $\pi \rightarrow \mu, e\,\bar{\nu_s}$ and  $K   \rightarrow \mu, e\,\bar{\nu_s}$ respectively. For $\pi$ (DAR), the \emph{electron} channel offers a larger window simply because of the larger amount of phase space available whereas the maximum mass available for a heavy sterile neutrino in the muon channel is $\sim 33.92\,\mathrm{MeV}$.

\vspace{2mm}

\textbf{A Stern-Gerlach experiment:}

 In meson (DAR), the presence of a heavy sterile neutrino is manifest as a monochromatic line in the charged lepton spectrum at kinetic energy \be T_l(q^*) = \frac{1}{2m_M}\Big[(m_M-m_l)^2-m^2_s\Big]~~;~~m_s \leq (m_M-m_l)\,. \label{kiner}\ee

 The negative helicity component of the charged lepton in (DAR) provides another manifestation of a massive sterile neutrino which can be exploited in an experiment to complement the search of monochromatic peaks in the charged lepton spectrum. The experimental setup to exploit the negative helicity component (or positive helicity for the opposite charged meson and charged lepton) should be akin to the original Stern-Gerlach experiment to separate spin components. In this case, the relevant quantity is helicity; therefore, consider collimating the charged leptons in (DAR) along a $z-axis$ and setting up a magnetic field with \emph{a gradient along this direction } so that the direction of motion of the collimated charged leptons coincide with the direction of the gradient of the magnetic field. Under these circumstances, there is a magnetic force acting on the charged leptons
 \be F_z \propto -h \frac{d B_z}{dz}\,, \label{Force} \ee where $h$ is the helicity component; thus, opposite helicity components \emph{separate spatially} and the fraction of negative helicity charged leptons is measured by the branching ratio (\ref{BRmm}). Therefore, searching for spatially separated domains of charged leptons \emph{in combination} with a monochromatic line, \emph{may} provide a more robust signature of heavy sterile neutrinos and allow extraction of the mixing matrix element $|U_{ls}|$: the mass of the sterile neutrino is inferred from the peak in the monochromatic spectrum while the ratio of abundances of the helicity components is determined by the branching ratio (\ref{BRmm}); therefore, with the input for $q^*$ obtained from the peak in the monochromatic line and the measurement of the ratio of abundances of helicity states, the branching ratio (\ref{BRmm}) yields $|U_{ls}|$.

 An estimate for the upper bound on the branching ratios, $Br^{--}$, given by (\ref{BRmm}) can be obtained from the summary of the bounds on the mixing matrix elements, $|U_{ls}|^2$,  provided in ref.\cite{kuseexpt2} for $l=\mu$:   the exclusion region for $\pi$ (DAR) from the $\mu$ spectrum yields an upper bound
\be |U_{\mu s}|^2 \lesssim 10^{-5} ~~;~~ 3\,\mathrm{MeV} \lesssim m_s \lesssim 33 \,\mathrm{MeV} \label{U2pimu}\ee and for $K$ (DAR)
\be |U_{\mu s}|^2 \lesssim 10^{-6}-10^{-5} ~~;~~ 30\,\mathrm{MeV} \lesssim m_s \lesssim 330 \,\mathrm{MeV} \,. \label{U2Kmu}\ee The experiments\cite{heavyexpts1,heavyexpts2,heavyexpts3,heavyexpts4,heavyexpts5,heavyexpts6,heavyexpts7,heavyexpts8} on which the bounds in ref.\cite{kuseexpt2} are based, search for monochromatic peaks in the muon spectrum, both from $\pi,K$ (DAR). Ref.\cite{britton} reported an upper limit $|U_{es}|^2 < 10^{-7}$ ($90\% C.L.$) for $30 < m_s < 130\, \mathrm{MeV}$,  therefore  we find from   fig. (\ref{fig:brs}) that the upper bound for $Br^{--}_{\pi \rightarrow \mu,e \bar{\nu_s}} $
\bea &&  Br^{--}_{\pi \rightarrow \mu  \, \bar{\nu_s}} \lesssim 10^{-8}-10^{-7} ~~;~~  3\,\mathrm{MeV} \lesssim m_s \lesssim 33 \,\mathrm{MeV}  \label{upboundpimu2} \\
&&Br^{--}_{\pi \rightarrow e \, \bar{\nu_s}} \lesssim 10^{-9}-10^{-7} ~~;~~  30\,\mathrm{MeV} \lesssim m_s \lesssim 130 \,\mathrm{MeV}  \label{upboundpiel} \eea

  The small $m_s$ region is obviously suppressed by the  $ m^2_{s}/m^2_l$ factor whereas the region near the kinematic edge is suppressed by phase space. For $\pi$ decay, the electron channel is the  most favorable  to study the intermediate mass region $\simeq 3 \, \mathrm{MeV} \lesssim m_s \lesssim 135 \,\mathrm{MeV}$ with typical upper bounds on the branching ratios $10^{-8}-10^{-6}$.

   For $K$ decay, both $\mu,e$ channels yield similar branching ratios with upper bounds
   in the range
  \be   Br^{--}_{K \rightarrow \mu,e \, \bar{\nu_s}} \lesssim 10^{-9}-10^{-6}~
   \mathrm{for}~\Bigg\{ \begin{array}{c}
 4\,\mathrm{MeV} \lesssim m_s \lesssim 360 \,\mathrm{MeV}~(\mu-\mathrm{channel})   \\
   4\, \mathrm{MeV} \lesssim m_s \lesssim 414 \,\mathrm{MeV}~(e-\mathrm{channel})                                                                                             \end{array}   \label{upboundk}\,.\ee

The ``low'' mass region of cosmological interest, $m_s \simeq \mathrm{few}~\mathrm{keV}$, is much more challenging. The experimental results of refs.\cite{heavyexpts1,heavyexpts2,heavyexpts3,heavyexpts4,heavyexpts5,heavyexpts6,heavyexpts7,heavyexpts8} and the analysis of ref.\cite{kuseexpt2} do not provide reliable upper bounds; however, bounds for this mass range emerge from cosmology: a ``heavy'' sterile neutrino can decay into a photon and a light active neutrino, which, for $m_s\simeq \mathrm{keV}$, leads to an X-ray line. Cosmological constraints are summarized in the review articles in refs.\cite{kuse,shapo} with an upper bound $|U_{ls}|^2 \simeq 10^{-10}-10^{-9}$ which would make the branching ratios exceedingly small, even for the high intensity sources envisaged.

To the best of our knowledge, Shrock\cite{shrock}\footnote{We thank R. Shrock for making us aware of his early work on these aspects.}  provided an early proposal to use \emph{polarization} in combination with monochromatic line searches to obtain an assessment of neutrino masses and mixing. Our study differs from this earlier study in two main aspects: i) we advocate using combinations of magnetic fields in a Stern-Gerlach-type setup to \emph{separate} the different helicity components. The relative abundance of the ``wrong'' helicity is determined by the branching ratio obtained above. This is important, while the polarization will be dominated by the lighter active-like neutrinos because they mix with larger mixing angles the \emph{separation} of helicity components by magnetic fields, if experimentally feasible, could result in a clearer signal. ii) Separating the helicity components via magnetic field configurations does not require searching for monochromatic lines and is an independent and \emph{complementary} method. The proposal of ref.\cite{shrock} requires first identifying the monochromatic lines and after this identification measuring the polarization, both aspects must be combined in this proposal to extract information perhaps increasing the challenge from the observational perspective.

A firmer assessment of whether the Stern-Gerlach type experiments, combined with searches of monochromatic peaks in $\pi,K$ (DAR), are feasible in determining the masses of ``heavy'' sterile neutrinos calls for a detailed understanding of backgrounds which is a task that is beyond the scope of this article. Furthermore the above results only apply for $V-A$ weak interactions, therefore if sterile neutrinos feature non-standard weak interactions a re-assessment of the results is required\cite{shrock}.

\section{Oscillations in short-baseline experiments}

For short baseline oscillation experiments, the relevant range of neutrino mass differences is $\delta m^2 \simeq (\mathrm{eV})^2$. A detailed analysis  of oscillation phenomena requires an understanding of the production and detection process. In ref.\cite{boya} a quantum field theoretical generalization of the Wigner-Weisskopf method\cite{ww,cp} was introduced to obtain the correct quantum state arising from the decay of the parent particle. A previous treatment of the correlations of the decay product within a Wigner-Weisskopf approach to semiclassical wave packets was originally studied in ref.\cite{nauenberg} and the dynamics of propagation were studied in ref.\cite{patkos} in simple models. In ref.\cite{boya}, the method was implemented in a simple quantum field theory model of charged current interactions and several aspects were found to be much more general, such as the decoherence effects associated with the lifetime of the decaying parent particle as well as the observation (or stopping) of the charged lepton produced as partner of the neutrino in a charged current interaction vertex.

Many of these aspects were found also in refs.\cite{hernandez} in a different formulation but without explicitly obtaining the quantum mechanical state that describes the decay products.

 Meson decay leads to a correlated state of the charged lepton and the neutrino, a quantum entangled state\cite{glashow,boya,boywu}, the entanglement being a consequence of the kinematics and conservation laws pertinent to the decay\cite{goldman}. As originally observed in ref.\cite{glashow} and analyzed in detail in refs.\cite{boya,boywu},   quantum entanglement leads to decoherence in neutrino oscillations which is a result that has been confirmed  more recently in \cite{hernandez,akmeent} within a different approach.

In this article, we generalize the quantum field theoretical Wigner-Weisskopf method introduced in ref.\cite{boya} to describe (pseudoscalar) meson decay via charged current interactions \emph{in the standard model}, including all aspects of the interactions both for \emph{Dirac and Majorana neutrinos}.
An alternative formulation is offered in ref.\cite{hernandez}; however, the full quantum field theoretical Wigner-Weisskopf method not only illuminates clearly the \emph{quantum entanglement and correlations} between the charged lepton and neutrino states both in momentum and helicity, but also allows a systematic study of Dirac and Majorana fermions including the dynamics of $\nu \leftrightarrow \overline{\nu}$ oscillations and $|\Delta L|=2$ processes discussed in detail in section (\ref{sec:majorana}).

\subsection{Production from meson decay:}

\vspace{2mm}

In appendix  \ref{app:ww} (see also ref.\cite{desiternuestro} for more details), we implement  a quantum field theoretical version of Wigner-Weisskopf theory and we find the Schroedinger picture quantum state that results from pseudoscalar meson decay which is given by
\bea   |M^{-}_{\vp}(t))\rangle  & = &    e^{-iE_M(p)\,t}\,e^{-\Gamma_M(p) \,\frac{t}{2}}\,|M^{-}_{\vec{p}}(0)\rangle  -    \sum_{\vec{q},\alpha j,h,h'}\Bigg\{ U_{\alpha j} \, \Pi^{\produ}_{\alpha j} \,\mathcal{M}^{\produ}_{\alpha j}(\vk,\vq,h,h')\,   \mathcal{F}_{\alpha j}[\vk,\vq;t]  \nonumber \\ &\times &    \,e^{-i(E_\alpha(k)+E_j(q))t}\, |l^-_\alpha(h,\vk)\rangle\,|\overline{\nu}_j(h',-\vq)\rangle  \Bigg\}~~;~~\vk=\vp+\vq\, , \label{wwpistate}\eea where
\be \Pi^{\produ}_{\alpha j} = \frac{1}{[8VE_M(p)E_\alpha(k)E_j(q)]^\frac{1}{2}} \,.
\label{Piprodu}\ee

Although we consider plane wave states, the generalization to wave-packets is straightforward and we comment on the wave-packet approach in section (\ref{wavepack}). The production matrix element $ \mathcal{M}^{\produ}_{\alpha j}(\vk,\vq,h,h')  $ is given by (\ref{mfiprod}), (see eqn. (\ref{finaresul}) in appendix (\ref{app:ww}))  $\Gamma_M(p) = m_M\Gamma_M/E_M(p)$ where $\Gamma_M$ is the decay width in the rest frame of the meson, and
\be \mathcal{F}_{\alpha j}[\vk,\vq;t] = \frac{ 1-e^{-i\big(E_M(p)-E_\alpha(k)-E_j(q)\big)t}\,e^{-\Gamma_M(p) \,\frac{t}{2}} }{ E_M(p)-E_\alpha(k)-E_j(q)-i\frac{\Gamma_M(p)}{2}  }\,. \label{Fs}\ee

  The second term in (\ref{wwpistate}) reveals that the emerging charged lepton and neutrino are entangled both in momentum and in helicity.

   The factor $\mathcal{F}_{\alpha j}[\vk,\vq;t]$ encodes the time dependence of the production process. In order to understand the content of this factor, consider the case $\Gamma_M =0$. In this case,
   \be \mathcal{F}_{\alpha j}[\vk,\vq;t] = e^{-i(E_M-E_\alpha-E_j)\frac{t}{2}}~ \frac{2i\,\sin\Big[(E_M-E_\alpha-E_j)\frac{t }{2} \Big]}{\Big[E_M-E_\alpha-E_j\Big] }~ \stackrel{t\rightarrow \infty}{\rightarrow}  ~2\pi i \delta (E_M-E_\alpha-E_j) \label{delF}\ee namely, in the long time limit, this function describes \emph{energy conservation at the production vertex}. The width of the decaying meson state determines a time (or energy) uncertainty and, either for a narrow width or large time, the function $\mathcal{F}_{\alpha j}[\vk,\vq;t]$ is strongly peaked at $E_\alpha+E_j \simeq E_M$ which describes \emph{approximate energy conservation} within the time or width uncertainty.

   In a typical experiment, the charged lepton produced by pion (kaon) decay is stopped shortly after the end of the pion decay pipe, at which point the correlated quantum state after the neutrino state is \emph{disentangled} by the observation, capture or absorption of the charged lepton at $t_c$.

   If the charged lepton $l_\alpha$  is observed, or absorbed with momentum $\vk$ and helicity projection $h_i$ at time $t_c$, the wave function is projected onto the state $\langle l^-_\alpha(h_i,\vk) |$ and the correct (anti) neutrino state that propagates is given by
  \be |\widetilde{\overline{\nu}}(\vq;h_i)\rangle = - e^{-iE_\alpha(k)t_c} \,\sum_{j,h'} U_{\alpha j}\,\Pi^{\produ}_{\alpha j}\,\mathcal{M}^{\produ}_{\alpha j}(\vk,\vq,h_i,h')  \,\mathcal{F}_{\alpha j}[\vk,\vq;t_c] \,e^{-i E_j(q)t_c}\, |\overline{\nu}_j(h',-\vq)\rangle\,, \label{corneut}\ee where $\vq = \vp-\vk$. This neutrino state still carries the label $h_i$ as a consequence of the helicity entanglement with the measured charged lepton.

  We note that \emph{if} $\mathcal{M}^{\produ}_{\alpha j},\mathcal{F}_{\alpha j},E_j$ are all independent of the mass of the neutrino, $j$, these factors can be taken out of the sum and the resulting (anti) neutrino state is proportional to the familiar Pontecorvo coherent superposition of mass eigenstates. We will analyze this
  approximation below after assessing the total transition amplitude from production to detection; however, before doing so, it proves illuminating to understand the \emph{normalization} of the state (\ref{corneut}).
  \be \mathcal{N}_\nu (\vq;h_i) \equiv \langle \widetilde{\overline{\nu}}(\vq;h_i)|\widetilde{\overline{\nu}}(\vq;h_i)\rangle =   ~\,\sum_{j,h'} \big|U_{\alpha j}\big|^2\,\big|\Pi^{\produ}_{\alpha j}\,\mathcal{M}^{\produ}_{\alpha j}(\vk,\vq,h_i,h')\big|^2 \,\Big|\mathcal{F}_{\alpha j}[\vk,\vq;t_c]\Big|^2\,. \label{norma}\ee

  In the narrow width limit, the function $\Big|\mathcal{F}_{\alpha j}[\vk,\vq;t_c]\Big|^2$ becomes $\propto \delta\big( E_M(p)-E_\alpha(k)-E_j(q)\big)$ and the proportionality constant can be obtained by integrating this function in the variable $\mathcal{E}= E_M(p)-E_\alpha(k)-E_j(q)$, from which we find
  \be \Big|\mathcal{F}_{\alpha j}[\vk,\vq;t_c]\Big|^2 = \frac{2\pi}{\Gamma_M(p)}~ \Big[1- e^{-\Gamma_M(p)t_c} \Big] ~\delta\big( E_M(p)-E_\alpha(k)-E_j(q)\big)\,.  \label{F2}\ee Therefore
  \be \mathcal{N}_\nu (\vq;h_i) = \frac{\Big[1- e^{-\Gamma_M(p)t_c} \Big]}{\Gamma_M(p)}  \sum_{j,h'} \big|U_{\alpha j}\big|^2\,\big|\Pi^{\produ}_{\alpha j}\,\mathcal{M}^{\produ}_{\alpha j}(\vk,\vq,h_i,h')\big|^2 \, 2\pi\,\delta\big( E_M(p)-E_\alpha(k)-E_j(q)\big)\,. \label{normalnu}\ee

   In appendix (\ref{app:norma}), we obtain the relation between the normalization   (\ref{normalnu}), the partial and total decay width of the meson and the number density of charged leptons produced by meson decay during a time $t_c$. While ref.\cite{book5} discusses the normalization of the neutrino state\footnote{See section (8.1.1), pages 285,286 in ref.\cite{book5}.}, to the best of our knowledge, the relation of the neutrino normalization to the number density of charged leptons produced has not been recognized previously.

\vspace{2mm}

 \subsection{Detection via a charged current vertex:}

  \vspace{2mm}

  In what follows we assume the neutrino to be described by a Dirac fermion,   extending the discussion to Majorana fermions in section (\ref{sec:majorana}). We note here that the Dirac or Majorana nature is irrelevant for the $\Delta L = 0$ process considered here but
  plays a nontrivial role in section (\ref{sec:majorana}).

\begin{figure}[ht!]
\begin{center}
\includegraphics[keepaspectratio=true,width=5in,height=5in]{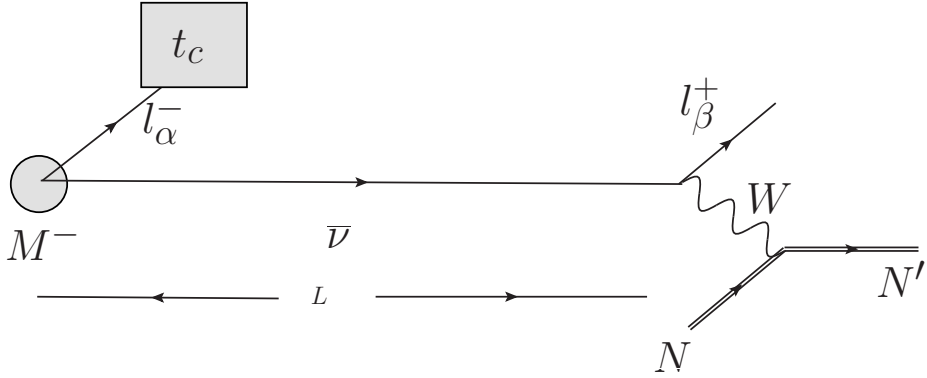}
\caption{Production via $M^-\rightarrow l^-_{\alpha}\overline{\nu}$ detection via a charged current vertex $\overline{\nu}\,N \rightarrow l^+_\beta\,N'$ at a baseline $L$ with $N,N'$ nucleons or nuclear targets. The charged lepton $l^-_{\alpha}$ produced with the antineutrino is observed, absorbed or decays at a time $t_c$. }
\label{fig:expt}
\end{center}
\end{figure}

  Consider   the case in which the (anti)neutrino is
  detected via a charged current event $\overline{\nu}\,N \rightarrow l^+_\beta \,N' $ at a detector situated at a baseline $L$ (fig.(\ref{fig:expt})).

    The Schroedinger picture quantum states that describe  the initial and final states are
  \be |i\rangle = |\widetilde{\overline{\nu}}\,;N\rangle = |\widetilde{\overline{\nu}}\rangle \otimes |N\rangle_{\mathcal{D}}~~;~~|f\rangle = |l^+_\beta;N'\rangle = |l^+_\beta\rangle \otimes |N'\rangle \label{inifin}\ee where $|\widetilde{\overline{\nu}}\rangle$ is given by (\ref{corneut}), the state $|N\rangle_{\mathcal{D}}$ describes a nucleon or nuclear target localized at the detector and the outgoing charged lepton is measured with helicity $h_f$. The transition amplitude in the Schroedinger picture is given by
  \be \mathcal{T}_{i\rightarrow f} = \langle f|e^{-iH(t_D-t_c)}|i\rangle \simeq -ie^{-iE_Ft_D}\int^{t_D}_{t_c}e^{iE_Ft'}\langle f|H_{i}\, e^{-iH_0t'}\,e^{iH_0t_c}|i\rangle  \, dt'\label{inoutamp}\ee where $E_F = E_{l_\beta}+E_{N'}$ is the total energy of the final state, and $H_0,H_i,H$ are the unperturbed, interaction and total Hamiltonians respectively. To obtain this expression we have used   $e^{-iH(t_D-t_c)}= e^{-iH_0 t_D}\, U(t_D,t_c) \,e^{iH_0t_c}$ and $U(t_D,t_c)$ is the usual time evolution operator in the interaction picture.

  Up to an irrelevant overall phase we find
  \be \mathcal{T}_{i\rightarrow f} =  \sum_{j,h'}\Pi^{\mathcal{P}}_{\alpha j} U_{\alpha j}\,\mathcal{M}^{\produ}_{\alpha j} \,\mathcal{F}_{\alpha j}\,G_{\beta j}~ \langle l^+_\beta\;;N'|H_{i} |\overline{\nu}_{j,h'}\;;N\rangle \label{Tfifin}\ee where  $\Pi^{\mathcal{P}}_{\alpha j}$ is given by eqn. (\ref{Piprodu}), we have suppressed the indices to avoid cluttering the notation and introduced
  \be  G_{\beta j} = e^{\frac{i}{2}(E_F-E_N-E_j)(t_D+t_c)}~ \frac{2\,\sin\Big[\frac{1}{2}(E_F-E_N-E_j)(t_D-t_c) \Big]}{\Big[E_F-E_N-E_j \Big]}\,.\label{Gs}\ee

  The relevant interaction Hamiltonian is given by
  \be H_{i} =  U^*_{\beta j} ~\sqrt{2}G_F\,\int   \overline{\Psi}_{\nu_j}(\vx)\gamma^\mu \mathbbm{L}\, \Psi_{l_\beta}(\vx) \,\mathcal{J}^{(N,N')}_\mu(\vx) d^3x +h.c.\label{Hintdet}\ee where $\mathcal{J}^{(N,N')}_\mu(\vx)$ is the hadron current with matrix element\footnote{This matrix element may be written in terms of vector and axial vector form factors, but such expansion is not necessary in our analysis.}
  \be \langle N|\mathcal{J}^{(N,N')}_\mu(\vx)|N'\rangle \equiv  \sum_P\, \frac{j^{N,N'}_\mu(P)}{\sqrt{4VE_NE_{N'}}}\, e^{i\vec{P}\cdot\vx}\,. \label{mtxJNN}\ee leading to the matrix element
  \be \langle l^+_\beta\;;N'|H_{int} |\overline{\nu}_j\;;N\rangle = U^*_{\beta j}\, \Pi^{\mathcal{D}}_{\beta j}\,\mathcal{M}^{\mathcal{D}}_{j \beta} \label{mtxdet}\ee where
  \bea && \Pi^{\mathcal{D}}_{\beta j} = \frac{1}{[16VE_NE_{N'}E_\beta(k') E_j(q)]^\frac{1}{2}}\label{pidet}\\
  &&  \mathcal{M}^{\mathcal{D}}_{j \beta} =   \sqrt{2}G_F\,   \overline{\mathcal{V}}_{j,h'}(\vq) \,\gamma^\mu \mathbbm{L} \, \mathcal{V}_{\beta,h_f}(\vk') \, j^{N,N'}_\mu(P) ~~;~~ \vq = \vk-\vp = \vec{P}+\vk'\,. \label{Mdet} \eea Therefore, the total transition amplitude from production to detection
  is given by
  \be \mathcal{T}_{i\rightarrow f} =  \sum_{j,h'}U_{\alpha j}\,\Pi^{\mathcal{P}}_{\alpha j}\, \mathcal{M}^{\produ}_{\alpha j} \,\mathcal{F}_{\alpha j}\,G_{\beta j}\,U^*_{\beta j }\,\Pi^{\mathcal{D}}_{\beta j}\,\mathcal{M}^{\mathcal{D}}_{j \beta} \label{Tfifin2}\ee where we have suppressed all the arguments to simplify notation. The factors $\mathcal{F}_{\alpha j}$ and $G_{\beta j}$ encode the \emph{time dependence} of the production, measurement of the charged lepton produced with the (anti) neutrino  and final detection processes and the energy uncertainty from the finite lifetime of the parent meson. As noted above (see eqn.\ref{delF}  ), $\mathcal{F}_{\alpha j}$ describes nearly energy conservation in the long time narrow width limit but includes the energy uncertainty from the width of the decaying state. Similarly
  \be G_{\beta j}\stackrel{t\rightarrow \infty}{\rightarrow}  ~2\pi   \delta (E_F-E_N-E_j) \label{ecdet}\ee describes energy conservation at the detection vertex in the long time limit.

  The phases in these factors encode the information of interference effects between the different mass eigenstates.

   In order to isolate the contribution of these factors there are several approximations that are dictated by the experimental aspects:

  \vspace{2mm}

  \textbf{Approximations:}

  \begin{enumerate}\label{approxi}
  \item For neutrino masses consistent with oscillation experiments utilizing baselines of a few hundred meters, namely $m_j \simeq  \mathrm{eV}$, and typical neutrino energy from meson decay, $\gtrsim 30~ \mathrm{MeV}$, the neutrinos are ultrarelativistic so we can approximate $E_j(q) = E(q) + m^2_j/2E(q)$ with $E(q) = q$. Obviously, this approximation is valid for even higher energies and longer baselines so that the results may be extrapolated appropriately.
  \item We  neglect the neutrino masses in the factors $E_j(q)$ in  the denominators in $\Pi^{\mathcal{P}}_{\alpha j}\, ,\Pi^{\mathcal{D}}_{\beta j} $ (eqns.\ref{Piprodu},\ref{pidet}).
  \item We also neglect the mass dependence of neutrino spinors $\mathcal{V}$, which depend upon the mass through the factor $\varepsilon_j(q) = m_j/(E_j(q)+q)$ (see (\ref{Vspi},\ref{Neps})). Neglecting the neutrino masses in the spinors leads to the production and detection matrix elements $\mathcal{M}^{\produ}\,,\mathcal{M}^{\mathcal{D}}$ to be independent of the neutrino masses, therefore independent of the label $j$.
 \item Neglecting the neutrino mass, the negative chirality (anti) neutrino only features a positive helicity component, therefore only $h'=+$ remains in the sum. This is, obviously, a consequence of $\varepsilon_j \ll 1$.
  \end{enumerate}

  Under these approximations and the unitarity of the mixing matrix $U$, the normalization $ \mathcal{N}_\nu (q)  $  (\ref{norma}) of the neutrino state $|\widetilde{\overline{\nu}(\vq)}\rangle $ becomes
  \be \mathcal{N}_\nu (q) =  \frac{ \big| \Pi^{\mathcal{P}}_\alpha\,\mathcal{M}^{\produ}_{\alpha}\big|^2 }{\Gamma_M(p)}~ \Big[1- e^{-\Gamma_M(p)t_c} \Big] ~2\pi\,\delta\big( \mathcal{E}_{\produ})~~;~~\mathcal{E}_{\produ}=E_M-E_\alpha-E\,. \label{norma2}\ee

  The time dependent factors $\mathcal{F}_{\alpha j},G_{\beta j}$ feature phases whose interference leads to the oscillations in the transition probabilities, therefore the terms $m^2_j/2E(q)$ must be kept in these phases.

Under these approximations, the factors $\Pi^{\mathcal{P}}\mathcal{M}^{\produ}$ and $\Pi^{\mathcal{D}}\mathcal{M}^{\mathcal{D}}$ can be taken out of the sum and the final result for the transition amplitude \emph{factorizes} into production, propagation with oscillations, and detection contributions:
\be \mathcal{T}_{i\rightarrow f} = \underbrace{\Bigg[ \Pi^{\mathcal{P}}_\alpha\,\mathcal{M}^{\produ}_{\alpha}\Bigg]}_{\mathrm{Production}}  \, \underbrace{\Bigg[\sum_{j} U_{\alpha j}\,\mathcal{F}_{\alpha j} \,G_{\beta j} \,U^*_{\beta j}\Bigg]}_{\mathrm{Propagation-Oscillations}} \,   \underbrace{\Bigg[\Pi^{\mathcal{D}}_\beta \, \mathcal{M}^{\mathcal{D}}_\beta \Bigg]}_{\mathrm{Detection}} \label{factorTfi}\ee The transition probability is given by
\be |\mathcal{T}_{i\rightarrow f}|^2 =  \Big| \Pi^{\mathcal{P}}_\alpha\,\mathcal{M}^{\produ}_{\alpha}\Big|^2  \,  \Big[\sum_{j}\sum_{i} U_{\alpha j}U^*_{\alpha i}\,\mathcal{F}_{\alpha j}\mathcal{F}^*_{\alpha i} \,G_{\beta j} G^*_{\beta i} \,U^*_{\beta j}U_{\beta i}\Big]  \,    \Big|\Pi^{\mathcal{D}}_\beta \, \mathcal{M}^{\mathcal{D}}_\beta \Big|^2 \label{factorTfi2}\ee where $\mathcal{F}_{\alpha j}\equiv \mathcal{F}_{\alpha j}[\vk,\vq,t_c]$ is given by (\ref{Fs}) evaluated at $t=t_c$ and $G_{\beta j}$ is given by (\ref{Gs}).

It proves convenient to introduce:
\be \mathcal{E}_{\produ} =  E_M(p)-E_\alpha(k)-E(q)~~;~~\mathcal{E}_{\mathcal{D}}= E_F-E_N-E(q)\,. \label{defisE} \ee For $m_j \ll E(q)$ and narrow width $ \Gamma_M \ll E_M$, the products $\mathcal{F}_{\alpha j}\mathcal{F}^*_{\alpha i}$ are sharply peaked at $\mathcal{E}_{\produ}$, becoming nearly energy conserving delta functions in the long time and small width limit (see (\ref{delF})). Similarly, $G_{\beta j} G^*_{\beta i} $ is sharply peaked at $\mathcal{E}_{\mathcal{D}}$. Each term $\mathcal{F},G$ describe approximate energy conservation at the production and detection vertices respectively. In order to extract the coefficients of the energy conserving $\delta(\mathcal{E}_{\produ}),\delta(\mathcal{E}_{\mathcal{D}})$, we integrate the respective products with a smooth initial and final density of states  that are insensitive to $\Gamma_M$ and $\Delta_j$ (for details see ref.\cite{boya}). We find
\be \mathcal{F}_{\alpha j}\mathcal{F}^*_{\alpha i} = \frac{2\pi}{\Gamma_M(p)}~\frac{\Big[1-e^{-i\Delta_{ij}t_c}\,e^{-\Gamma_M(p)t_c} \Big]}{1+i\mathcal{R}_{ij}}~\delta({\mathcal{E}_{\produ}}) \label{Fprods}\ee were we have introduced
\be \Delta_{ij} = \frac{\delta m^2_{ij}}{2E(q)}~~;~~\mathcal{R}_{ij} = \frac{\Delta_{ij}}{\Gamma_M(p)} = \frac{\delta m^2_{ij}}{2 \Gamma_M M_M}~\frac{E_M(p)}{E(q)} ~~;~~\delta m^2_{ij}=m^2_i-m^2_j\,,\label{defisdel}\ee similarly
\be G_{\beta j}G^*_{\beta i} = 2\pi\, i\, e^{i\Delta_{ij}t_c}~\frac{\Big[1-e^{i\Delta_{ij}(t_D-t_c)} \Big]}{\Delta_{ij}}~\delta({\mathcal{E}_{\mathcal{D}}})\,. \label{Gprods}\ee

As usual, one is interested in obtaining the transition $\emph{rate}$; therefore, we focus on
 $ \frac{d}{dt_D}|\mathcal{T}_{i\rightarrow f}|^2$ for which we need
 \be  \frac{d}{dt_D}\Big( G_{\beta j}G^*_{\beta i}\Big)= 2\pi \,\delta({\mathcal{E}_{\mathcal{D}}})\,e^{i\Delta_{ij}t_D} \label{deriGprod}\ee
 Separating the diagonal $i=j$ from off-diagonal terms in the sums in (\ref{factorTfi2}), and using the result (\ref{norma2}) for the normalization of the (anti) neutrino state,  we   find the \emph{transition rate}
  \be \frac{d}{dt_D} |\mathcal{T}_{i\rightarrow f}|^2 =  \Big[\mathcal{N}_\nu \Big] \, {\mathcal{P}_{\alpha \rightarrow \beta}} \, \Bigg[ \frac{d\Gamma_{\overline{\nu}N\rightarrow l_\beta N'}}{(2\pi)^6 V^2\,d^3k'd^3P}\Bigg]  \label{finprobi}\ee where
  \be \frac{d\Gamma_{\overline{\nu}N\rightarrow l_\beta N'}}{(2\pi)^6 V^2\,d^3k'd^3P} = \Big|\Pi^{\mathcal{D}}_\beta \, \mathcal{M}^{\mathcal{D}}_\beta \Big|^2 \, 2\pi\,\delta({\mathcal{E}_{\mathcal{D}}})\label{2diffsig} \ee is the   double differential detection rate for $\overline{\nu} N \rightarrow l^+_{\beta} N'$ for an incoming \emph{massless} neutrino, and $\mathcal{P}_{\alpha \rightarrow \beta}$ is the flavor transition probability
  \be \mathcal{P}_{\alpha \rightarrow \beta} =   \sum_{ j,i }  U_{\alpha j} U^*_{\beta j} U^*_{\alpha i} U_{\beta i} I_{ij} \label{finalprobfla}\ee where $I_{ij}$ are the interference terms
  \be I_{ij} = e^{i\Delta_{ij}t_D} \,\Bigg[\frac{ 1-e^{-i\Delta_{ij}t_c}\,e^{-\Gamma_M(p)t_c}  }{ 1- e^{-\Gamma_M(p)t_c} }\Bigg]\,\Bigg[\frac{1-i\mathcal{R}_{ij}}{1+ \mathcal{R}^2_{ij}}\Bigg] ~~;~~I_{ji}=I^*_{ij}\,. \label{inteference}\ee   Unitarity of the $U$ matrix allows to write
  \bea \mathcal{P}_{\alpha \rightarrow \beta} = \delta_{\alpha,\beta}~~ - &&  2\sum_{j>i}\mathrm{Re}\Big[ U_{\alpha j} U^*_{\beta j} U^*_{\alpha i} U_{\beta i} \Big]\, \mathrm{Re}\Big[1-  I_{ij} \Big]  \nonumber \\&& - 2\sum_{j>i}\mathrm{Im}\Big[ U_{\alpha j} U^*_{\beta j} U^*_{\alpha i} U_{\beta i} \Big]\, \mathrm{Im}\Big[I_{ij} \Big]\,. \label{finprobab}\eea

  In the above expressions we have implicitly assumed Dirac neutrinos, the case of Majorana neutrinos is obtained by the replacement (see eqns. (\ref{mayomat},\ref{mayomatels})) $U\rightarrow \widetilde{U}~~;~~ \widetilde{U}_{\alpha j} = U_{\alpha j}\,e^{i\phi_j/2}\, \forall \alpha $ from which it is  obvious that the CP-violating Majorana phases do not play any role in $\overline{\nu}_\alpha  \rightarrow \overline{\nu}_\beta$ oscillations.

  The possibility of $CP$ violation in the neutrino sector from Dirac phases is encoded in the imaginary part in (\ref{finprobab}) since for the transition probabilities for $\nu_\alpha \rightarrow \nu_\beta$ it follows that $U_{\alpha i} \rightarrow U^*_{\alpha i}$. Therefore decoherence effects in the \emph{imaginary part} of $I_{ij}$ lead to possible \emph{suppression of CP-violating contributions in the transition probabilities}.

  The transition rate (\ref{finprobi}),  along with (\ref{finprobab}), are some of the important results of this article; the factorized form  of (\ref{finprobi})   is a consequence of the approximations described above. The origin of the prefactor $\mathcal{N}_\nu$ is clear, it is the normalization of the neutrino state that emerges from disentangling the charged lepton in the production process, since this is the correct neutrino state that propagates to the detector and triggers the charged current reaction that yields the measured charged lepton in the final state. The interference terms (\ref{inteference}) encode the decoherence effects arising from the finite lifetime of the source \emph{and} the energy uncertainty associated with the time scale in which the charged lepton produced in a correlated quantum state with the neutrino is observed (or captured). This decoherence can be understood clearly in two limits:
  \begin{itemize}
  \item When $\Delta_{ij} \ll \Gamma_M$  it follows that  $\mathcal{R}_{ij} \rightarrow 0$ and $I_{ij}$ is the usual interference term. In this limit the energy uncertainty associated with the lifetime of the source does not allow to separate the mass eigenstates and the coherence of the superposition of mass eigenstates is maintained. However, in the opposite limit, $\Delta_{ij} \gg \Gamma_M$, the factor $\mathcal{R}_{ij} \gg 1$ and the interference term is suppressed. In this limit, the lifetime of the source is long, the corresponding energy uncertainty is small and the mass eigenstates are separated in the time evolution and coherence between them in the superposition is suppressed.
  \item In the limit $\Gamma_M \rightarrow 0$ it follows that

 \be  I_{ij} \rightarrow e^{i\Delta_{ij}(t_D-t_c/2)}~  \frac{\sin[\Delta_{ij}t_c/2]}{[\Delta_{ij}t_c/2]} \,. \label{gammaeq0}\ee   There are two effects in this expression: 1) a shortening of the baseline by the distance travelled  by the charged lepton produced with the (anti) neutrino and 2) a suppression factor associated with the time uncertainty: if $\Delta_{ij}t_c > 1$, then the interference term is suppressed, this is because  if the charged lepton produced with the (anti) neutrino is entangled all throughout the evolution at long time $t_c \gg 1/\Delta_{ij}$ the energy uncertainty becomes much smaller than the difference in energy between  mass eigenstates and these  are projected out by energy conservation which leads to their decoherence in the superposition. This is another manifestation of energy conservation as encoded in Fermi's Golden rule. In terms of the oscillation length, $L^{osc}_{ij}$, defined as
 \be \Delta_{ij} = \frac{\delta m^2_{ij}}{2E} \equiv \frac{2\pi}{L^{osc}_{ij}} \label{Lij}\ee the suppression factor $2 \sin[\Delta_{ij}t_c/2]/\Delta_{ij}t_c < 1$ when the stopping length scale $L_c \equiv t_c \simeq L_{ij}$.  The case $\Gamma\rightarrow 0$ is relevant for reactor experiments. See the discussion in section(\ref{sec:reactors}).

  \end{itemize}

  The suppression factor associated with the lifetime is relevant in the case of possible new generation of (sterile) neutrinos with masses in the $\mathrm{eV}$ range when produced in the decay of pions or kaons.

  For pion decay at rest, the typical energy of a (nearly massless) neutrino is $E^* \sim 30\,\mathrm{MeV}$, the pion width at rest  $\Gamma_\pi = 2.5\,\times  10^{-8}\,\mathrm{eV}$ and for one generation of sterile neutrino with $m_4 \gg m_{1,2,3}$ we find
  \be \mathcal{R} \simeq \frac{m^2_4}{2E^*\,\Gamma_\pi} \simeq \frac{2}{3}\Big(\frac{m_4}{\mathrm{eV}}\Big)^2 \,, \label{R4}\ee therefore, for $m_4 \geq 1\,\mathrm{eV}$, the suppression factor can be substantial and the transition probability is suppressed. For the decay of a pion in flight with a large Lorentz $\gamma$ factor, the result only changes by a factor $2$ as can be seen as follows: consider a neutrino that is emitted   collinear with the direction of the pion in the laboratory frame (say along the $z-axis$), its energy in the laboratory frame is
  \be E = \gamma E^* \big(1+ {V}_\pi\big) \label{LTL}\ee where $ {V}_\pi$ is the pion's velocity, for $\gamma \gg 1$ it follows that $E \sim 2 \gamma E^*$. The width  of the pion in the laboratory frame is $\Gamma_\pi/\gamma$; therefore, for neutrinos produced by pion decay in flight with a large Lorentz factor \be \mathcal{R} \simeq \frac{1}{3} \,\Big(\frac{m_4}{\mathrm{eV}}\Big)^2\,. \label{decayflight}\ee

  In conclusion, for new generations of (sterile) neutrinos with masses in the $\mathrm{eV}$ range, experiments in which oscillations are probed with neutrinos from pion decay feature the suppression factors associated with the pion width. For Kaons, the situation improves  because in this case
 $$\Gamma_K \simeq 5\,\times 10^{-8}\,\mathrm{eV}~~;~~E^* \simeq 235.5 \, \mathrm{MeV}$$
 and  for Kaon (DAR)
 $$\mathcal{R} \simeq \frac{1}{25}\,\Big(\frac{m_4}{\mathrm{eV}}\Big)^2, $$ thus $\mathcal{R} <1$  for  $m_4 \simeq  \mathrm{few}~\mathrm{eV}$.

 \section{$3+2$ and $3+1$ cases in the ``short-baseline approximation'':}

 \vspace{2mm}

In the ``short-baseline'' approximation, we assume that there are sterile neutrinos $j = 4,5 \cdots$ with $m_{4},m_5 \cdots \gg m_1,m_2,m_3$ so that $\delta m^2 L/E \simeq \mathcal{O}(1)$ for $L \simeq 10-1000 \,\mathrm{mts}$ corresponding to short baseline experiments.

We begin by considering the $3+2$ scenario from which we will extract the case $3+1$.

\textbf{3+2 case:} In this case, $m_5, m_4 \gg m_1,m_2,m_3$ so that
\be I_{ij} \simeq 1 ~~,~ i,j = 1,2,3~~;~~ I_{i4} = I_{14} ~~;~~I_{i5} = I_{15} ~~,~~ i=1,2,3 \,. \label{Iijotas}\ee Unitarity of the $U$ matrix entails
\be \sum_{i=1}^{3} U^*_{\alpha i} U_{\beta i} = \delta_{\alpha \beta} - U^*_{\alpha4} U_{\beta 4} -U^*_{\alpha 5} U_{\beta 5}\,. \label{unitcons}\ee
Separating the terms with $j=4,5$ in (\ref{finprobab}), we find for $\alpha \neq \beta$ (appearance)
\bea \mathcal{P}_{\alpha \rightarrow \beta} & = & 4|U_{\alpha 4}| |U_{\beta 4}| \Big[ |U_{\alpha 4}| |U_{\beta 4}|+|U_{\alpha 5}|  |U_{\beta 5}| \cos \phi_{54} \Big]\,\frac{1}{2}\,\mathrm{Re}[1-I_{41}]   \nonumber \\  & - & 4  |U_{\alpha 4}| |U_{\beta 4}||U_{\alpha 5}|  |U_{\beta 5}| \cos \phi_{54}\,\frac{1}{2}\,\mathrm{Re}[1-I_{54}] \nonumber \\ & + &  4|U_{\alpha 5}| |U_{\beta 5}| \Big[ |U_{\alpha 5}| |U_{\beta 5}|+|U_{\alpha 4}|  |U_{\beta 4}| \cos \phi_{54} \Big]\,\frac{1}{2}\,\mathrm{Re}[1-I_{51}] \nonumber \\ & + & 2 \Big[ |U_{\alpha 4}| |U_{\beta 4}||U_{\alpha 5}|  |U_{\beta 5}| \sin \phi_{54} \Big]\mathrm{Im}\Big[ \mathrm{I}_{41}-\mathrm{I}_{51}+\mathrm{I}_{54}\Big] \label{32Pneq}\,.
\eea where following \cite{conrad} we have defined
\be \phi_{54} = \mathrm{Arg}\Big[  U_{\alpha 5}   U^*_{\beta 5} U^*_{\alpha 4}  U_{\beta 4}\Big] ~,~\mathrm{for}~\alpha = e \,,\beta= \mu \,,\label{phidef}  \ee and used $\mathrm{Im}[I_{ij}]=-\mathrm{Im}[I_{ji}]$. We note that interchanging $\alpha \leftrightarrow \beta$ ($ e \leftrightarrow \mu$) is equivalent to the exchange $ 4 \leftrightarrow 5$, namely $\phi_{54} \rightarrow - \phi_{54}=\phi_{45}$ which leaves the result (\ref{32Pneq}) invariant since $\mathrm{Re}[I_{ji}]$ is even and $\mathrm{Im}I_{ji}$ odd respectively under $ i \leftrightarrow j$. If  $\phi_{54}\neq 0$, there is CP-violation in the neutrino sector because $\phi_{54} \rightarrow -\phi_{54}$ for $\nu  \rightarrow \nu$ oscillations since this implies that the elements of the mixing matrix $U_{\alpha i} \rightarrow U^*_{\alpha i}$.

The $3+2$ case effectively describes mixing between \emph{three} species; consequently, it features only one \emph{effective} CP-violating angle.

For $\alpha = \beta$ (disappearance), we find

\bea  \mathcal{P}_{\alpha \rightarrow \alpha} & = & 1- 4\Bigg\{ |U_{\alpha 4}|^2 \Big[1-|U_{\alpha 4}|^2-|U_{\alpha 5}|^2 \Big]\,\frac{1}{2}\,\mathrm{Re}[1-I_{41}] \nonumber \\ & + & |U_{\alpha 4}|^2 |U_{\alpha 5}|^2 \,\frac{1}{2}\,\mathrm{Re}[1-I_{54}] \nonumber \\ & + & |U_{\alpha 5}|^2 \Big[1-|U_{\alpha 4}|^2-|U_{\alpha 5}|^2 \Big]\,\frac{1}{2}\,\mathrm{Re}[1-I_{51}]\Bigg\}\,, \label{Palal32}\eea which does not feature a contribution from the CP-violating angle.

 \textbf{3+1 case:} This case is obtained from the $3+2$ case above by setting $U_{\alpha 5}=0 \,\forall \alpha$, leading to the    appearance probability ($\alpha \neq \beta$) ,
 \be  \mathcal{P}_{\alpha \rightarrow \beta}   =  4|U_{\alpha 4}|^2 |U_{\beta 4}|^2  \,\frac{1}{2}\,\mathrm{Re}[1-I_{41}]   \label{31Pneq}\,,\ee and the disappearance (survival) probability ($\alpha = \beta$)
 \be  \mathcal{P}_{\alpha \rightarrow \alpha}   =   1- 4 |U_{\alpha 4}|^2 \Big[1-|U_{\alpha 4}|^2  \Big]\,\frac{1}{2}\,\mathrm{Re}[1-I_{41}]\,. \label{Palal31}\ee The $3+1$ case effectively describes mixing between two generations and, consequently, does not feature any CP-violating contribution. For this case, it is often convenient\cite{book5} to introduce the effective mixing angles
 \bea  \sin^2 2\theta_{\alpha \beta} & \equiv & 4|U_{\alpha 4}|^2 |U_{\beta 4}|^2 ~~,~~\alpha \neq \beta \label{tetaab}\\
 \sin^2 2\theta_{\alpha \alpha} & \equiv & 4|U_{\alpha 4}|^2 \Big[1- |U_{\alpha 4}|^2\Big] ~~,~~\alpha =  \beta \,.\label{tetaaa}\eea

\section{Majorana sterile neutrinos and $|\Delta L|=2$  $\nu \leftrightarrow \overline{\nu}$ oscillations:}\label{sec:majorana} In the previous section we have assumed that sterile neutrinos are of the Dirac variety; however, if neutrinos are Majorana fermions,  new processes, such as neutrino-less double beta decay (see\cite{bilenkygiunti} for  recent reviews) and $\nu \leftrightarrow \overline{\nu}$ oscillations, are available.  As discussed in ref.\cite{moha}, $\nu \leftrightarrow \overline{\nu}$ oscillations have the potential to reveal CP-violating Majorana phases and, to make clear the Majorana nature of the mixing matrix, we write it as $\widetilde{U}$ following eqn. (\ref{mayomat}).

\begin{figure}[ht!]
\begin{center}
\includegraphics[keepaspectratio=true,width=5in,height=5in]{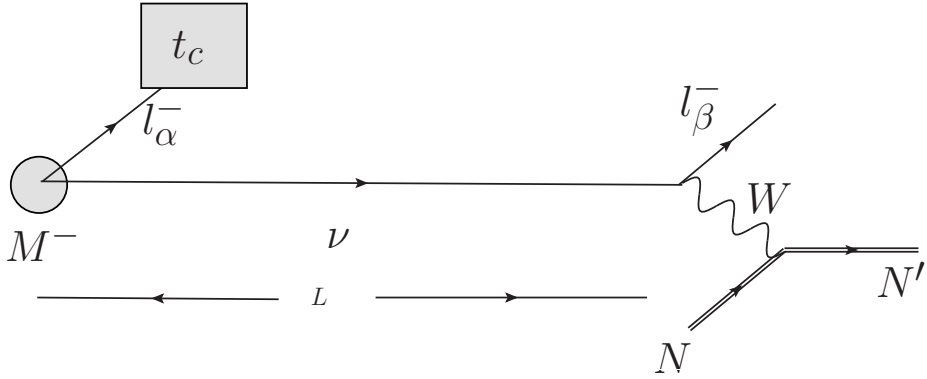}
\caption{$|\Delta L|=2$ process from Majorana neutrinos.   The charged lepton $l^-_{\alpha}$ produced with the  neutrino is observed, absorbed or decays at a time $t_c$ and another charged lepton $l^-_{\beta}$ is detected. }
\label{fig:dl2expt}
\end{center}
\end{figure}

These processes can be understood by considering the full interaction Hamiltonian including the  the hermitian conjugate of the one displayed in  (\ref{Hintdet}), namely

 \bea H_{i} & = &  \widetilde{U}^*_{\beta j} ~\sqrt{2}\,G_F\,\int   \overline{\Psi}_{\nu_j}(\vx)\gamma^\mu \mathbbm{L}\, \Psi_{l_\beta}(\vx) \,\mathcal{J}^{(N,N')}_\mu(\vx) d^3x \nonumber
   \\ & + & \widetilde{U}_{\beta j} ~\sqrt{2}\,G_F\,\int   \overline{\Psi}_{l_\beta}(\vx)\gamma^\mu \mathbbm{L}\, \Psi_{\nu_j}(\vx) \,\mathcal{J}^{\dagger\,(N,N')}_\mu(\vx) d^3x \,.\label{Hintdethc}\eea The first line yields the $\Delta L =0$ $\overline{\nu} \leftrightarrow \overline{\nu}$ oscillations just as for the Dirac case discussed in the previous section. The second line contributes to the detection process \emph{only} for Majorana neutrinos and yields the $|\Delta L|=2$ contribution, as can be simply understood from the following argument pertaining to Majorana fermions: the production Hamiltonian (\ref{Hint}) is determined by charge conservation: a $\pi^-$ decays into a negatively charged lepton $l^-_{\alpha}$, thus requiring the $\overline{\Psi}_{l_\alpha}$ in (\ref{Hint}), the $\Psi_{\nu_j}$ \emph{creates} a neutrino (same as an \emph{antineutrino} for Majoranas) with an operator $\hat{b}^\dagger_{\vk,h}$ that multiplies a charge conjugate spinor $\mathcal{U}^{\,c}_h(\vk)$ (see the expansion (\ref{majorana}) in appendix \ref{app:quant}). Using the first line in (\ref{Hintdethc}), the neutrino is destroyed at the detection vertex using a $\hat{b}_{\vk,h}$ of the $\overline{\Psi}_{\nu_j}$ which \emph{also} multiplies the spinor $\mathcal{U}^{\,c}_h(\vk)$ along with the creation of positively charged lepton $l^+_\beta$. Therefore,  this  $\Delta L=0$ contribution   is the same as that for a Dirac neutrino and features the product $\widetilde{U}_{\alpha j} \widetilde{U}^*_{\beta j}$, which is insensitive to the Majorana phase. However, the neutrino in the intermediate state can also be annihilated by using a $\hat{b}_{\vk,h}$ from   ${\Psi}_{\nu_j}$ in the second line, which now multiplies the spinor $\mathcal{U}_h(\vk)$,  along with the creation of a \emph{negatively charged lepton} $l^-_\beta$ from $\overline{\Psi}_\beta$. This contribution features the product $\widetilde{U}_{\alpha j} \widetilde{U}_{\beta j}= {U}_{\alpha j} {U}_{\beta j}\,e^{i \theta_j }$ and manifestly displays the Majorana phase. This process is depicted in fig.(\ref{fig:dl2expt}).

The CP-conjugate process $\pi^+ \rightarrow l^+_{\alpha}\, \nu \rightarrow \nu N \rightarrow N' \, l^+_{\beta}$ with $|\Delta L|=2$ features the product $\widetilde{U}^*_{\alpha j} \widetilde{U}^*_{\beta j}$ showing that the Majorana phase is also CP-violating. It is convenient to introduce
\be \widetilde{\sigma}^\mu = (1, -\vec{\sigma})\, \label{tilsig}\ee we now find for the transition matrix element $ \mathcal{T}_{i\rightarrow f}$ from the initial ($i=\pi^-\,N$) to the final ($f= l^-_\alpha l^-_\beta N'$)  state
\bea \mathcal{T}_{i\rightarrow f} & = &  \sqrt{2}F_M G_F \sum_{h}\sum_{j}\widetilde{U}_{\alpha j} \widetilde{U}_{\beta j} \Big(\mathcal{U}^\dagger_{\,l_\alpha,h'} \Big)_L \,\Big(\widetilde{\sigma}\cdot J^M\Big)\Big(\mathcal{U}^{\,c}_{h,j}(\vq) \Big)_L \, \mathcal{F}_{\alpha j}\, \Pi^{\mathcal{P}}_{\alpha j} \nonumber \\ & \times &  \Big(\mathcal{U}^\dagger_{\,l_\beta,h''} \Big)_L \,\Big(\widetilde{\sigma}\cdot J^{(N,N')}\Big)\Big(\mathcal{U}_{h,j}(\vq) \Big)_L   G_{\beta j} \, \Pi^{\mathcal{D}}_{\beta j}\label{Tifdl2}\eea where again we suppressed arguments to simplify notation. The sum over helicity states $h$ can be carried out straightforwardly using the results of appendix (\ref{app:quant}), we find (no sum over $\alpha, \beta$)
\be \mathcal{T}_{i\rightarrow f} = \Big[T^{\alpha \beta}_{-+}-T^{\alpha \beta}_{+-}\Big]\, \sum_{j}\widetilde{U}_{\alpha j} \widetilde{U}_{\beta j}  \,\frac{m_j}{2E_j(q)}\,\mathcal{F}_{\alpha j}\,     G_{\beta j} \, , \label{sumhT}\ee where we have introduced
\be T^{\alpha \beta}_{ab} \equiv \sqrt{2}F_M G_F\,2E(q)\, \Pi^{\mathcal{P}}_{\alpha }\,\Pi^{\mathcal{D}}_{\beta}\,\Bigg[\Big(\mathcal{U}^\dagger_{\,l_\alpha,h'} \Big)_L \,\Big(\widetilde{\sigma}\cdot J^M\Big)v_a(\vq)\Bigg]\Bigg[\Big(\mathcal{U}^\dagger_{\,l_\beta,h''} \Big)_L \,\Big(\widetilde{\sigma}\cdot J^{(N,N')}\Big)v_b(\vq) \Bigg]~;~a,b= -,+ \,\label{bigTi}\ee  where the Weyl spinors $v_\pm(q)$ are the helicity eigenstates  (\ref{repspinors}). Here, $E(q)=q$ for \emph{massless neutrinos} and $T^{\alpha \beta}_{ab}$ do not depend on the mass eigenstate label $j$. In arriving at expressions (\ref{sumhT},\ref{bigTi}) we have written $ \Pi^{\mathcal{P}}_{\alpha j}\,\Pi^{\mathcal{D}}_{\beta j} = \Pi^{\mathcal{P}}_{\alpha }\,\Pi^{\mathcal{D}}_{\beta }\, 2E(q)/2E_j(q) $ where the $\Pi^{\mathcal{P}}_{\alpha }\,\Pi^{\mathcal{D}}_{\beta }$ now correspond to the phase space factors (\ref{Piprodu},\ref{pidet}) for \emph{massless neutrinos}, namely $E_j(q) \rightarrow E(q)=q$.

The amplitudes $T^{\alpha \beta}_{ab}$ have a simple   interpretation: $T^{\alpha \beta}_{-+}$ is the amplitude for the combined process $\pi^- \rightarrow l^-_\alpha \nu ~,~ \overline{\nu} N \rightarrow l^-_\beta N'$ and $T^{\alpha \beta}_{+-}$ for the process $\pi^- \rightarrow l^-_\alpha  \overline{\nu} ~,~ {\nu} N \rightarrow l^-_\beta N'$ where $\nu, \overline{\nu}$ are \emph{massless} left handed neutrinos (and right handed antineutrinos), corresponding to the $\nu \leftrightarrow \overline{\nu}$ mixing that violates lepton number by two units. These two amplitudes contribute coherently to the process $\pi^- N \rightarrow l^-_\alpha l^-_\beta N'$ and are added (with their respective signs) in the total amplitude. The mass dependence in the transition amplitude is a consequence of a helicity change in the $\Delta L = 2$ process $\nu \leftrightarrow \bar{\nu}$.

The expression (\ref{sumhT}) is generally valid for arbitrary masses of Majorana sterile neutrinos and, with a simple modification of the final state, also describes the $\Delta L = 2$ processes studied in ref.\cite{dibmajo}.

 Proceeding as in the $\Delta L =0 $ case, we finally find for the transition \emph{rate}
\be  \frac{d}{dt_D} |\mathcal{T}_{i\rightarrow f}|^2 = \Upsilon^{\alpha \beta}\,P^{|\Delta L|=2}_{\alpha \rightarrow \beta}~~,~~\mathrm{no~sum~over} \,\alpha,\beta \, \label{finTdl2rate}  \ee where

\be P^{|\Delta L|=2}_{\alpha \rightarrow \beta} =  \sum_{j,i} \widetilde{U}_{\alpha j} \widetilde{U}_{\beta j} \widetilde{U}^*_{\alpha i} \widetilde{U}^*_{\beta i} \,\frac{m_j m_i}{4E_jE_i} \,I_{ij} \,, \label{probadl2} \ee is the $\nu  \leftrightarrow \overline{\nu}$ transition probability with $|\Delta L|=2$ and

\be \Upsilon^{\alpha \beta} =   \Big|T^{\alpha \beta}_{-+}-T^{\alpha \beta}_{+-}\Big|^2\, \frac{ \big[\Pi^{\mathcal{P}}_\alpha \big]^2 }{\Gamma_M(p)}~ \Big[1- e^{-\Gamma_M(p)t_c} \Big] ~2\pi\,\delta\big( \mathcal{E}_{\produ}) \,\Big[\Pi^{\mathcal{D}}_\beta \, \Big]^2 \, 2\pi\,\delta({\mathcal{E}_{\mathcal{D}}})\,  \ee encodes the transition matrix elements for production and detection. We note that unlike the $\Delta L=0$ case, here there is \emph{no factorization} of production and detection, this is a consequence of the fact that $\nu \leftrightarrow \overline{\nu}$ oscillation implies helicity change (and a mass insertion) and both helicity changing contributions contribute coherently to the total amplitude as explained above.  A similar observation was pointed out in ref.\cite{dele}. We are not concerned here with $\Upsilon^{\alpha \beta}$ but with the transition probability $P^{|\Delta L|=2}_{\alpha \beta}$, which can be written as
\bea P^{|\Delta L|=2}_{\alpha \rightarrow \beta} =  \sum_{j} | {U}_{\alpha j}|^2 \,| {U}_{\beta j}|^2  \,\frac{m^2_j}{4E^2_j}  & + &  2 \sum_{j>i}   \mathrm{Re}[ \widetilde{U}_{\alpha j} \widetilde{U}_{\beta j} \widetilde{U}^*_{\alpha i} \widetilde{U}^*_{\beta i}] \,\frac{m_j m_i}{4E_jE_i} \,\mathrm{Re}[I_{ji}]\nonumber \\ & + &  2 \sum_{j>i}   \mathrm{Im} [\widetilde{U}_{\alpha j} \widetilde{U}_{\beta j} \widetilde{U}^*_{\alpha i} \widetilde{U}^*_{\beta i}] \,\frac{m_j m_i}{4E_jE_i} \,\mathrm{Im}[I_{ji}] \,, \label{probadl2reim} \eea

  Just as in the $\Delta L=0$ case, the main difference with the usual quantum mechanical case is the replacement
\be  e^{i\Delta_{ij}L} \rightarrow  I_{ij} = e^{i\Delta_{ij}L} \,\Bigg[\frac{ 1-e^{-i\Delta_{ij}L_c}\,e^{-\Gamma_M(p)L_c}  }{ 1- e^{-\Gamma_M(p)L_c} }\Bigg]\,\Bigg[\frac{1-i\mathcal{R}_{ij}}{1+ \mathcal{R}^2_{ij}}\Bigg]\,, \label{repla}  \ee where $\Delta_{ij};\mathcal{R}_{ij}$ are given by eqn. (\ref{defisdel}) where the extra factors describe decoherence effects associated with the lifetime of the decaying meson and the measurement of the charged lepton partner of the produced neutrino.

To be sure \emph{if} the absolute mass scale of the new generation of sterile neutrinos is $\simeq \mathrm{eV}$ then the factor $\simeq m^2/E^2 \lesssim 10^{-14}$ makes the $|\Delta L|=2$ contribution all but unobservable with the current (and foreseeable) facilities for short-baseline experiments with $m \simeq \mathrm{eV}$. However, oscillation experiments measure the squared mass \emph{differences}; therefore, in absence of a determination of the absolute scale of masses, there remains the possibility that new generation of sterile neutrinos may be heavy  but nearly degenerate so that the difference in squared masses is small and lead to interference and oscillations on the length scales of short baseline experiments and $P^{|\Delta L|=2}_{\alpha \rightarrow \beta}$ is not negligible .

\textbf{3+2 and 3+1 schemes:} Under the assumption that $m_4,m_5 \gg m_i, i=1,2,3$, the contribution from active-like mass eigenstates is clearly subleading for the $|\Delta L|=2$ transitions; therefore, keeping only the two largest mass eigenstates

\bea P^{|\Delta L|=2}_{\alpha \rightarrow \beta}  & = &    | {U}_{\alpha 5}|^2 \,| {U}_{\beta 5}|^2  \,\frac{m^2_5}{4E^2_5} +  | {U}_{\alpha 4}|^2 \,| {U}_{\beta 4}|^2  \,\frac{m^2_4}{4E^2_4}+\cdots\nonumber \\
 & + &  2   | {U}_{\alpha 5}| | {U}_{\beta 5}| | {U}_{\alpha 4}| | {U}_{\beta 4}|\,\cos(\delta_{54}+\theta_{54}) \,\frac{m_5 m_4}{4E_5E_4} \,\mathrm{Re}[I_{54}] +\cdots \nonumber \\
&+&  2   | {U}_{\alpha 5}| | {U}_{\beta 5}| | {U}_{\alpha 4}| | {U}_{\beta 4}|\,\sin(\delta_{54}+\theta_{54}) \,\frac{m_5 m_4}{4E_5E_4} \,\mathrm{Im}[I_{54}]+\cdots\label{Pdell2fin}\eea where the dots stand for the contributions from $i=1,2,3$,  $U$ is the Dirac mixing matrix (\ref{mayomat}) and
\be \delta_{54}= \mathrm{Arg}\Big[  U_{\alpha 5}   U_{\beta 5} U^*_{\alpha 4}  U^*_{\beta 4}\Big] ~,~\mathrm{for}~\alpha = e \,,\beta= \mu  \label{delta54} \ee is a \emph{Dirac CP-violating phase
 different} from the $\phi_{54}$ that enter in the $\Delta L=0$ case (\ref{phidef}) and $\theta_{54}= \theta_5-\theta_4$ with $\theta_j$ the Majorana CP-violating phases (\ref{mayomat}).

 The $3+1$ scheme is obtained simply by setting $U_{\alpha 5}=0\,\forall \alpha$ in which case there are no oscillations to leading order in $m/E$.

\section{Analysis of decoherence effects in accelerator experiments:}

The decoherence effects associated with the lifetime of the source and the measurement (or capture) length scale of the charged lepton emitted with the (anti) neutrino are encoded in the quantities $\mathrm{Re}[I_{ji}]~,~ \mathrm{Im}[I_{ji}]$ given by eqns. (\ref{realpartIij},\ref{imagpartIij}) the latter one determines the suppression of the CP violating contributions from these decoherence effects. In this section we compare these terms to the familiar ones obtained from the quantum mechanical description of neutrino oscillations (\ref{realimagusual}) as a function of the neutrino energy for fixed baselines.

  Introducing
  \be \Delta_{ji}(E) = \frac{\delta m^2_{ji}}{2E} ~~;~~ \mathcal{R}_{ji}= \frac{\delta m^2_{ji}}{2E\Gamma_M(p)} ~~;~~\delta m^2_{ji}= m^2_j-m^2_i\,, \label{deltaR}\ee  and replacing as usual
  \be t_D \rightarrow L ~~;~~ t_c \rightarrow L_c \label{distances}\ee  we find
  \bea  \mathrm{Re}[I_{ji}]  & = &  \frac{1}{1+\mathcal{R}^2_{ji}}~\frac{1}{1-e^{-\Gamma_M(p)L_c}}~\Bigg[ \Bigg(\cos\Big[ \frac{\delta m^2_{ji}}{2E}\,L\Big]+\mathcal{R}_{ji}\sin\Big[\frac{\delta m^2_{ji}}{2E}\,L\Big]\Bigg)-\nonumber \\ &&
e^{-\Gamma_M(p)L_c}\Bigg(\cos\Big[\frac{\delta m^2_{ji}}{2E}\,(L-L_c)\Big]+\mathcal{R}_{ji}\sin\Big[ \frac{\delta m^2_{ji}}{2E}\,(L-L_c)\Big]\Bigg) \Bigg]   \,,\label{realpartIij}\eea

 \bea   \mathrm{Im}[I_{ji}]  & = &  \frac{1}{1+\mathcal{R}^2_{ji}}~\frac{1}{1-e^{-\Gamma_M(p)L_c}}~\Bigg[ \Bigg(\sin\Big[ \frac{\delta m^2_{ji}}{2E}\,L\Big]-\mathcal{R}_{ji}\cos\Big[ \frac{\delta m^2_{ji}}{2E}\,L\Big]\Bigg)-\nonumber \\ &&
e^{-\Gamma_M(p)L_c}\Bigg(\sin\Big[ \frac{\delta m^2_{ji}}{2E}\,(L-L_c)\Big]-\mathcal{R}_{ji}\cos\Big[ \frac{\delta m^2_{ji}}{2E}\,(L-L_c)\Big]\Bigg) \Bigg]\,.   \label{imagpartIij}\eea we note that
\be \frac{\delta m^2_{ji}}{2E}\,L_c \equiv \mathcal{R}_{ji}\,\Gamma_M(p)L_c\label{equivalencia}\ee this relation  highlights that there are only two combination of parameters that determine the corrections, namely $\mathcal{R}_{ji}$ and $\Gamma_M(p)L_c$; furthermore, $\Gamma_M(p)L_c \equiv L_c/l_M(p)$ where $l_M(p)$ is the decay length of the meson in the laboratory frame. We would like to point out that similar results have been obtained in refs\cite{hernandez} in which wave packets are analyzed throughout the production/detection process whereas our results are obtained in a completely different manner. In our treatment, we did not attempt to include localization wavepackets for the pion and, in the WW treatment, the pions would be the only source where an introduction of wavepackets would be appropriate. The usual decay matrix elements for pion decay from quantum field theory were used and a full discussion of wavepackets is available in refs\cite{hernandez}.

In absence of the decoherence contributions, the usual expressions emerge, namely

\be   \mathrm{Re}[ I_{ji}]    =   \cos\Big[\frac{\delta m^2_{ji}\,L}{2E}\Big] ~~;~~ \mathrm{Im}[I_{ji}]    =   \sin \Big[ \frac{\delta m^2_{ji}\,L}{2E}\Big] \,, \label{realimagusual}\ee with
\be \frac{\delta m^2_{ji}\,L}{2E} = 2.54 \Bigg( \frac{\delta m^2_{ji}}{\mathrm{eV}^2}\Bigg)\,\Bigg( \frac{L}{\mathrm{km}}\Bigg)\,\Bigg( \frac{\mathrm{GeV}}{E}\Bigg)\,.\label{equiv}\ee

Whereas the length scale $L_c$ is determined by the particular experimental setting and is therefore a parameter, the width of the parent particle is a function of the neutrino energy through the Lorentz factor as follows.

In the rest frame of the decaying meson, its width  is $\Gamma_M$ and the antineutrino (neutrino) is emitted isotropically with an energy $  E^*_j = \sqrt{{q^*}^2+m^2_j}$ with $q^*$ given by (\ref{qstar}); in the laboratory frame, where the meson is moving with velocity $V_M$, the width is $\Gamma_M/\gamma$ and the energy of an anti (neutrino) collinear with the meson is blue shifted to
\be E  = \gamma E^* (1 + V_M) \label{Elab}\ee where we have neglected the mass of the neutrino. Therefore \be \gamma(E) = \frac{E^2+{E^*}^2}{2 E E^*}~~;~~E^* < E \label{gammaE}\ee hence
\be \mathcal{R}_{ji}(E) =   \frac{\delta m^2_{ji}}{4E^*\,\Gamma_M}\, \Bigg(1+\frac{ {E^*}^2}{E^2}\Bigg)\,. \label{ratiolab} \ee In the analysis below, we focus on neutrinos from Pion decay and the analysis for Kaon decay is similar. Using the Pion decay width, $\Gamma_\pi= 2.5\times 10^{-8}\,\mathrm{eV}$, as a benchmark, we obtain
\be \mathcal{R}_{ji}(E) = \frac{1}{3}\Bigg(\frac{\delta m^2_{ji}}{\mathrm{eV}^2}\Bigg)\,\Bigg(\frac{30\,\mathrm{MeV}}{E^*}\Bigg)\,
\Bigg(\frac{\Gamma_\pi}{\Gamma_M}\Bigg)\,\Bigg(1+\frac{ {E^*}^2}{E^2}\Bigg)\,. \label{ratiolabpion} \ee

An illuminating interpretation of the results (\ref{realpartIij},\ref{imagpartIij}) emerges by defining\footnote{Note that $\mathrm{sign}\big(\theta_{ij}\big) = \mathrm{sign}\big(\delta m^2_{ij}\big)$.}
\be \cos[\theta_{ji}(E)] = \frac{1}{\sqrt{1+\mathcal{R}^2_{ji}}}~~;~~\sin[\theta_{ji}(E)] = \frac{\mathcal{R}_{ji}}{\sqrt{1+\mathcal{R}^2_{ji}}}\,,  \label{thetaji}\ee in terms of which we find
\be  \mathrm{Re}[I_{ji}] = \frac{1}{\sqrt{1+\mathcal{R}^2_{ji}}}~\frac{1}{1-e^{-\Gamma_M(p)L_c}}~\Bigg\{ \cos\Big[ \frac{\delta m^2_{ji}}{2E}\,L-\theta_{ji}(E)\Big]-e^{-\Gamma_M(p)L_c}\,\cos\Big[\frac{\delta m^2_{ji}}{2E}\,(L-L_c)-\theta_{ji}(E)\Big]\Bigg\} \label{renew}\ee

\be  \mathrm{Im}[I_{ji}] = \frac{1}{\sqrt{1+\mathcal{R}^2_{ji}}}~\frac{1}{1-e^{-\Gamma_M(p)L_c}}~\Bigg\{ \sin\Big[ \frac{\delta m^2_{ji}}{2E}\,L-\theta_{ji}(E)\Big]-e^{-\Gamma_M(p)L_c}\,\sin\Big[\frac{\delta m^2_{ji}}{2E}\,(L-L_c)-\theta_{ji}(E)\Big]\Bigg\} \,.\label{imnew}\ee

While the general case must be studied numerically, the limit $\Gamma_M L_c \gg 1$ provides a most clear assessment: as compared to the usual quantum mechanical expression (\ref{realimagusual}), the decoherence factors result in i) a suppression of the transition probabilities $\simeq 1/\sqrt{1+\mathcal{R}^2_{ji}}$ and ii) an overall energy dependent phase shift $\theta_{ji}(E)$.

For  example, for   a sterile neutrino   mass $m_s \gtrsim 1\,\mathrm{eV} \gg m_{1,2,3}$  from $\pi$ decay, it follows that $1 \lesssim \mathcal{R}$ thus  from (\ref{ratiolabpion},\ref{thetaji}) $\pi/4 \lesssim\theta_{ji}\lesssim \pi/2$. Trying to fit the mass (and mixing angles) by using the usual expression  (\ref{realimagusual}) would imply an effective  $\delta m^2_{eff} = \delta m^2 - 2  E\,\theta(E)/L$. For example, for accelerators experiments with $E \simeq \mathrm{GeV}~,~L\simeq 1\,\mathrm{Km}$,   such a fit  would lead to $2  E\,\theta(E)/L \simeq \mathrm{eV}^2$ and a \emph{large underestimate of   the sterile neutrino mass and the mixing and CP-violating angle}.

A similar interpretation holds for the  imaginary part (\ref{imnew}), which is associated with CP-violating amplitudes, the suppression factor would result in an \emph{underestimate} of CP-violation if the usual quantum mechanical expression (\ref{realimagusual}) is used in fitting experimental data. For both cases, if the product $\Gamma_M L_c \gtrsim 1$ then the usual quantum mechanical formulae will not be valid and decoherence effects must be considered.

This simpler case illustrates that for short baseline accelerator experiments in which neutrinos are produced from the decay of pions and are designed to reveal oscillations of new generations of sterile neutrinos with masses in the $\mathrm{eV}$ range,  the decoherence aspects associated with the pion lifetime and the stopping length scale of the muon comparable to the decay length of the pion may lead to substantial corrections to the quantum mechanical oscillation probabilities. A more reliable assessment is obtained numerically below for different experimental situations and, in these investigations, we focus on sterile mass ranges that are relevant for current accelerator searches rather than masses relevant to structure formation.

\vspace{1mm}

\begin{figure}[h!]
\begin{center}
\includegraphics[keepaspectratio=true,width=3.2 in,height=3.5  in]{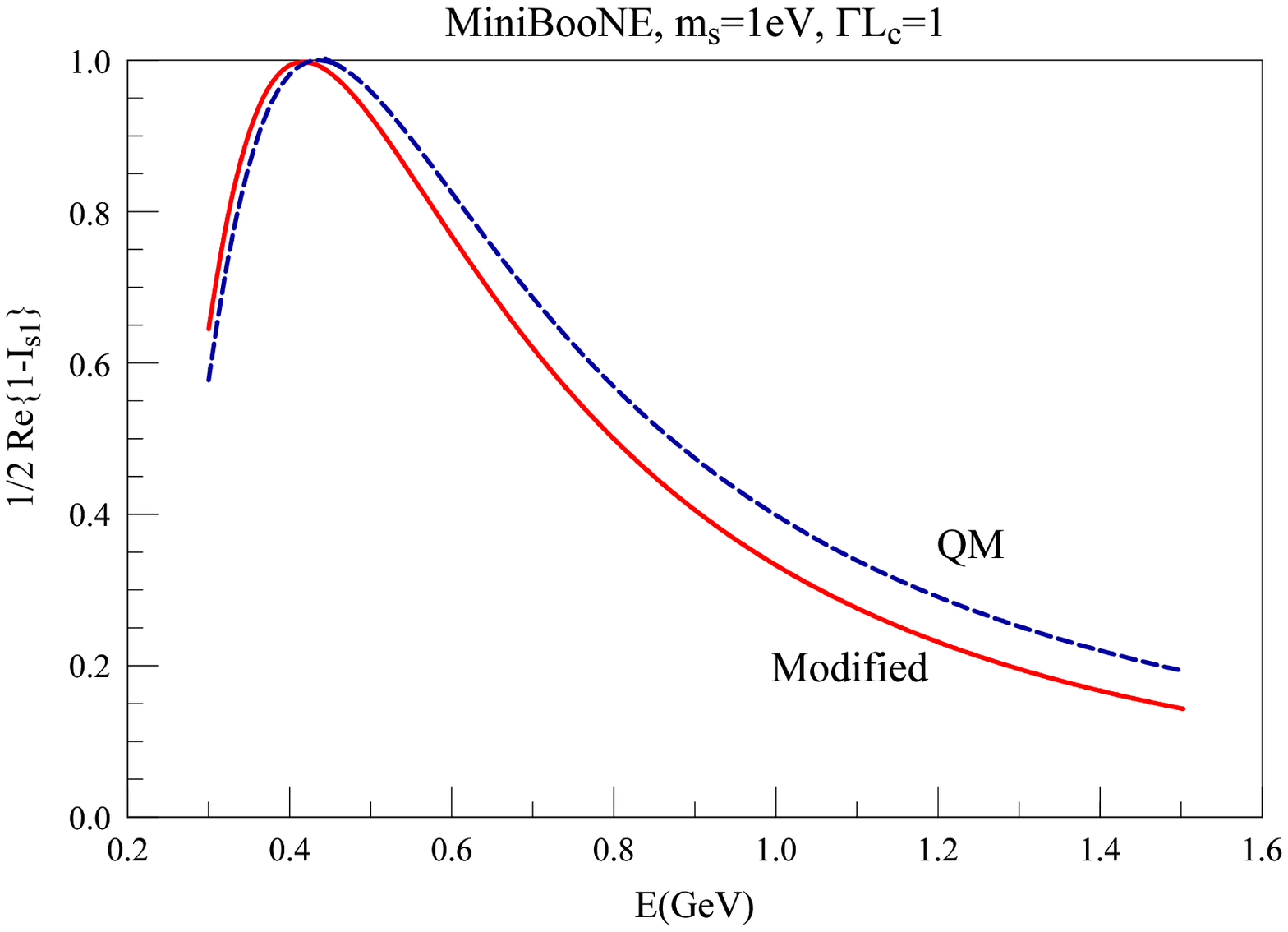}
\includegraphics[keepaspectratio=true,width=3.2  in,height=3.5  in]{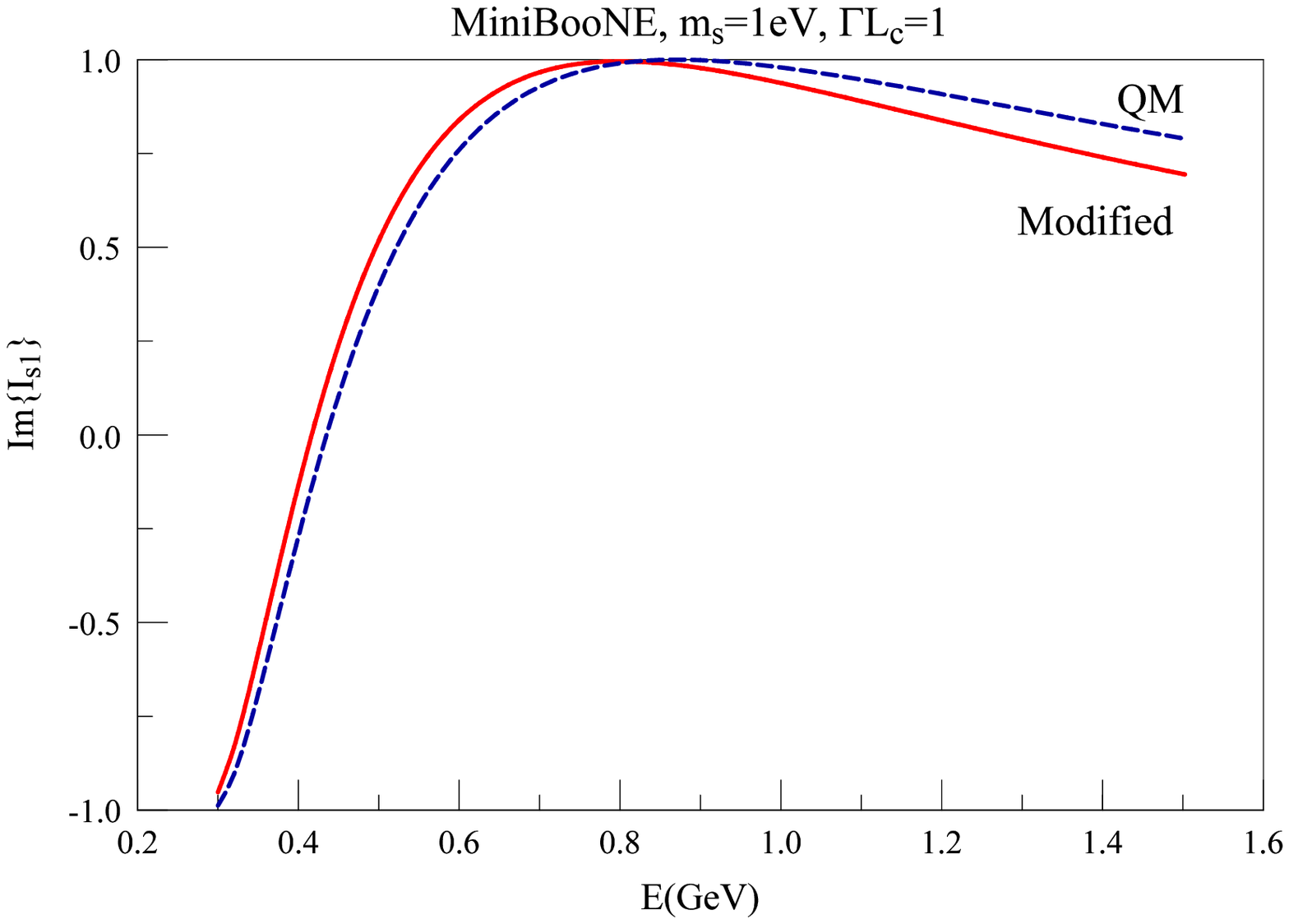}
\caption{CP-even/odd parts of transition probability for MiniBooNE parameters: $L=540 \mathrm{m}, \Gamma_\pi L_c \simeq 1$ for $m_s = 1  \mathrm{eV}$. Solid line (modified) $\mathrm{Re}[1-I_{s1}]/2$ dashed line (Qm) is the quantum mechanical result $\sin^2[m^2_s/4E]$. }
\label{fig:mini1eV}
\end{center}
\end{figure}

\begin{figure}[h!]
\begin{center}
\includegraphics[keepaspectratio=true,width=3.2 in,height=3.5 in]{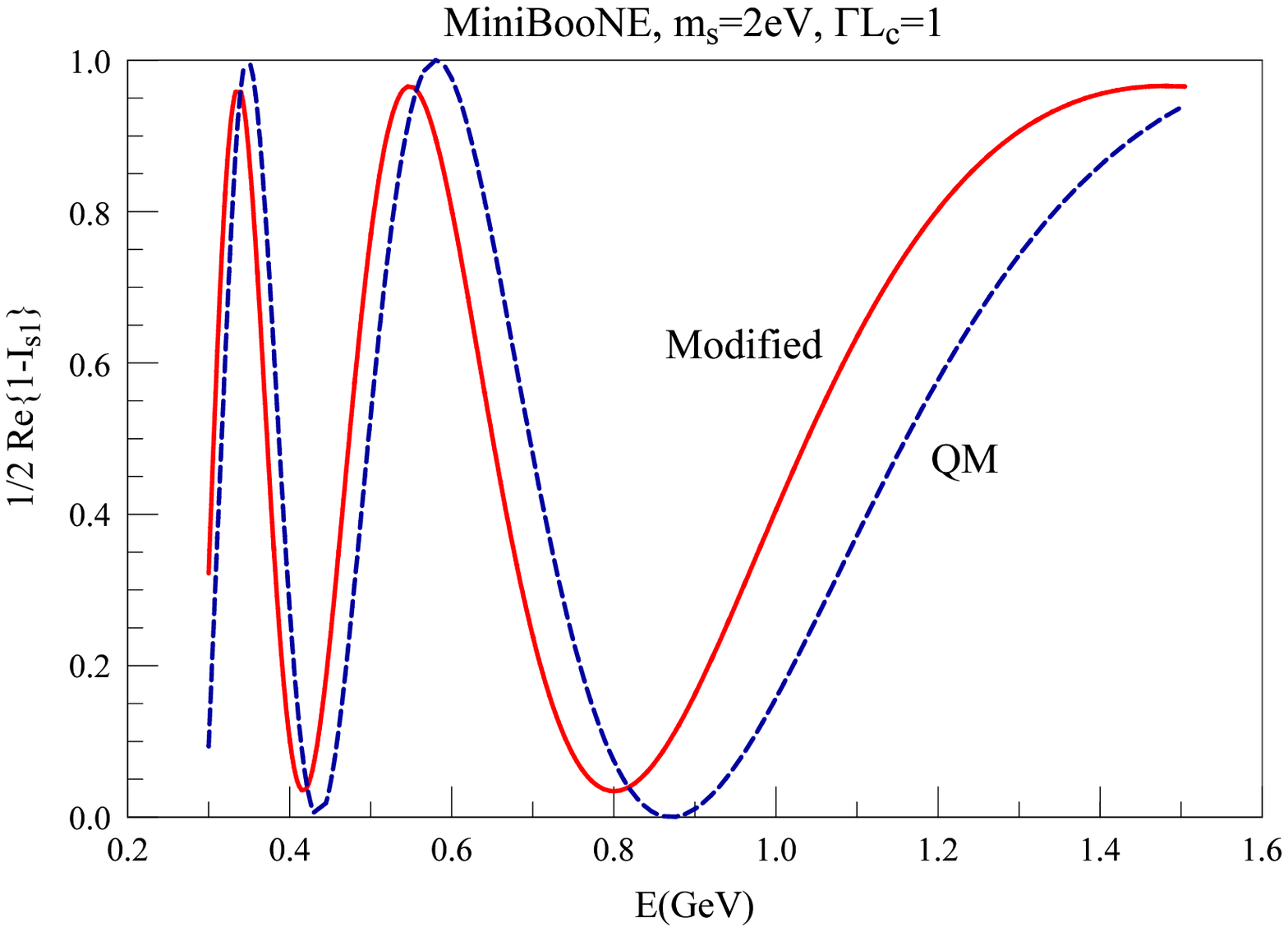}
\includegraphics[keepaspectratio=true,width=3.2 in,height=3.5 in]{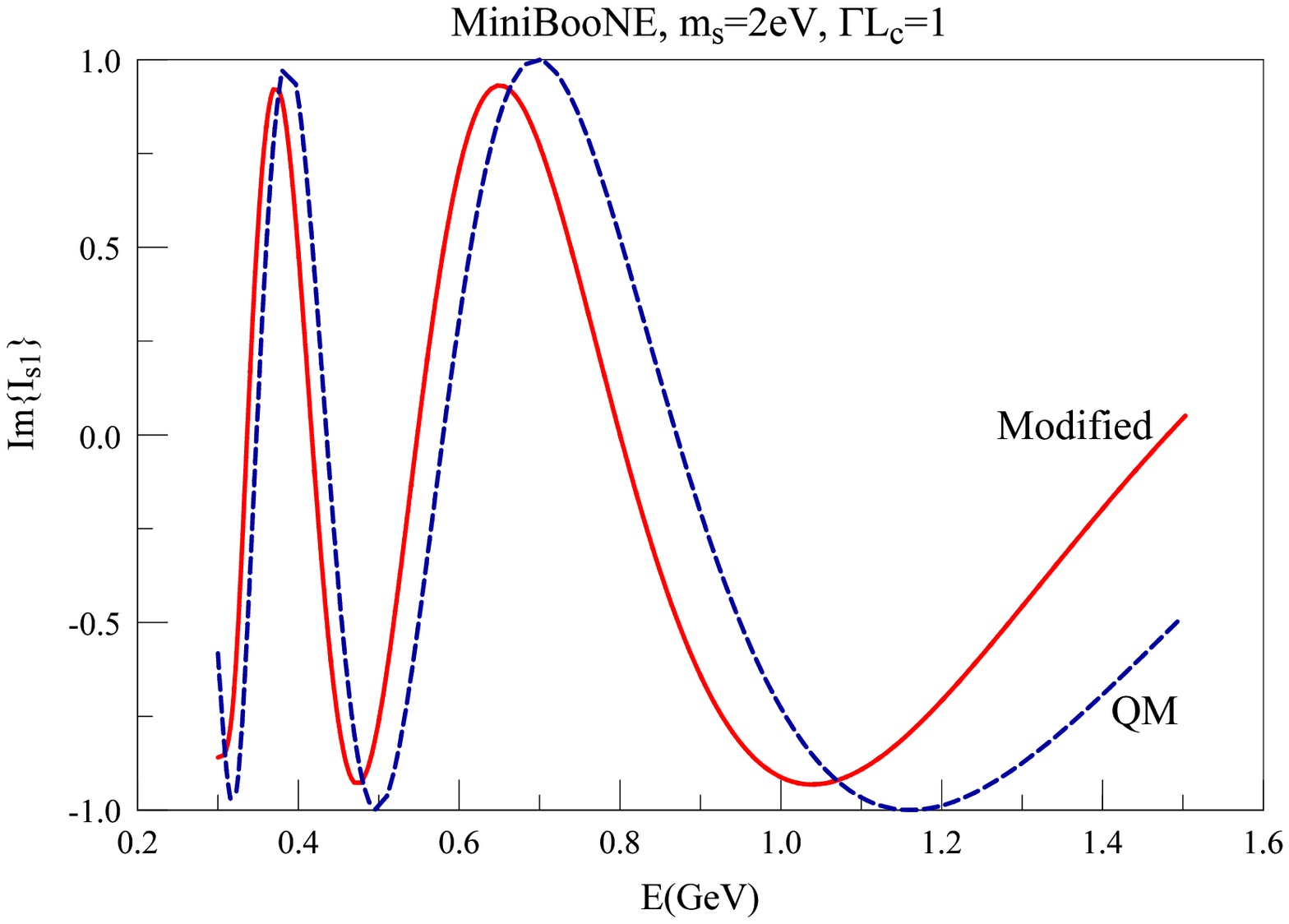}
\caption{Same as Fig.(\ref{fig:mini1eV}) for MiniBooNE for $m_s = 2  \mathrm{eV}$.   }
\label{fig:mini2eV}
\end{center}
\end{figure}

\begin{figure}[h!]
\begin{center}
\includegraphics[keepaspectratio=true,width=3.2 in,height=3.5 in]{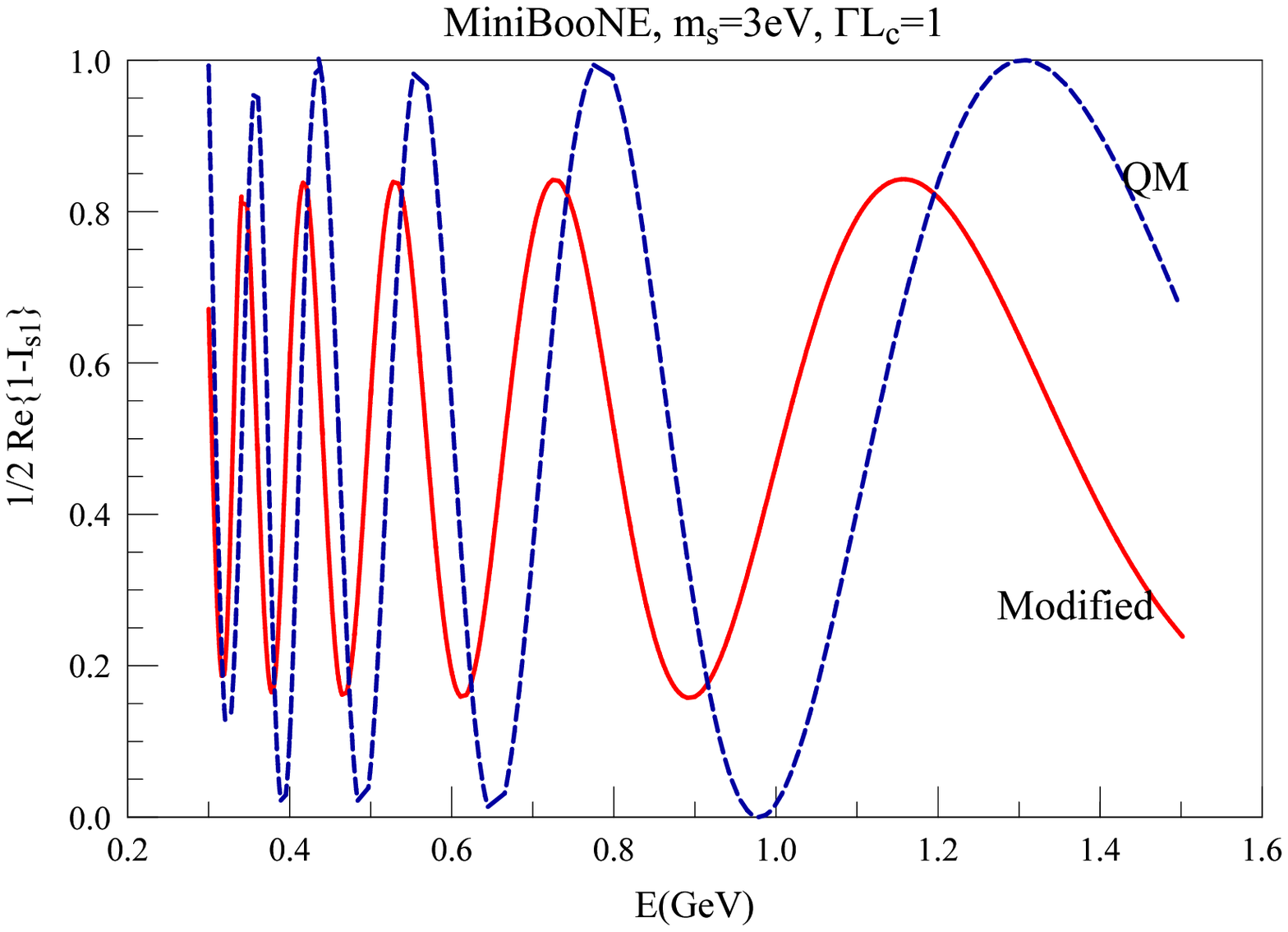}
\includegraphics[keepaspectratio=true,width=3.2 in,height=3.5 in]{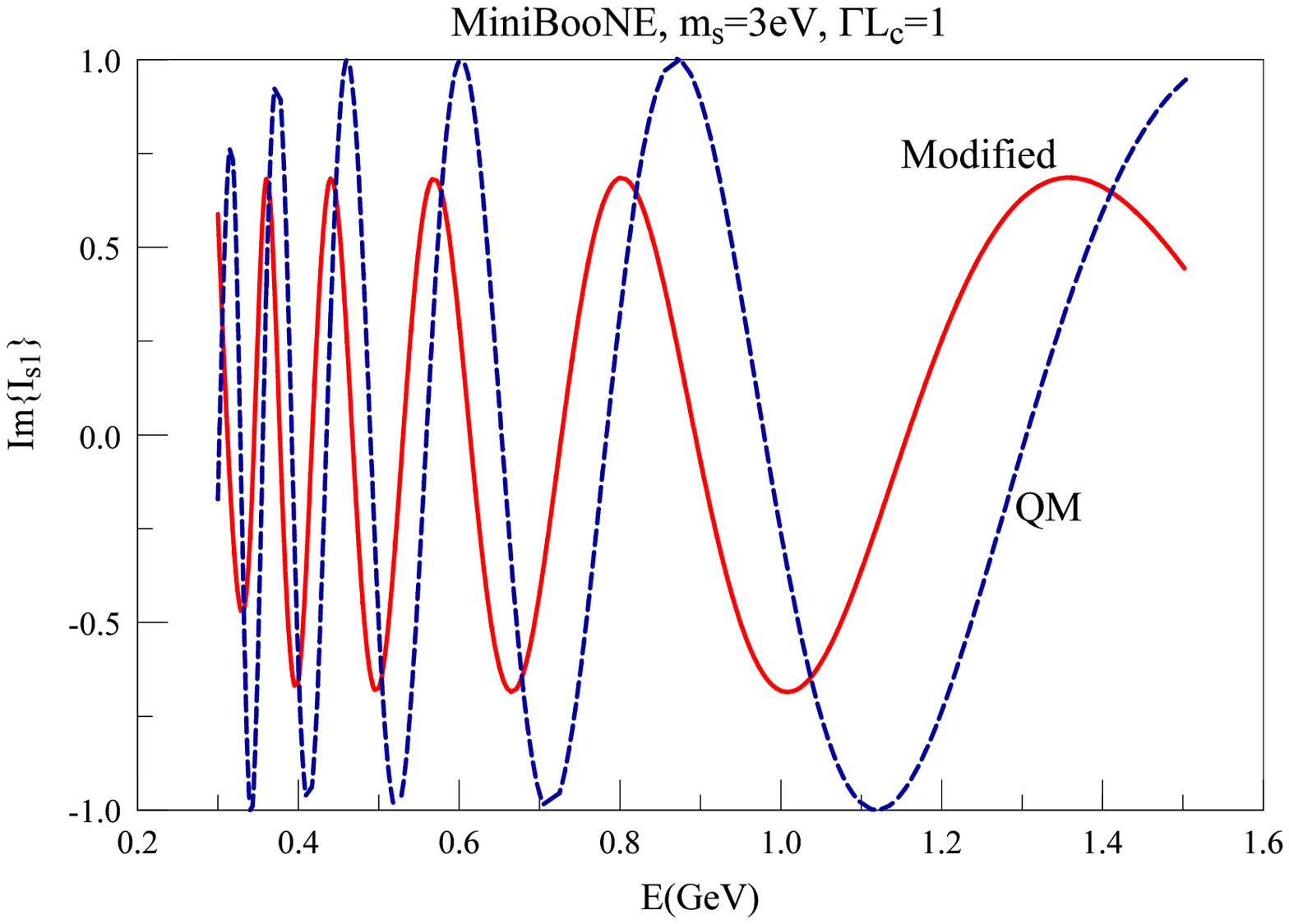}
\caption{Same as Fig.(\ref{fig:mini1eV}) for MiniBooNE for $m_s = 3  \mathrm{eV}$.  }
\label{fig:mini3eV}
\end{center}
\end{figure}

\begin{figure}[h!]
\begin{center}
\includegraphics[keepaspectratio=true,width=3.2 in,height=3.5  in]{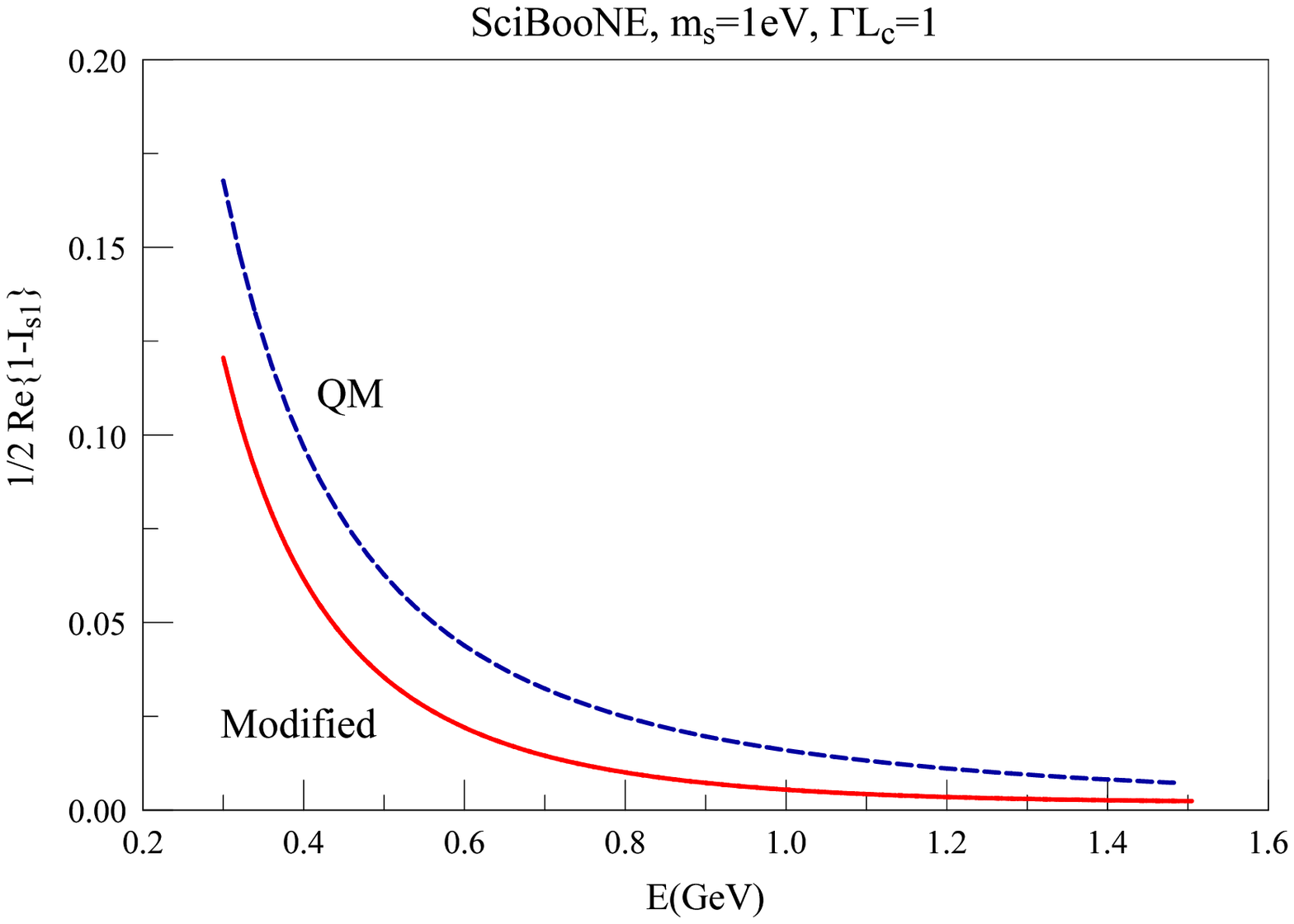}
\includegraphics[keepaspectratio=true,width=3.2  in,height=3.5  in]{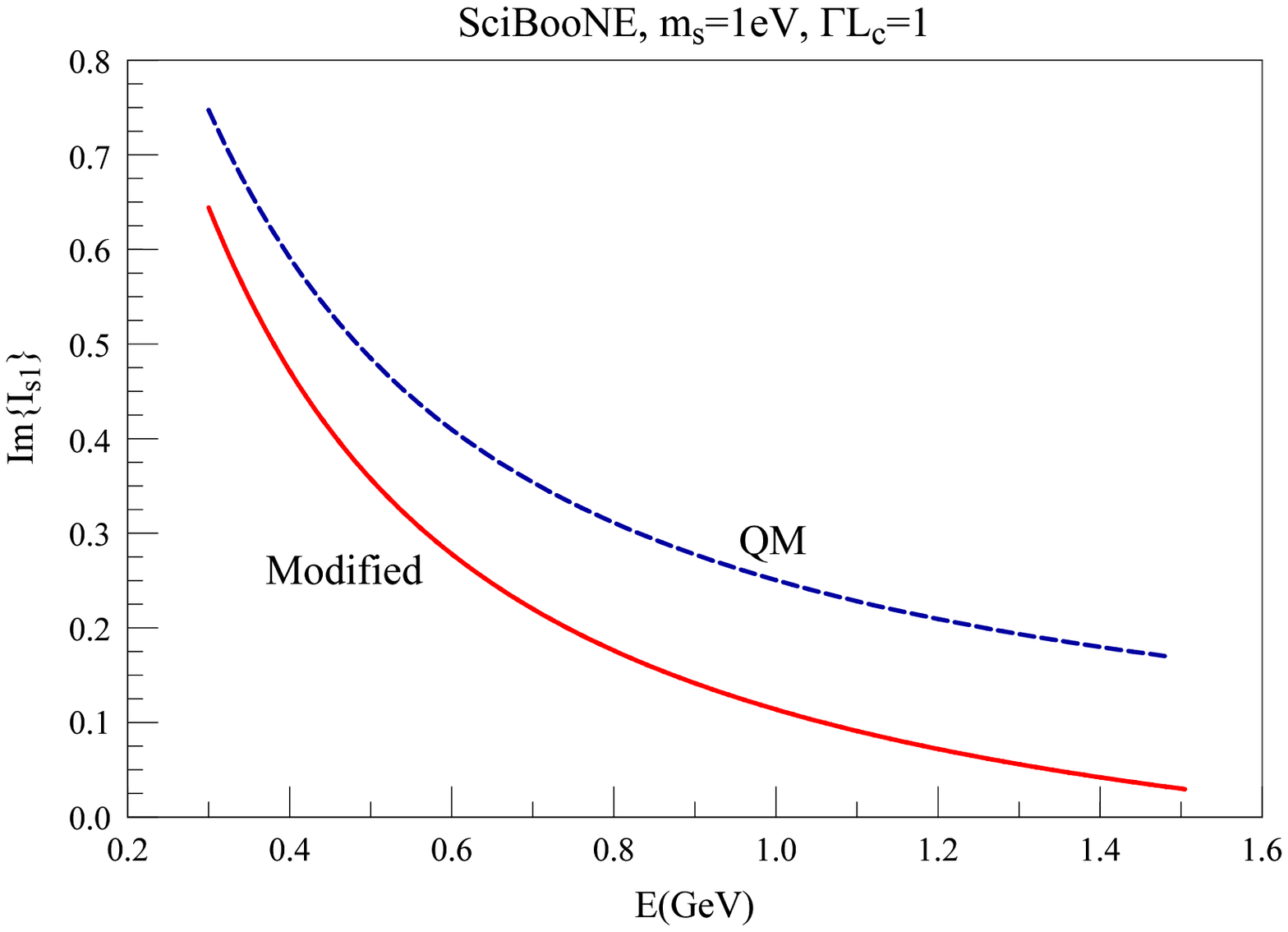}
\caption{CP-even/odd parts of transition probability for SciBooNE parameters: $L=100 \mathrm{m}, \Gamma_\pi L_c \simeq 1$ for $m_s = 1  \mathrm{eV}$. Solid line (modified) $\mathrm{Re}[1-I_{s1}]/2$ dashed line (Qm) is the quantum mechanical result $\sin^2[m^2_{\nu_s}/4E]$. }
\label{fig:sci1eV}
\end{center}
\end{figure}

\begin{figure}[h!]
\begin{center}
\includegraphics[keepaspectratio=true,width=3.2 in,height=3.5  in]{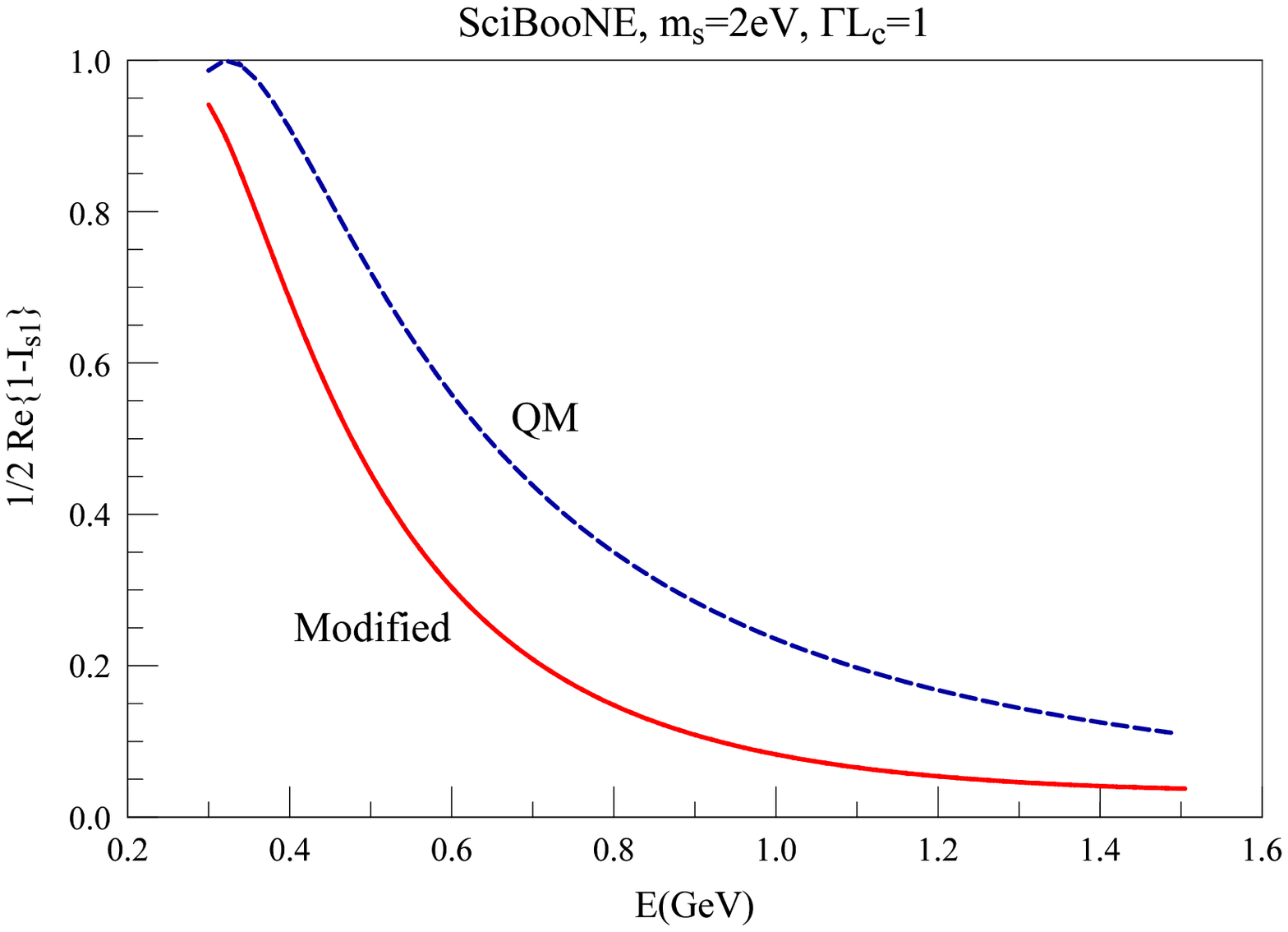}
\includegraphics[keepaspectratio=true,width=3.2  in,height=3.5  in]{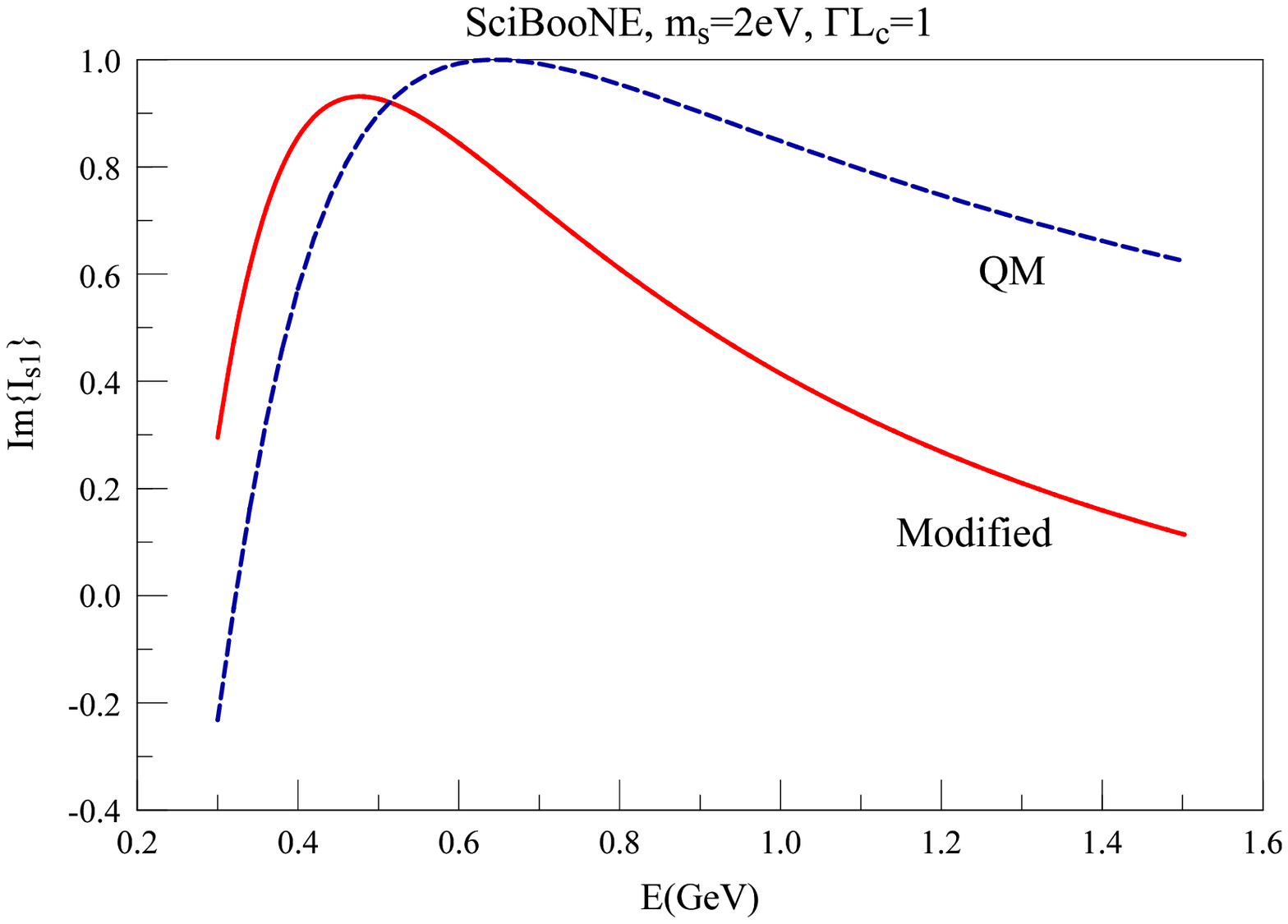}
\caption{Same as Fig. (\ref{fig:sci1eV}) for SciBooNE for $m_s = 2  \mathrm{eV}$.  }
\label{fig:sci2eV}
\end{center}
\end{figure}

\begin{figure}[h!]
\begin{center}
\includegraphics[keepaspectratio=true,width=3.2 in,height=3.5  in]{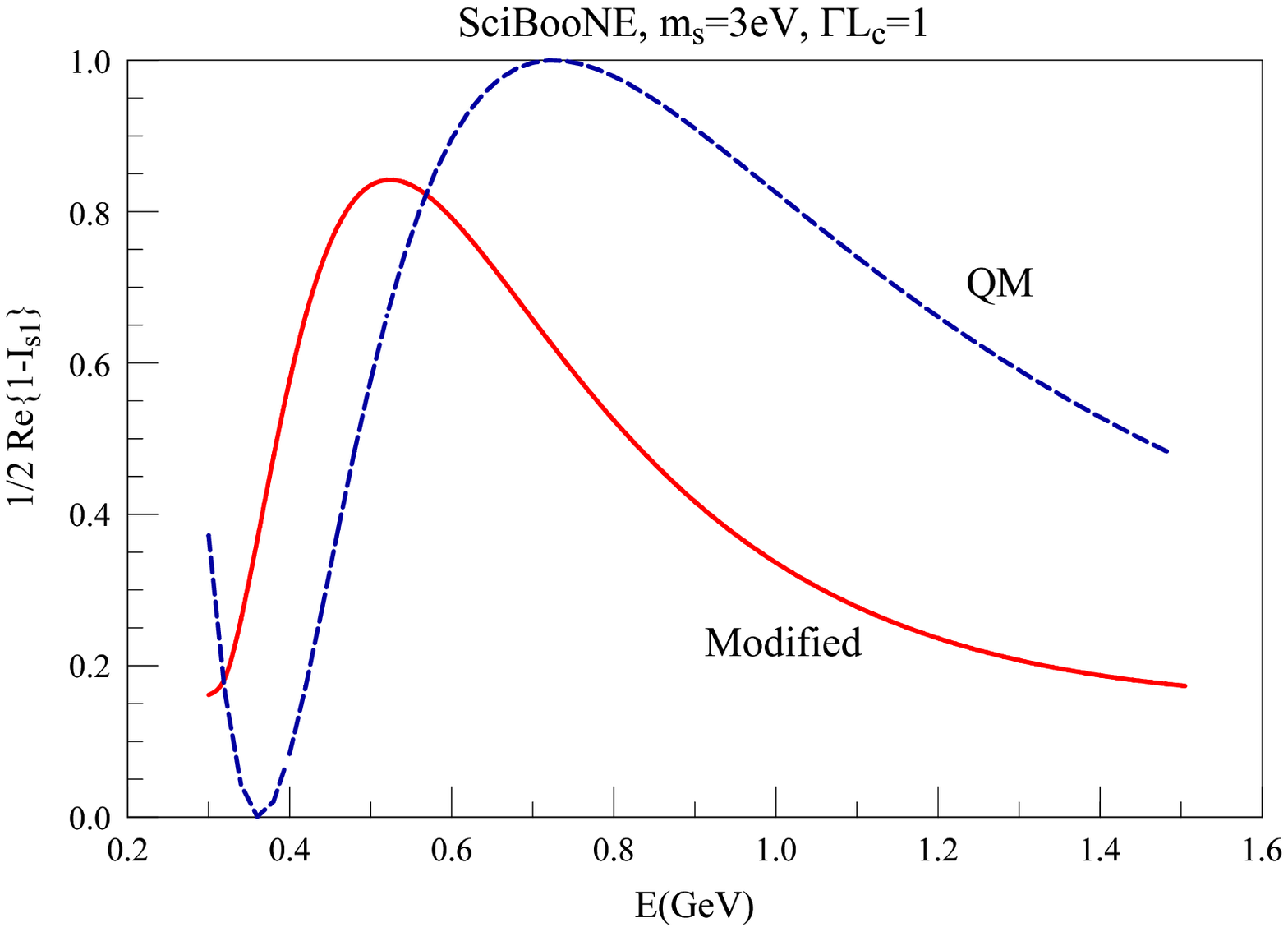}
\includegraphics[keepaspectratio=true,width=3.2  in,height=3.5  in]{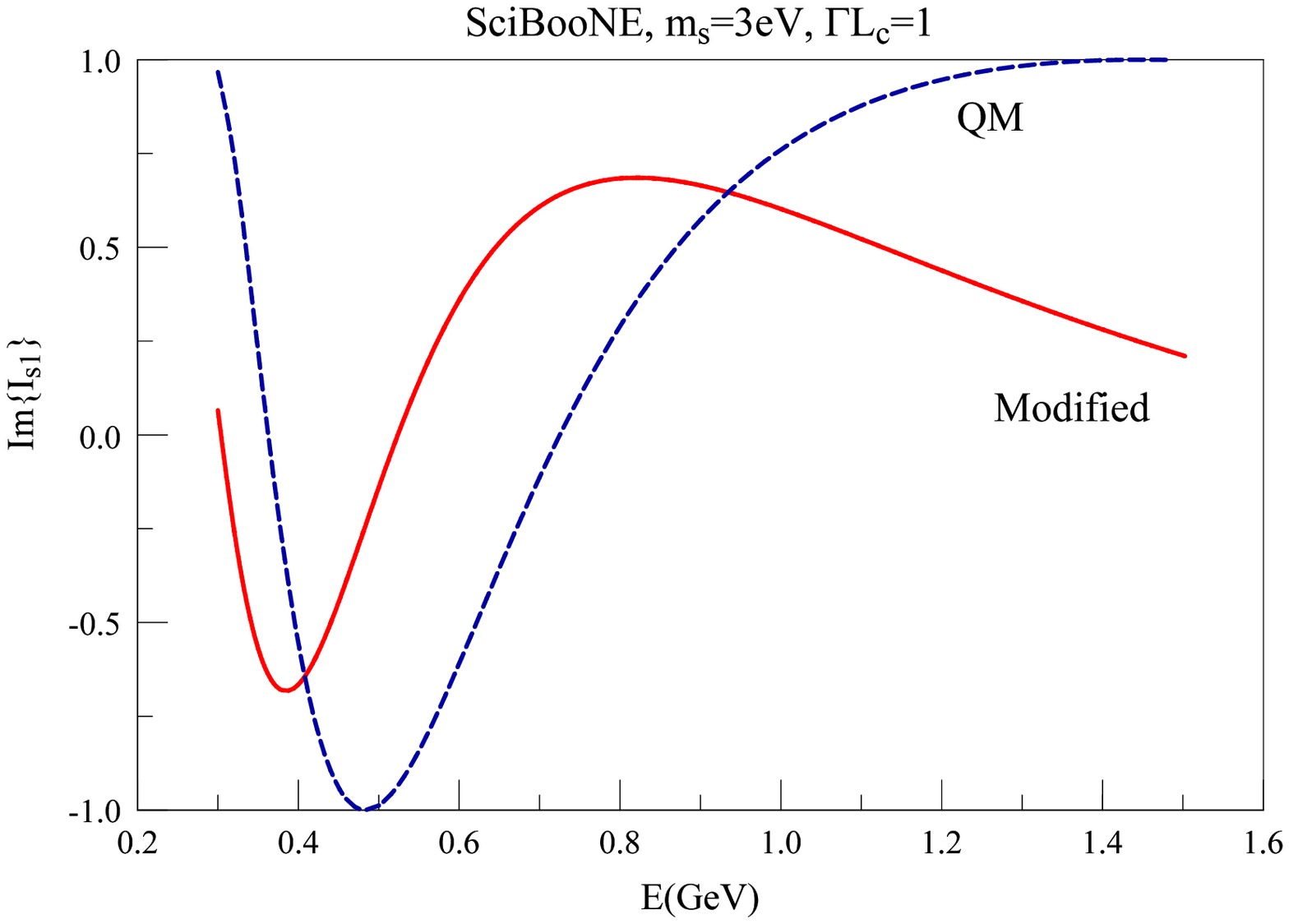}
\caption{Same as Fig. (\ref{fig:sci1eV}) for SciBooNE for $m_s = 3 \mathrm{eV}$. }
\label{fig:sci3eV}
\end{center}
\end{figure}

\textbf{MiniBooNE/SciBooNE:} For MiniBooNE/SciBooNE, antineutrinos are produced primarily from $\pi^- \rightarrow \mu^- \overline{\nu}_\mu$, Pions decay in a decay tunnel $\simeq 50 \,\mathrm{mts}$ long and muons are stopped in the ``dirt'' at a typical distance $\simeq 4 \, \mathrm{mts}$ beyond the decay tunnel\footnote{D.B. is indebted to William C. Louis III for correspondence clarifying these aspects.}, therefore in this situation $\Gamma_\pi(p) L_c \simeq 1$. The SciBooNE detector is at a distance $L = 100 \,\mathrm{mts}$ from the production region, in between the end of the decay tunnel and MiniBooNE, whose detector is at a  baseline $L = 540 \,\mathrm{mts}$, and the neutrino energy range (for both) is $ 0.3 \leq E \lesssim 1.6\,\mathrm{GeV}$. Figs. (\ref{fig:mini1eV}-\ref{fig:mini3eV}) show the comparison between the CP-even and odd parts with (modified) and without (QM) the decoherence corrections for MiniBooNE for $m_s = 1,2,3\,\mathrm{eV}$ respectively and figs. (\ref{fig:sci1eV}-\ref{fig:sci3eV}) show the same comparison  for SciBooNE parameters with $L=100 m$ and same energy range and values of $m_s$.

These figures confirm the interpretation of the decoherence modifications in terms of an overall suppression of the amplitude and a phase-shift that leads to an offset in the position of the peaks with respect to the quantum mechanical result. Since the mixing angle is extracted from the maximum amplitude of the probability and the mass from the position of the peaks, a fit with the quantum mechanical formula would \emph{underestimate} both the mixing angle and the mass as analyzed above.  A similar conclusion applies to the CP-violating angle. For MiniBooNE, the suppression and off-set are small when $\delta m^2  \lesssim 1\, \mathrm{eV}^2$, resulting in an underestimate of about $3-5\%$ in amplitude and mass as shown in fig. (\ref{fig:mini1eV}) but is larger at SciBooNE as shown in fig. (\ref{fig:sci1eV}), but for $\delta m^2 = m^2_s \sim 3\, \mathrm{eV}^2$, fig. (\ref{fig:mini3eV}) for MiniBooNE reveals $\sim 15\%$ suppression in the amplitude with a similar underestimate in the mass (off-set).

\vspace{2mm}

\textbf{Pions and Kaons (DAR):} Recent proposals \cite{anderson,spitz} for high intensity sources to study sterile-active oscillations with pions and kaons (DAR) motivate a study of the decoherence effects in these experiments. For (DAR) the energy is fixed at $E=E^*$ and presumably the baseline $L$ is also fixed,   we take   $L=30\,\mathrm{m}$ as a middle-range indicative value for the purpose of our analysis, other values can be explored numerically. What is less clear is the value of the product $\Gamma_M L_c$ which will ultimately depend on the experimental design. Namely, the muons (or charged leptons in general) must be stopped at distances much less than the baseline and that $\Gamma L_c \ll1$ in order for decoherence effects to be minimal.
We study the possibible ranges $\Gamma_M L_c \ll 1, \simeq 1, \gg 1$ respectively as a function of $m_s$. For $\pi-K$ (DAR) it follows that $E^*_\pi = 29.8 \,\mathrm{MeV}~~;~~ E^*_K =235.5 \,\mathrm{MeV} $  respectively for which we find the ratio (\ref{ratiolabpion}) to be
\be \mathcal{R}_\pi(E^*_\pi) = \frac{2}{3} \Bigg(\frac{  m^2_s}{\mathrm{eV}^2}\Bigg) ~~;~~  \mathcal{R}_K(E^*_K) = \frac{1}{25} \Bigg(\frac{  m^2_s}{\mathrm{eV}^2}\Bigg)\label{ratiospiK}\ee

\begin{figure}[h!]
\begin{center}
\includegraphics[keepaspectratio=true,width=3.2 in,height=3.5  in]{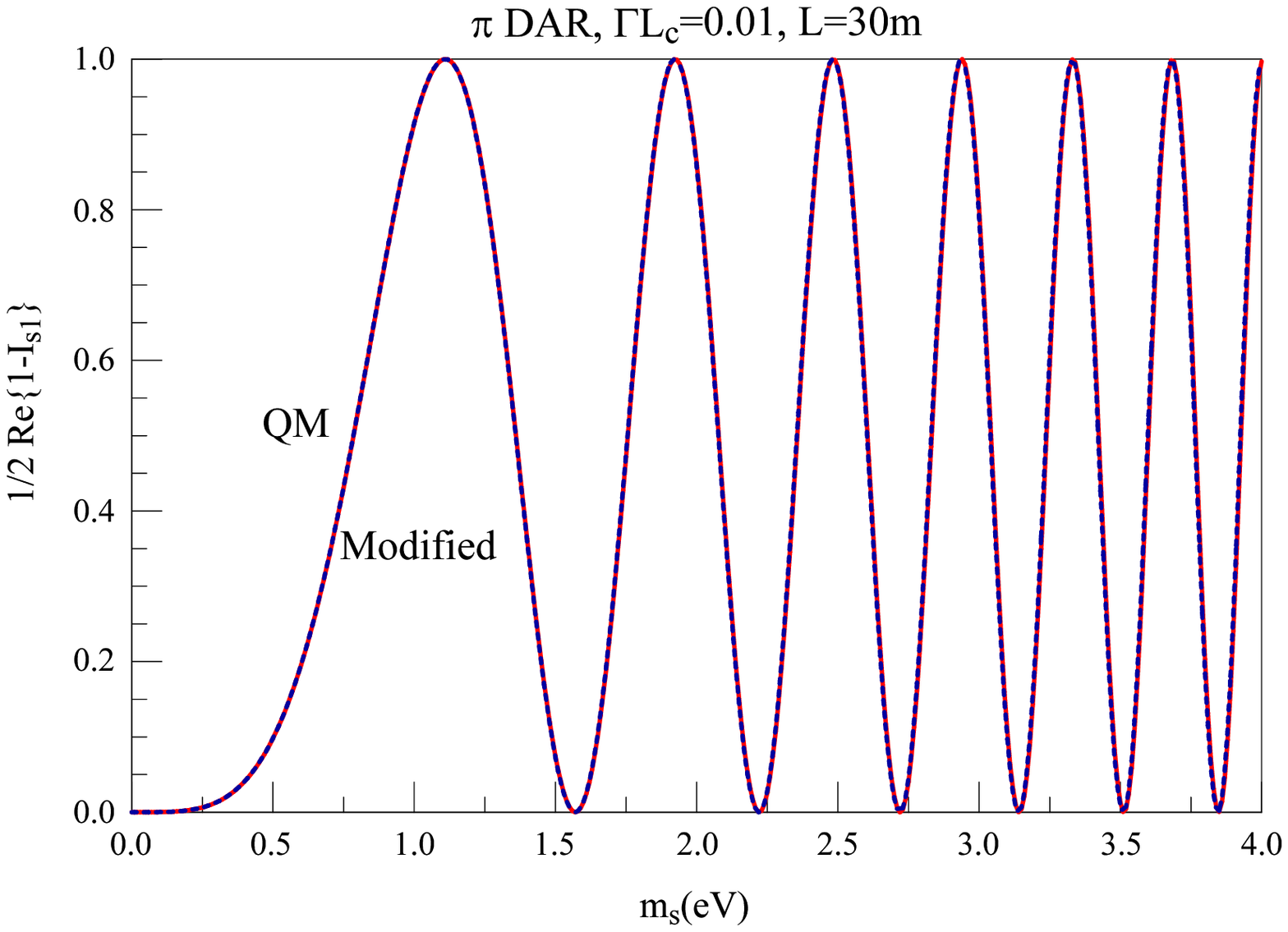}
\includegraphics[keepaspectratio=true,width=3.2  in,height=3.5  in]{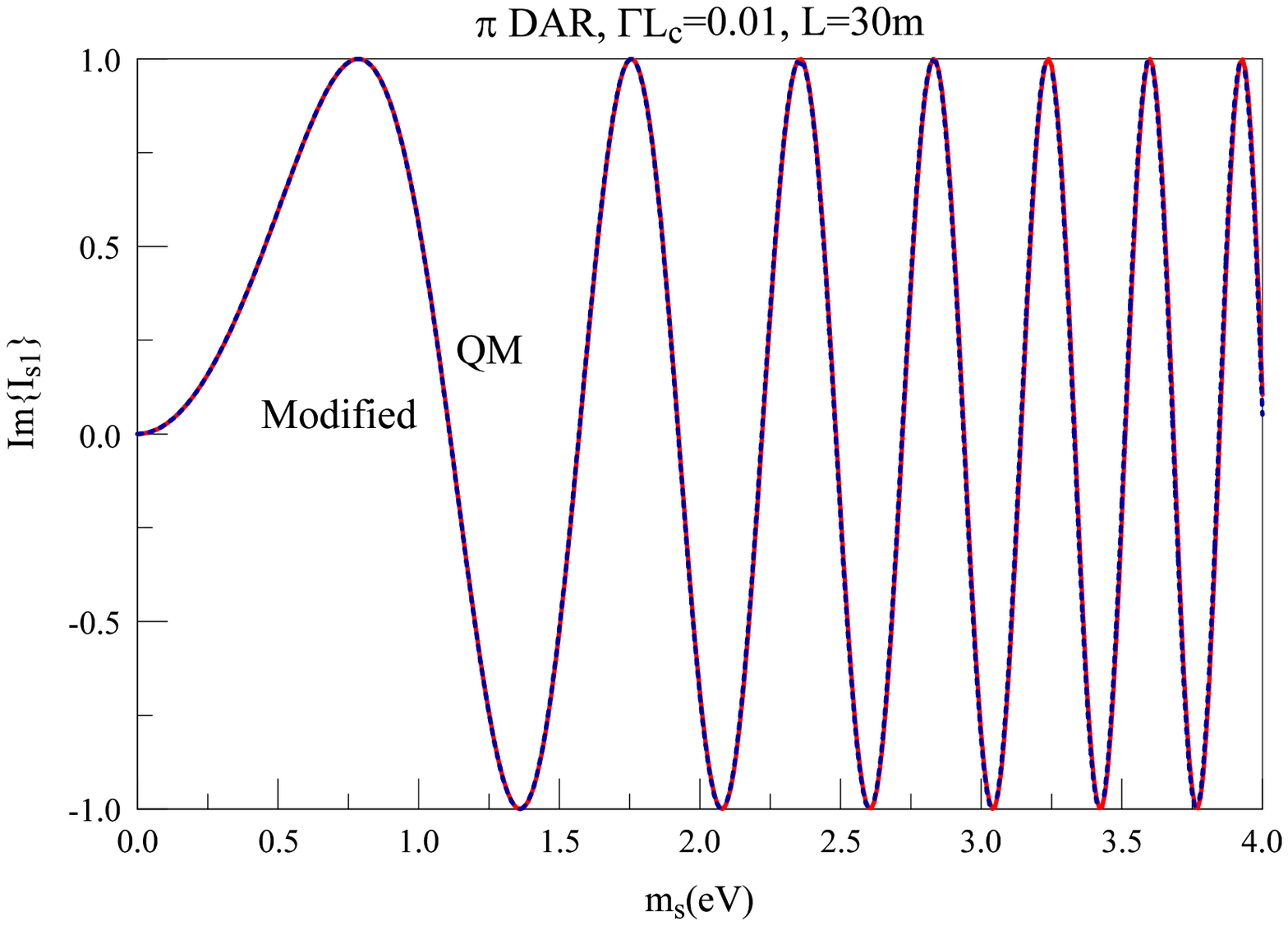}
\caption{$\pi$ (DAR) CP-even and CP-odd contributions for $\Gamma_\pi L_c =0.01$ vs $m_s$. The solid line (modified) shows $\mathrm{Re}[1-I_{s1}]$ where $\mathrm{Re}[I_{s1}]$ is given by  (\ref{realpartIij},\ref{imagpartIij}), the dashed line is the quantum mechanical    result (\ref{realimagusual}) for $\delta m^2_{s1}= m^2_s$. }
\label{fig:pidar01}
\end{center}
\end{figure}

\begin{figure}[h!]
\begin{center}
\includegraphics[keepaspectratio=true,width=3.2 in,height=3.5  in]{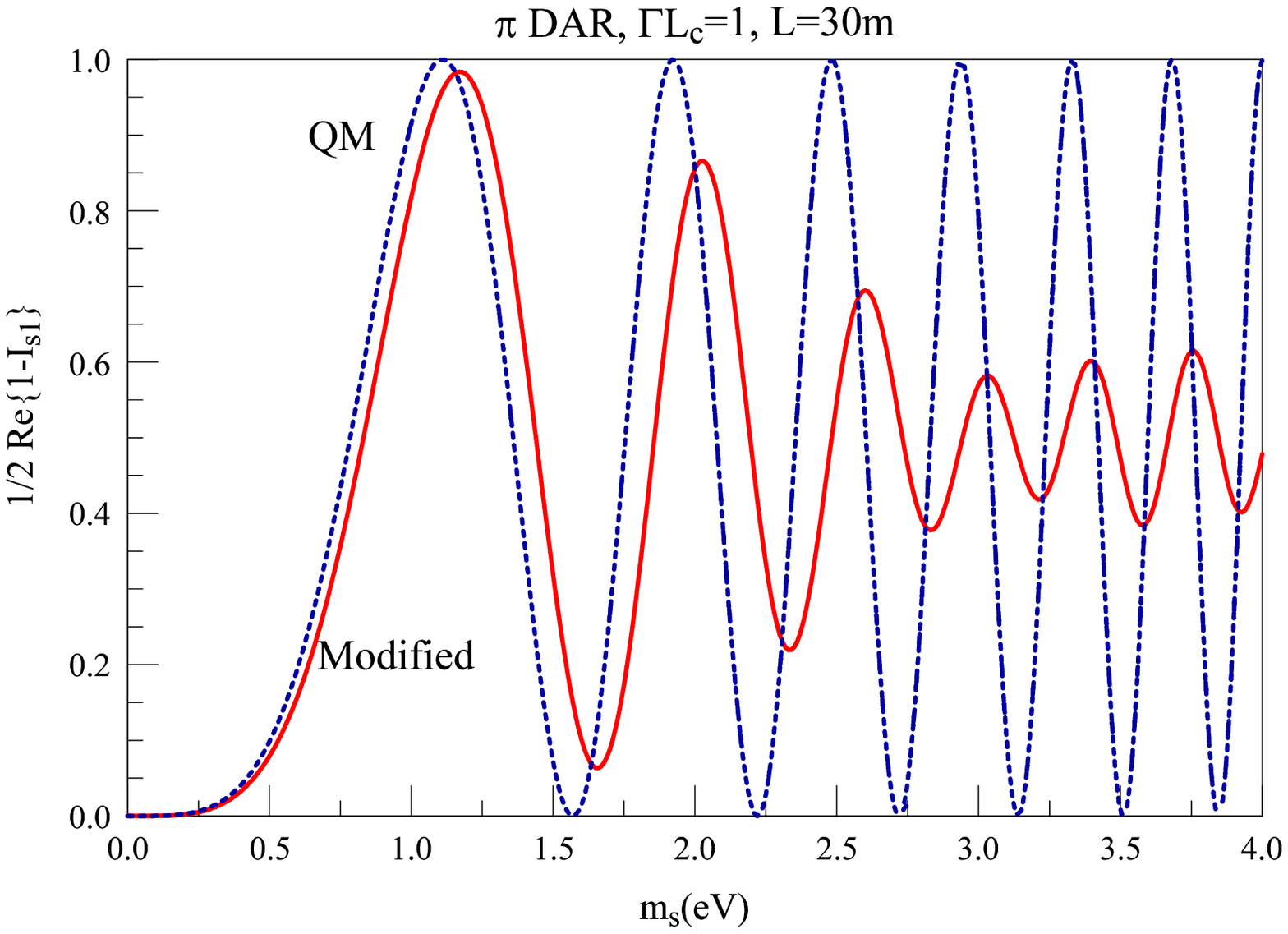}
\includegraphics[keepaspectratio=true,width=3.2  in,height=3.5  in]{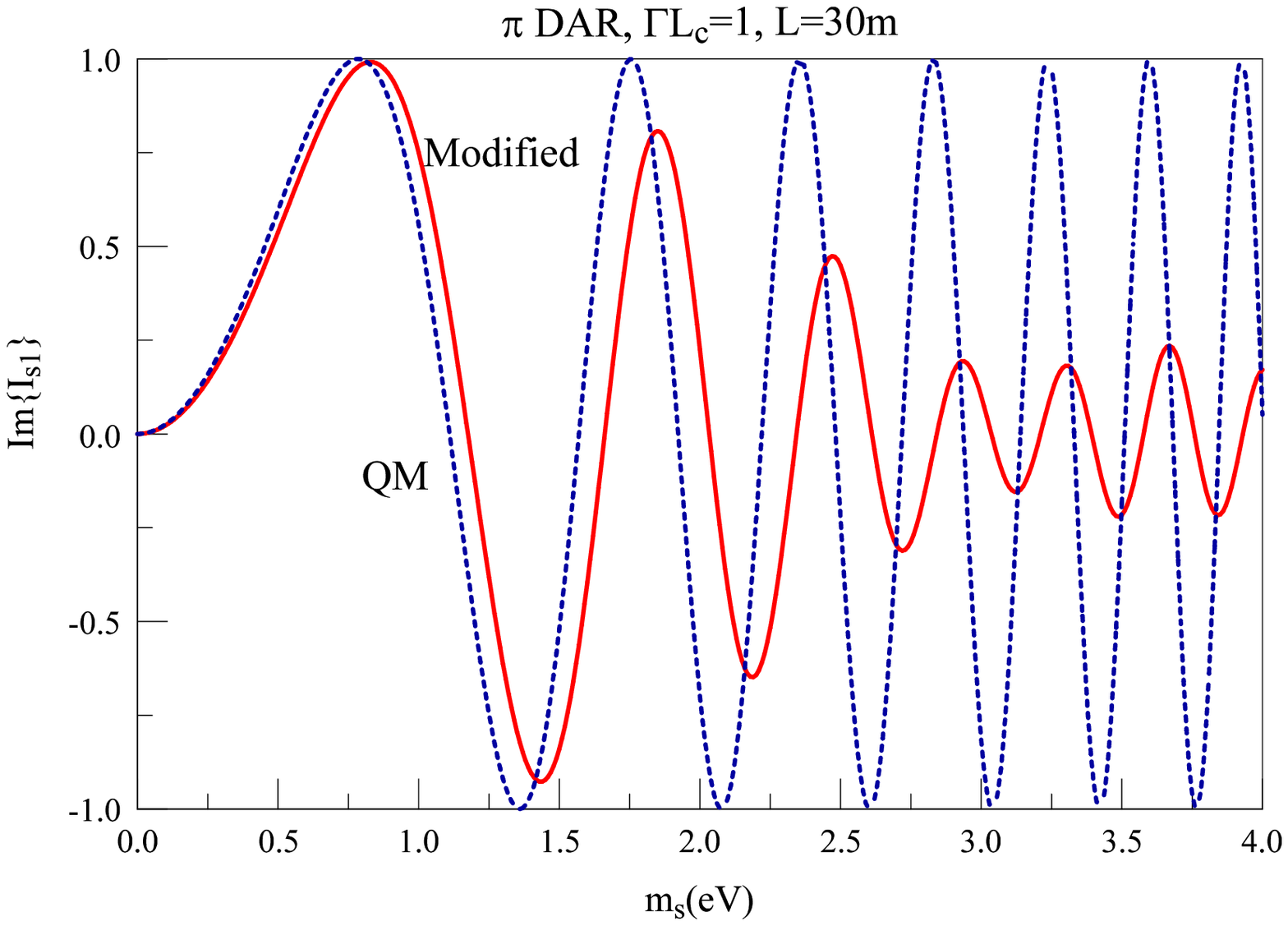}
\caption{Same as fig. (\ref{fig:pidar01}) with $\Gamma_\pi L_c =1$.    }
\label{fig:pidar1}
\end{center}
\end{figure}

\begin{figure}[h!]
\begin{center}
\includegraphics[keepaspectratio=true,width=3.2 in,height=3.5  in]{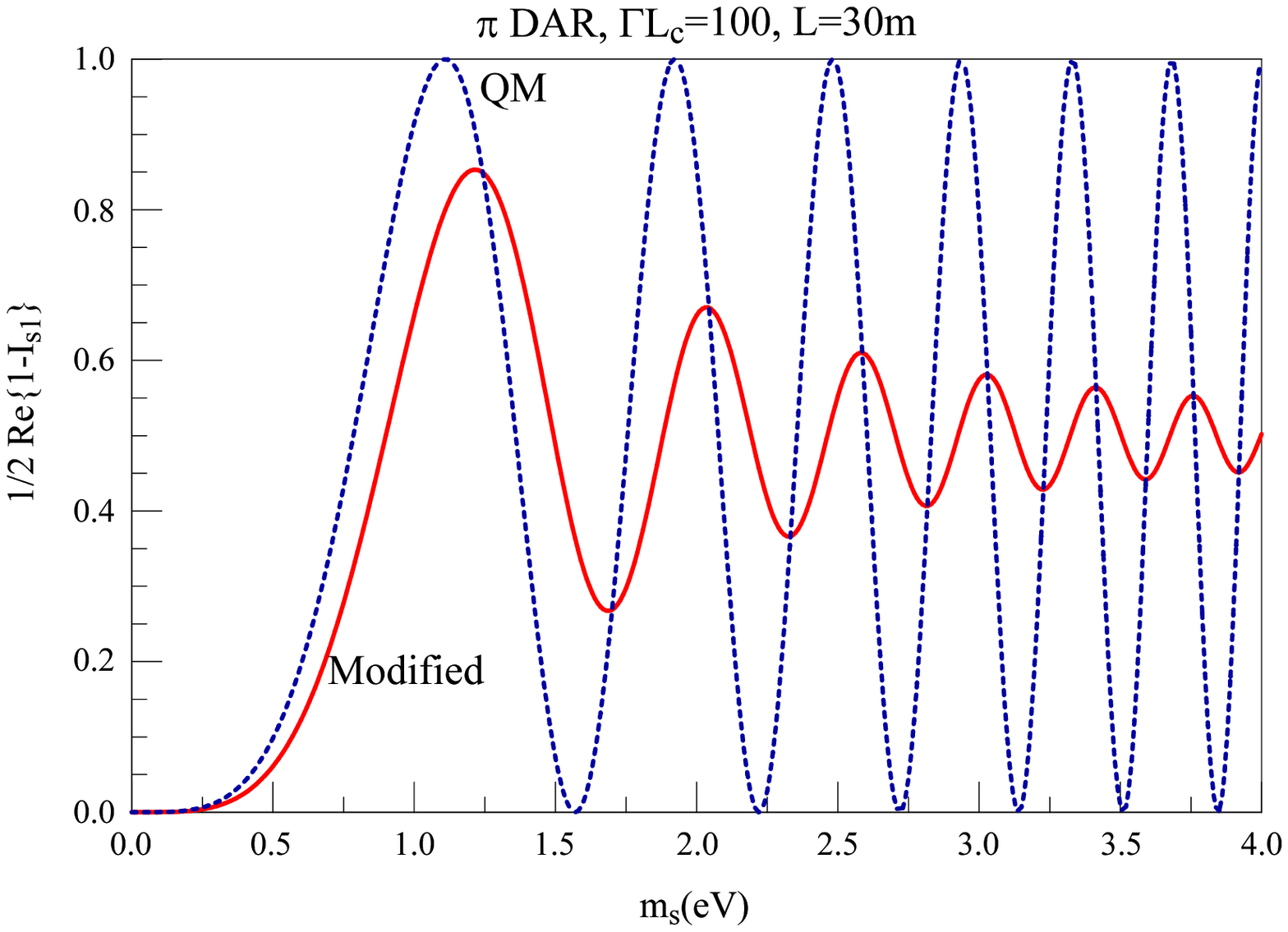}
\includegraphics[keepaspectratio=true,width=3.2  in,height=3.5  in]{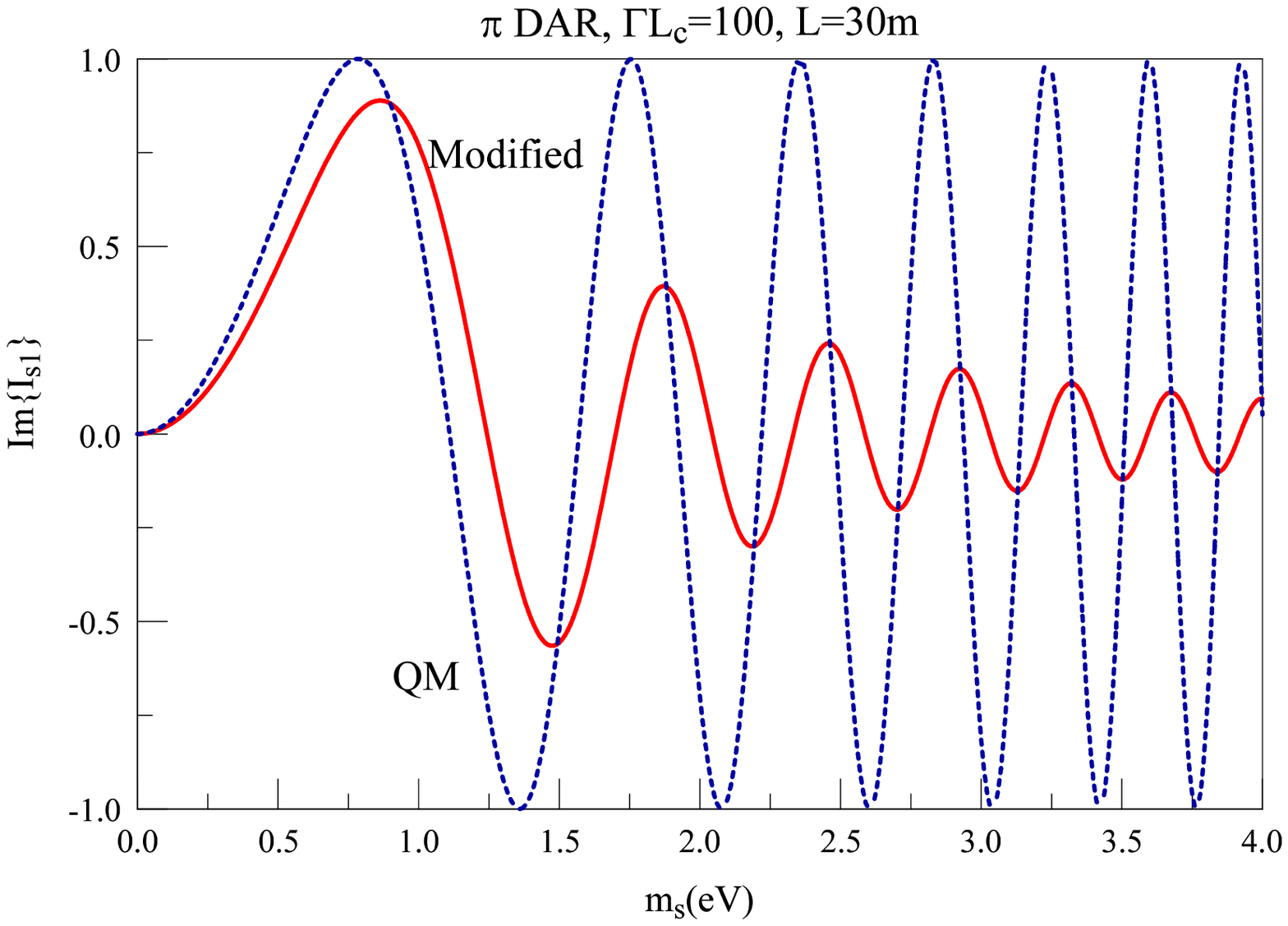}
\caption{Same as fig. (\ref{fig:pidar01}) with $\Gamma_\pi L_c =100$.    }
\label{fig:pidar100}
\end{center}
\end{figure}

\begin{figure}[h!]
\begin{center}
\includegraphics[keepaspectratio=true,width=3.2 in,height=3.5  in]{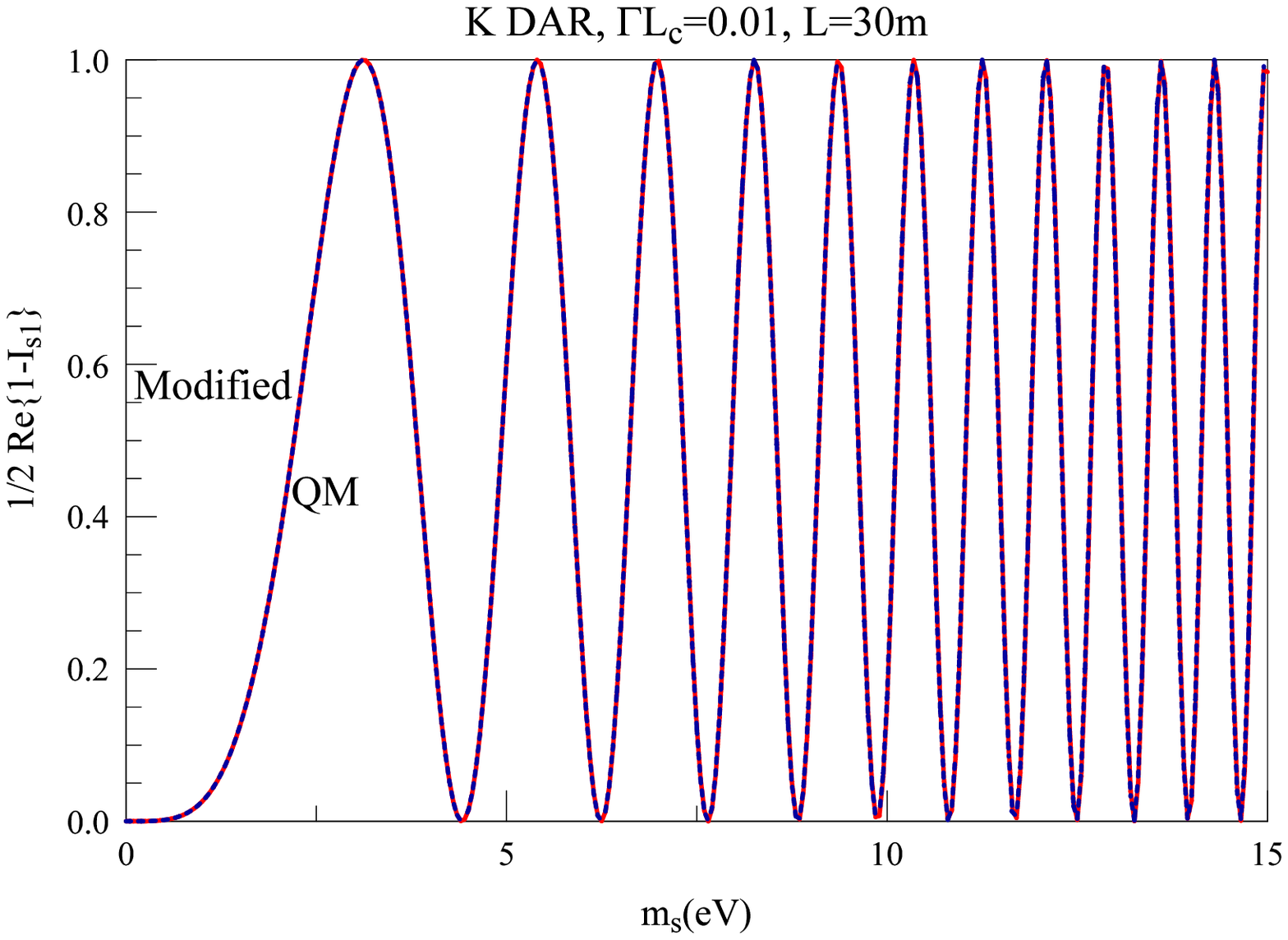}
\includegraphics[keepaspectratio=true,width=3.2  in,height=3.5  in]{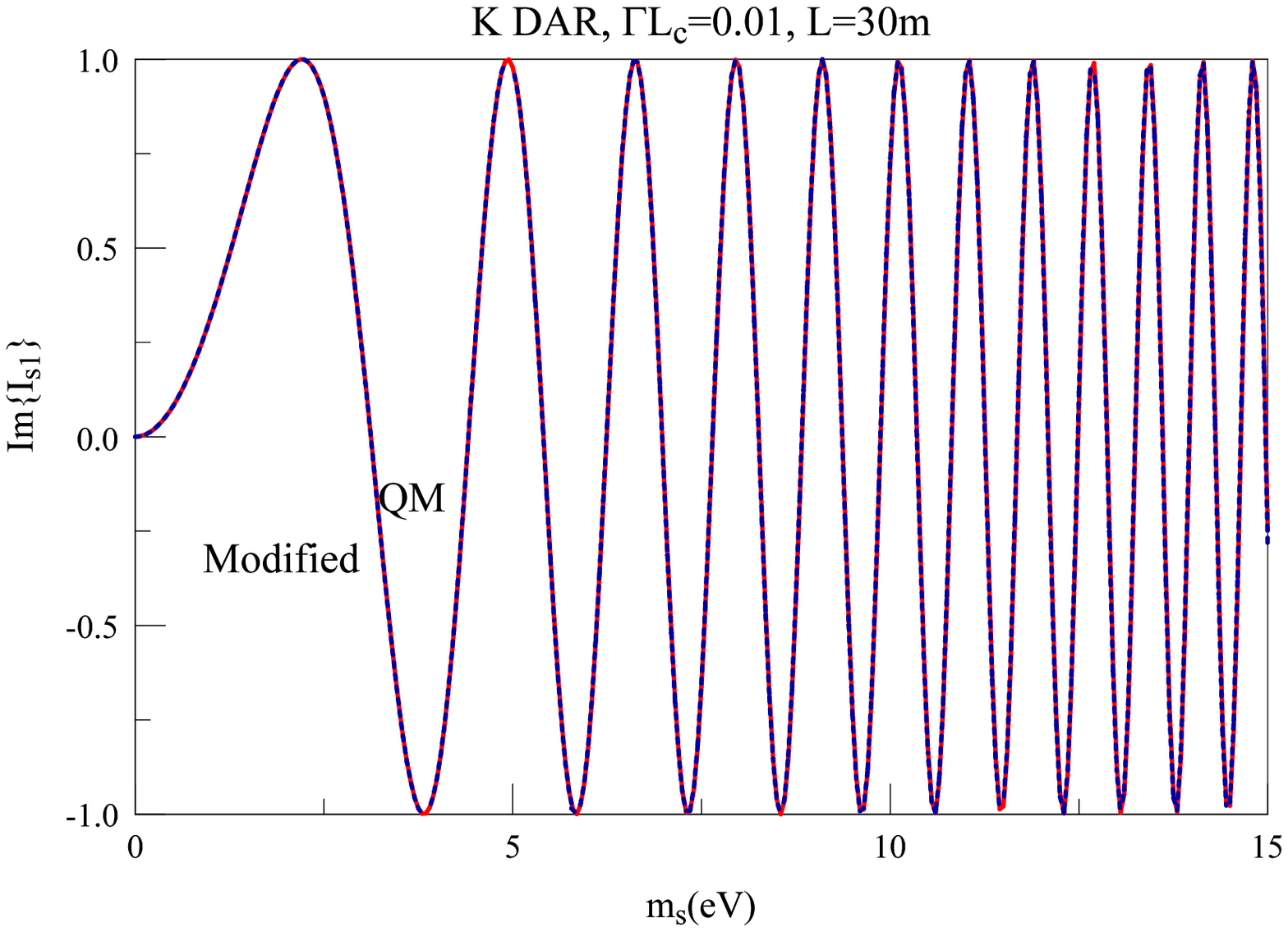}
\caption{$K$ (DAR) CP-even and CP-odd contributions for $\Gamma_K L_c =0.01$ vs $m_s$. The solid line (modified) shows $\mathrm{Re}[1-I_{s1}]$ where $\mathrm{Re}[I_{s1}]$ is given by  (\ref{realpartIij},\ref{imagpartIij}), the dashed line is the quantum mechanical    result (\ref{realimagusual}) for $\delta m^2_{s1}= m^2_s$.   }
\label{fig:kdar01}
\end{center}
\end{figure}

\begin{figure}[h!]
\begin{center}
\includegraphics[keepaspectratio=true,width=3.2 in,height=3.5  in]{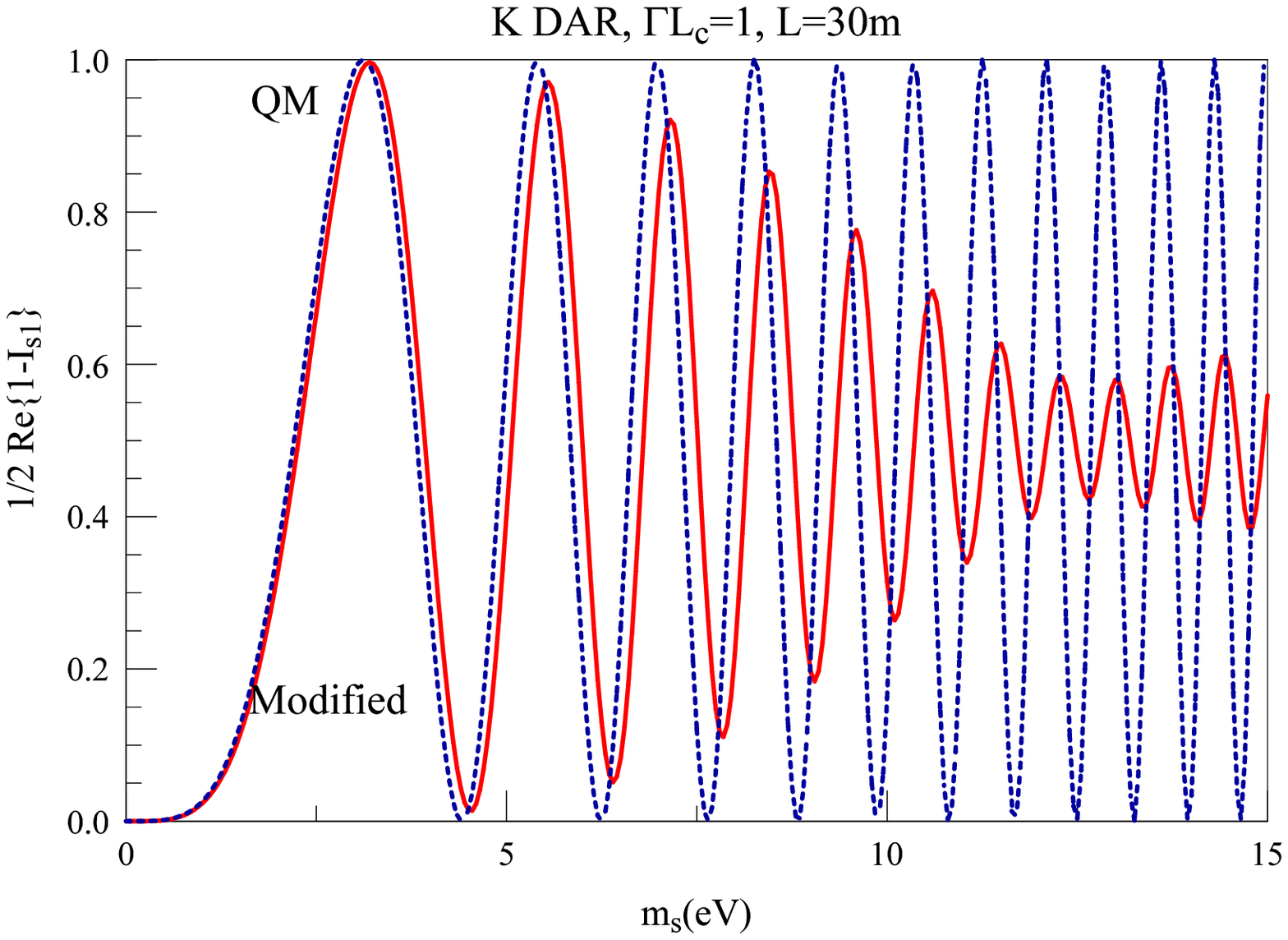}
\includegraphics[keepaspectratio=true,width=3.2  in,height=3.5  in]{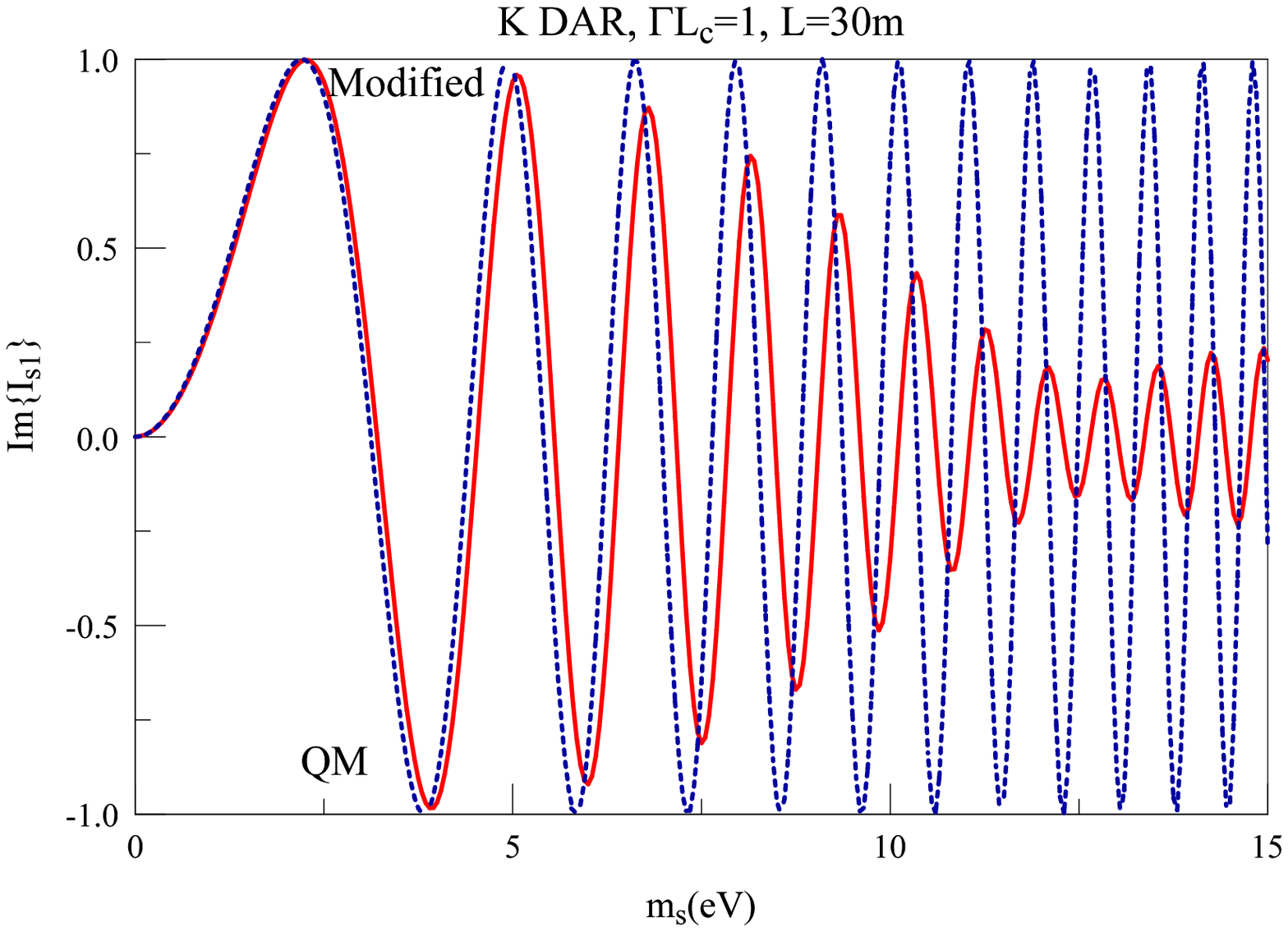}
\caption{Same as fig. (\ref{fig:kdar01}) with $\Gamma_K L_c =1$.    }
\label{fig:kdar1}
\end{center}
\end{figure}

\begin{figure}[h!]
\begin{center}
\includegraphics[keepaspectratio=true,width=3.2 in,height=3.5  in]{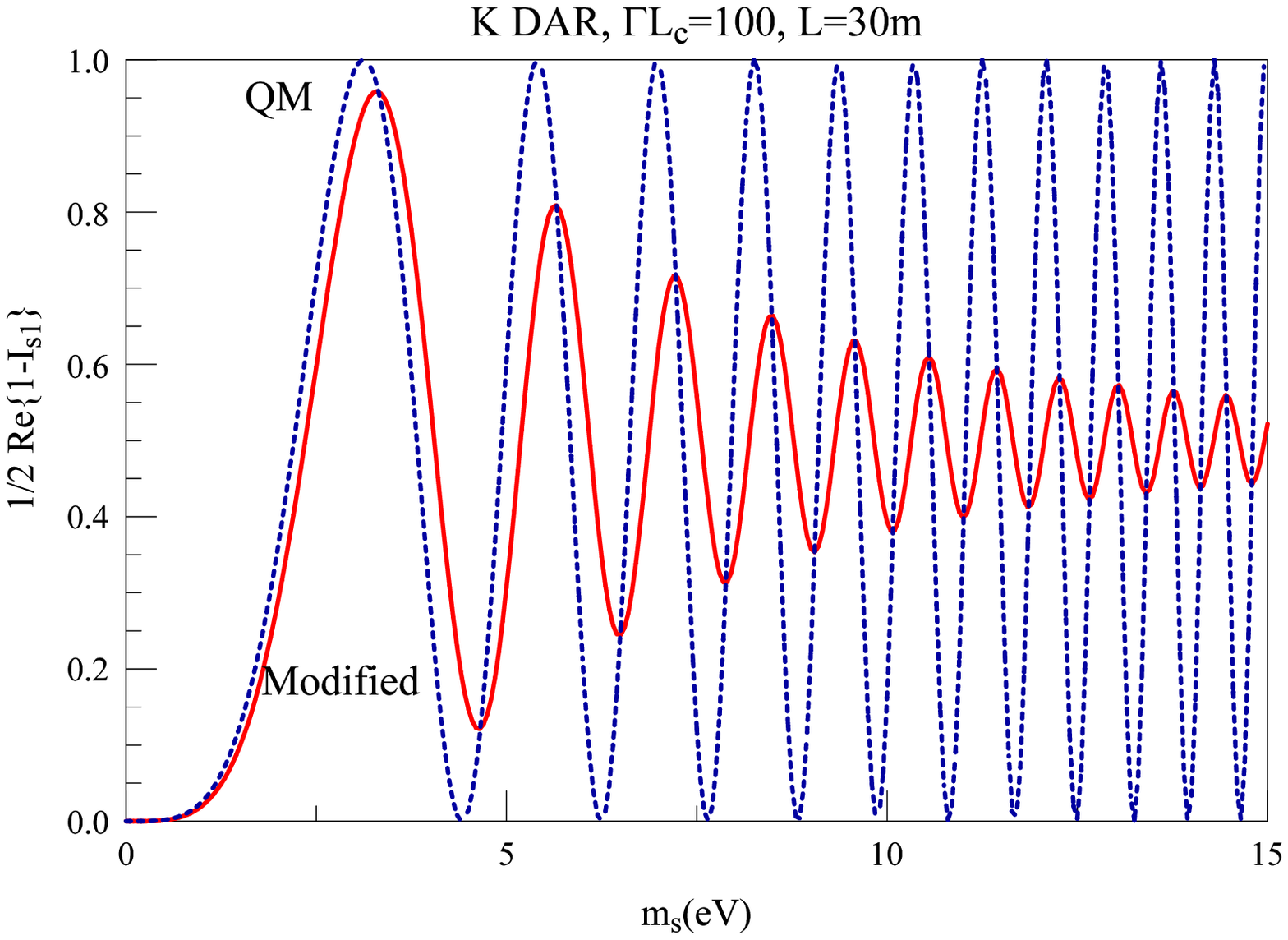}
\includegraphics[keepaspectratio=true,width=3.2  in,height=3.5  in]{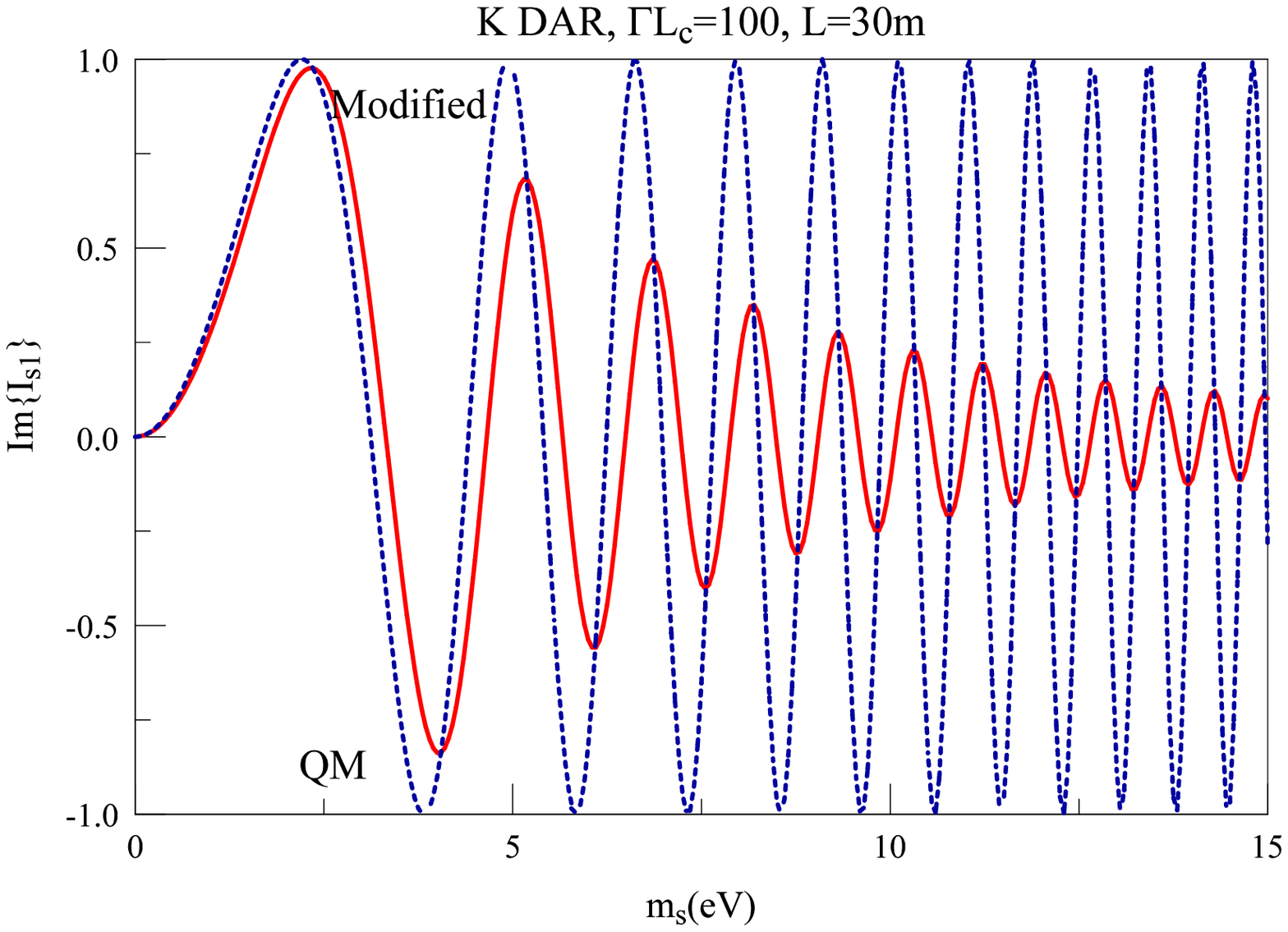}
\caption{Same as fig. (\ref{fig:kdar01}) with $\Gamma_K L_c =100$.    }
\label{fig:kdar100}
\end{center}
\end{figure}

The comparison between the modified results (\ref{realpartIij},\ref{imagpartIij}) and the usual quantum mechanical results (\ref{realimagusual}) are displayed in figs. (\ref{fig:pidar01}-\ref{fig:kdar100}) for $\pi,K$ (DAR) for a baseline representative $L=30\,\mathrm{m}$.

It is clear from this analysis, both for $\pi,K$ (DAR), that decoherence effects are very small and the modified result is indistinguishable from  the usual quantum mechanical results (\ref{realimagusual}) whenever $\Gamma_M L_c\ll 1$ but become large for $\Gamma_M L_c \gtrsim 1$. The decay length for $\pi,K$ are  $l_\pi = 7.8\,\mathrm{m}, l_K = 3.7 \,\mathrm{m}$ respectively; therefore, in order for the usual quantum mechanical results (\ref{realimagusual}) to describe a correct fit to the experimental data, the design must ensure that charged leptons (mainly $\mu$) \emph{be stopped at distances $L_c \ll l_{\pi},l_K$ respectively}, namely a few $\mathrm{cm}$ beyond the stopping target of the mesons.

\vspace{3mm}

\textbf{Long baseline experiments:} For long baseline experiments, the decoherence terms do not contribute. This is because these experiments study oscillations with $\delta m^2 \sim 10^{-3} \,\mathrm{eV}^2$ and $E \simeq \mathrm{few~GeV}$ for which $L \simeq 300-1000 \,\mathrm{km}$, an example of such experiment is  Minos in which pions produce neutrinos as in MiniBooNE/SciBoone. In these experiments, $\mathcal{R} \lesssim 10^{-3} ~;~ \Gamma_M L_c \lesssim 1$ so that $\Delta_{ij} L_c \ll 1$; therefore, decoherence effects are all but negligible generally for long baseline experiments.

\subsection{Reactors vs. accelerator experiments}\label{sec:reactors}

\vspace{3mm}

The suppression of the transition probabilities through the decoherence effects depend both on the lifetime of the parent particle and the stopping distance of the charged lepton which is produced along with the (anti) neutrino via the charged current interactions. This establishes a \emph{fundamental difference between accelerator and reactor oscillation experiments}: whereas in accelerator experiments neutrinos are produced via the decay of short lived mesons with typical lifetimes $\simeq 10^{-8}\mathrm{secs}$ and widths $\simeq 10^{-8}\,\mathrm{eV}$, in reactors  the (anti) neutrinos are produced via the   $\beta$ decay of long-lived unstable nuclei ${}^{235}\mathrm{U} \,, {}^{238}\mathrm{U} \,,{}^{239}\mathrm{Pu} \,, {}^{241}\mathrm{Pu}$\cite{flux1,reactor} with typical lifetimes in the range between hundreds and thousands of years. Furthermore, in current short baseline accelerator experiments such as MiniBooNE/SciBooNE, pions decay in a decay pipe and muons are stopped at short distances beyond the decay pipe so that $\Gamma_\pi L_c \simeq 1$; in reactor experiments, muons are stopped in the reactor core on distance scales so that $\Gamma L_c~,~\Delta_{ji} L_c \ll 1$. Our study above clearly shows that under this circumstance the modifications from decoherence are negligible and the transition probabilities are indistinguishable from the quantum mechanical result. Therefore, we conclude that the quantum mechanical fit to the oscillation probabilities in reactor experiments is always justified, whereas in accelerator experiments, decoherence effects both from the lifetime of the parent meson and the stopping length scale of the charged lepton partner are substantial for $m_s \gtrsim 1\,\mathrm{eV}$ and a fit to the usual quantum mechanical transition probabilities both for CP-even/odd contributions may substantially \emph{underestimate masses, mixing and CP-violating angles}.

\subsection{Wave packets:}\label{wavepack}

\vspace{3mm}

Our study is restricted to plane waves to exhibit the main results and conclusions in the clearest possible setting. As has been argued in the literature\cite{wavepacket,kayserwp,rich,giuntiwp,dolgovwp,beuthe,boyho,akmewp}, wave packet localization may be an important ingredient in the description of neutrino oscillations. The localization length both of the production and detection regions define momentum uncertainties that are important in the conceptual understanding of the interference phenomena. A wave packet description should also be implemented in the measurement  or stopping (disentanglement) of the charged lepton, which we treated as an event sharp in space time at a time scale $t_c$ and distance $L_c$, a wave packet treatment would smear these scales over a localization length scale of the wave packet, which is determined by the measurement process (or perhaps the mean free path of the charged lepton in the stopping material).

The typical  analysis of neutrino oscillations in terms of (Gaussian) wave packets \cite{wavepacket,kayserwp,rich,giuntiwp,dolgovwp,beuthe,boyho,akmewp}   clarifies that neutrino wave-packets evolve   semiclassically, the center moves as the front of a plane wave with the group velocity and is modulated by a Gaussian envelope which spreads via dispersion. Wave packets associated with the different mass eigenstates separate as they evolve with slightly different group velocities and, when their separation becomes of the order of or larger than the width of the wave packet, the overlap vanishes and oscillations are suppressed $\propto e^{-(L/L_{coh})^2}$, where $L_{coh} \simeq \sigma\,E^2_\nu/\delta m^2$ and $\sigma$ is the spatial localization scale of the wave packet. This suppression becomes important when $L_{coh} \lesssim L$, which for $\delta m^2 \simeq \mathrm{eV}^2, E \gtrsim 30 \,\mathrm{MeV}, L \sim 100-600 \,\mathrm{m}$ implies $\sigma \lesssim 1-5 \times 10^{-13} \,\mathrm{m}$ which, while much bigger than nuclear dimensions, is much smaller than atomic scales. If a firm assessment confirms that neutrino wavepackets are produced with such localization length or smaller, then this decoherence effect must be introduced in the oscillation probability.

 As discussed in \cite{wavepacket} the wave packet description also features another source of
decoherence in the \emph{localization term}, which suppresses coherence when    $\sigma > L_{osc}\sim E/\delta m^2$ which is unlikely to be relevant in short baseline accelerator experiments. A complementary interpretation of decoherence for $\Gamma_M \lesssim \delta m^2/2E$ in terms of wave packets is discussed in ref.\cite{wavepacket}:  \emph{if} a neutrino wavepacket produced by the decay of a parent particle of width $\Gamma_M$ is assigned a localization length, $\sigma \simeq 1/\Gamma_M$, then the condition for decoherence from the \emph{localization term}, $\sigma \simeq L_{osc}$, becomes equivalent to $\Gamma_M \simeq \delta m^2/2E$ which is recognized as $\mathcal{R} \simeq 1$ in our discussion. Although we do not see an obvious relation between the results obtained above with the non-perturbative field theoretical Wigner-Weisskopf method  and the interpretation of a wavepacket with localization length $1/\Gamma_M$, our results are certainly in agreement with this interpretation; however, we emphasize that the analysis above also reveals another scale that is important for decoherence, namely $L_c$, which is the length scale at which the charged lepton that is emitted along with the neutrino is observed or absorbed. As pointed out above there are \emph{two} important dimensionless quantities that determine decoherence in the plane wave limit:
$\mathcal{R}$ and $\Gamma_M L_c$.

\section{Conclusions and further questions.} \label{sec:conclu}

Motivated by the cosmological importance of new generations of heavier sterile neutrinos and recent proposal for high intensity sources, this article focuses on two different aspects related to the search of sterile neutrinos: 1) a proposal  to search for heavy ($\simeq \mathrm{MeV}$-range) sterile neutrinos by studying the production of \emph{negative helicity} charged leptons in $\pi^-,K^-$ decay at rest (or positive helicity in the decay of $\pi^+,K^+$) as a complement to the search for monochromatic lines in the muon (or electron) spectrum, and 2) an assessment of the impact of decoherence effects from the lifetime of the parent meson and the stopping distance scale of the charged lepton on the experimental fits for sterile neutrinos masses, mixing and CP-violating angles in short baseline experiments.

Massive sterile (anti) neutrinos  produced in $\pi^-,K^-$ decay at rest (DAR) lead to a \emph{negative} helicity (positive if the decay is $\pi^+,K^+$) component of the charged lepton produced in the decay. For searches of heavy sterile neutrinos from $\pi^-,K^-$ decay at rest, we obtain the branching ratio for charged leptons to be produced with \emph{negative helicity} (or positive helicity for the decay of $\pi^+,K^+$). This branching ratio determines the abundance of the negative helicity states in the production process and we suggest that a Stern-Gerlach type filter with a magnetic field with a gradient along the direction of the collimated charged lepton beam emitted back to back with the (anti) neutrinos allows to spatially separate the different helicity components. A combined measurement of the monochromatic line for the charged lepton and the ratio of abundances of the  spatial domains yield simultaneous information on the mass and the absolute value of the mixing matrix element. This setup is most sensitive for \emph{heavy} sterile neutrinos with mass $m_s$ in the $\mathrm{MeV}$ range. The ratio of abundances between the  negative and positive helicity states is determined by the branching ratio (\ref{BRmm}), shown in fig. (\ref{fig:brs}) (divided by $|U_{ls}|^2$), which,   in combination with the search for monochromatic lines allows, to extract both the mass and the element of the mixing matrix $|U_{ls}|^2$ by fitting both the energy and abundance with the branching ratios.

 Upper bounds on the sterile-active mixing matrix elements from previous experimental searches allow us to estimate the upper bounds for the branching ratios for the different processes, these are given by
\bea &&  Br^{--}_{\pi \rightarrow \mu  \, \bar{\nu_s}} \lesssim 10^{-8}-10^{-7} ~~;~~  3\,\mathrm{MeV} \lesssim m_s \lesssim 33 \,\mathrm{MeV}  \label{upboundpimu} \\
&&Br^{--}_{\pi \rightarrow e \, \bar{\nu_s}} \lesssim 10^{-8}-10^{-6} ~~;~~  3\,\mathrm{MeV} \lesssim m_s \lesssim 135 \,\mathrm{MeV}   \eea with the electron channel providing the largest window of opportunity because of the larger phase space. For K-(DAR), we find
 \be   Br^{--}_{K \rightarrow \mu,e \, \bar{\nu_s}} \lesssim 10^{-9}-10^{-6}~
   \mathrm{for}~\Bigg\{ \begin{array}{c}
 4\,\mathrm{MeV} \lesssim m_s \lesssim 360 \,\mathrm{MeV}~(\mu-\mathrm{channel})   \\
   4\, \mathrm{MeV} \lesssim m_s \lesssim 414 \,\mathrm{MeV}~(e-\mathrm{channel})                                                                                             \end{array}   \,.\ee
 These upper bounds estimates suggest that these searches could be implemented in the next generation of high intensity experiments.

 Short baseline experiments target new generation of sterile neutrinos in the mass range $\simeq \mathrm{eV}$ as suggested by the LSND, MiniBooNE results and reactor anomalies. In current accelerator experiments, (anti) neutrinos are produced from the decay of pions or kaons either in flight, as in MiniBooNE/SciBooNE, or at rest, as recent proposals suggest. We recognized \emph{two}  sources of decoherence that impact the interpretation of the data and experimental fits to extract masses, mixing and CP-violating angles: a) the width of the parent meson $\Gamma_M$ introduces an energy (or time) uncertainty and b) the stopping distance $L_c$ of the charged lepton that is produced in a quantum entangled state with the (anti) neutrino, decoherence effects are encoded in two different dimensionless quantities,
 \be \mathcal{R}_{ij}(E) = \frac{\delta m^2_{ij}}{2E\Gamma_M} ~~;~~ \Gamma_M L_c\,. \ee
 The usual quantum mechanical formula for the oscillation probabilities are modified as follows:
 \be e^{i\frac{\delta m^2_{ij}\,L}{2E}} \rightarrow  e^{i\frac{\delta m^2_{ij}\,L}{2E}}\, \Bigg[\frac{ 1- e^{-\Gamma_M(p)L_c\big(1+i\mathcal{R}_{ij}\big)}  }{ 1- e^{-\Gamma_M(p)L_c} }\Bigg]\,\Bigg[\frac{1-i\mathcal{R}_{ij}}{1+ \mathcal{R}^2_{ij}}\Bigg] \ee

 We study the impact of the decoherence effects both for Dirac and Majorana neutrinos, addressing in particular CP-violating effects as well as $\nu \rightarrow \overline{\nu}$ oscillations and $|\Delta L|=2$ transitions in the case of Majorana neutrinos. In all cases, we find that, for $\mathcal{R}_{ij},\Gamma_M L_c  \gtrsim 1$, the oscillation probabilities are suppressed and the oscillatory functions feature energy-dependent phase-shifts that results in an overall off-set that impacts the determination of the mass. If these decoherence effects are neglected in the experimental analysis and the data are fit with the usual quantum mechanical oscillation probabilities the masses, mixing and CP-violating angles are \emph{underestimated}.

 In particular, on MiniBooNE/SciBooNE, for example, neutrinos are produced from pion decay for which we find $\mathcal{R} \simeq 1/3 \big(\delta m^2/\mathrm{eV}^2\big)$ and $\Gamma_M L_c \simeq 1$, with one sterile neutrino with $m_s \sim 3 \, \mathrm{eV}$,  fitting with two-generation mixing underestimates $\sin^2(2\theta)$  and $\delta m^2$ by nearly $15\%$. Similar underestimates follow for CP-violating angles and $|\Delta L|=2$ processes in $3+2$ schemes.

 We also conclude that reactor and (current) accelerator experiments are fundamentally different in that the lifetime of the decaying parent particles in reactor experiments is hundreds to thousands of years, compared to pion or kaon lifetimes, and charged leptons (muons) are stopped within the core so that for reactors $\Gamma L_c, \Delta_{ij} L_c \ll 1$ and decoherence effects are all but negligible, unlike the situation for example for MiniBooNE/SciBooNE. We also suggest that next generation of high intensity experiments in which (anti) neutrinos are produced from $\pi,K$ (DAR), decoherence effects may be suppressed considerably by designing the experiment so that the charged leptons produced with the neutrinos (mainly muons) are stopped on distances much smaller than the decay length of the mesons, in which case the usual quantum mechanical oscillation probabilities furnish an accurate description of mixing and oscillations.

\acknowledgements The authors thank Robert Shrock for stimulating correspondence and for making them aware of his early work of ref.\cite{shrock}, they  acknowledge support from NSF through  grants  PHY-0852497,PHY-1202227.

\appendix
\section{Quantization: Mesons,   Dirac and Majorana neutrinos}\label{app:quant}
We quantize the (pseudo) scalar and fermion fields in a quantization volume $V$. The  charged (complex) (pseudo) scalar field is as usual

\be M(\vx,t)   =       \sum_{\vp} \frac{1}{\sqrt{2VE_M(p)}} \left[ \hat{\mathcal{A}}_{\vp}\, e^{-iE_M(p)\,t} + \hat{\mathcal{B}}^{\dagger}_{-\vp}\, e^{iE_M(p)\,t} \right] e^{i\vp \cdot \vx} \label{piKfields} \ee where $E^M_p= \sqrt{p^2+m^2_M}$ with $m_M$ the mass of the corresponding meson. It follows that
\be
J^{M}_{\mu} (\vx,t) =      \sum_{\vp} \frac{p_{\mu}}{\sqrt{2VE_M(p)}}  \left[     \hat{\mathcal{A}}_{\vp} \, e^{-iE_M(p)\,t} -   \hat{\mathcal{B}}^{\dagger}_{-\vp}\, e^{iE_M(p)\,t} \right] e^{i\vp \cdot \vx} \label{current}
\ee It proves convenient to introduce the combinations
\be M^{+}(\vp,t) \pm M^{-}(\vp,t)= \left(  \hat{\mathcal{A}}_{\vp}\, e^{-iE_M(p)\,t} \right) \pm \left( \hat{\mathcal{B}}^{\dagger}_{-\vp}\, e^{iE_M(p)t} \right)\, .  \label{combos}\ee

For Fermi fields we work in the chiral representation,
\be \gamma^0 =  \left[ \begin{array}{cc}
                         0 & -\mathbbm{1} \\
                         -\mathbbm{1} & 0
                       \end{array}
\right] ~~;~~ \gamma^i= \left[ \begin{array}{cc}
                         0 &  \sigma^i \\
                          -\sigma^i & 0
                       \end{array}
\right] ~~;~~ \gamma^5 = \left[ \begin{array}{cc}
                         \mathbbm{1} &  0 \\
                          0  & -\mathbbm{1}
                       \end{array}
\right]
\label{gammas}\ee and for a generic   Fermion $f$, either for charged lepton or Dirac neutrinos of mass $m_f$,  we write
\be  \Psi (\vx,t)    =      \sum_{h=\pm} \sum_{\vk} \frac{\psi (\vk,h,t)}{\sqrt{2VE_f(k)}} ~~e^{i\vk\cdot\vx} \label{psidirac}\ee

For \emph{Dirac} fermions  of mass $m_f$
\be \psi(\vk,h,t)  =
\Bigg[ \hat{b}_{\vk,h} \mathcal{U}_h(\vk)\, e^{-iE_f (k)\, t} + \hat{d}^{\dagger}_{-\vk,h} \mathcal{V}_h(\vk)\, e^{iE_f  (k)\, t}\Bigg]   \label{psis}\ee  where $E_f ( k) =\sqrt{k^2+m^2_f}$ and the spinors $\mathcal{U}_h,\mathcal{V}_h$ are eigenstates of helicity with eigenvalue $h=\pm 1$, these are given by
\bea \mathcal{U}_+(\vk)  & = &  N_f \Bigg( \begin{array}{c}
                       v_+(\vk) \\
                        -\varepsilon(k)\, v_+(\vk)
                      \end{array}
 \Bigg) ~~;~~ \mathcal{U}_-(\vk) = N_f \Bigg( \begin{array}{c}
                       -\varepsilon(k)\,v_-(\vk) \\
                        v_-(\vk)
                      \end{array}
 \Bigg) \label{Uspi}\\ \mathcal{V}_+(\vk) & = & N_f \Bigg( \begin{array}{c}
                        \varepsilon(k)\,v_+(\vk) \\
                        v_+(\vk)
                      \end{array}
 \Bigg) ~~;~~\mathcal{V}_-(\vk) =N_f \Bigg( \begin{array}{c}
                       v_-(\vk) \\
                         \varepsilon(k)\, v_-(\vk)
                      \end{array}
 \Bigg) \label{Vspi} \eea where
 \be N_f=\sqrt{E_f(k)+k} ~~;~~\varepsilon(k) = \frac{m_f}{E_f(k)+k} \label{Neps}\ee and
 $v_{\pm}(\vk)$ are helicity eigenstates Weyl spinors:
 \be v_+(\vk) = \Bigg( \begin{array}{c}
                         \cos\frac{\theta}{2} \\
                         \sin\frac{\theta}{2}\,e^{i\phi}
                       \end{array}
 \Bigg)~~;~~v_-(\vk) = \Bigg( \begin{array}{c}
                        -\sin\frac{\theta}{2}\,e^{-i\phi}   \\
                        \cos\frac{\theta}{2}
                       \end{array}
 \Bigg) \label{vpm}\ee where
 \be \vk = k \,\big(\sin\theta \cos\phi,\sin\theta \sin\phi,\cos\theta\big)\,. \label{veck}\ee A useful representation is
 \be v_+(\vk) = \frac{\big(1+\vec{\sigma}\cdot\hat{\vk} \big)}{\sqrt{2(1+\cos\theta)}}\,\Bigg( \begin{array}{c}
                                                                                        1 \\
                                                                                        0
                                                                                      \end{array}
 \Bigg)~~;~~v_-(\vk)=\frac{\big(1-\vec{\sigma}\cdot\hat{\vk} \big)}{\sqrt{2(1+\cos\theta)}}\,\Bigg( \begin{array}{c}
                                                                                        0 \\
                                                                                        1
                                                                                      \end{array}
 \Bigg)\,.\label{repspinors}\ee
 The Weyl spinors (\ref{vpm}) satisfy
 \be v^\dagger_h(\vk)\cdot v_{h'}(\vk) = \delta_{h,h'}
 \label{ortorel}\ee

 Majorana fields are charge self-conjugate and generally obey
 \be \psi^c = i\gamma^2\,\psi^* = e^{i\xi}\,\psi \label{majcond} \ee with $\xi$ an arbitrary (real) phase, which we choose $\xi=0$. In the chiral representation (\ref{gammas}) writing
 \be \psi = \Bigg(\begin{array}{c}
                    \psi_R \\
                    \psi_L
                  \end{array}
  \Bigg) \label{major}\ee it follows that
  \be \psi^c =  \Bigg(\begin{array}{c}
                    i\sigma^2\, \psi^*_L \\
                    -i\sigma^2\, \psi^*_R
                  \end{array}
  \Bigg) \label{ccpsi}\ee Therefore, a Majorana  field is obtained by combining the positive frequency component with its charge conjugate as the negative frequency, namely
  \be \chi(\vx,t) =  \sum_{h=\pm} \sum_{\vk} \frac{1}{\sqrt{2VE_f(k)}} \Bigg[\hat{b}_{\vk,h} \mathcal{U}_h(\vk)\, e^{-i(E_f(k)\,t-\vk\cdot\vx)} + \hat{b}^\dagger_{\vk,h} \mathcal{U}^{\,c}_h(\vk)\, e^{i(E_f(k)\,t-\vk\cdot\vx)}  \Bigg] \label{majorana}\ee where
  \be \mathcal{U}^{\,c}_+(\vk) =   N_f \Bigg( \begin{array}{c}
                        \varepsilon(k)\,v_-(\vk) \\
                        v_-(\vk)
                      \end{array}
 \Bigg)~~;~~\mathcal{U}^{\,c}_-(\vk)= N_f \Bigg( \begin{array}{c}
                       v_+(\vk) \\
                         \varepsilon(k)\, v_+(\vk)
                      \end{array}
 \Bigg)\label{ucs}\ee and we have used the property
 \be \big( i\sigma^2\big)v^*_+(\vk) = -v_-(\vk) ~~;~~  \big( i\sigma^2\big)v^*_-(\vk) = v_+(\vk)\,.\label{cprops}\ee In particular the negative chirality component of the Majorana neutrino is
 \bea \chi_L(\vx,t) = &&  \frac{1}{\sqrt{V}}  \sum_{\vk}  \Bigg[ \frac{E_f(k)+k}{2E_f(k)}\Bigg]^\frac{1}{2} \Bigg[\Bigg(-\hat{b}_{\vk,+}\,\varepsilon(k) \,v_+(\vk)+\hat{b}_{\vk,-} \,  v_-(\vk)\Bigg)e^{-i(E_f(k)\,t-\vk\cdot\vx)} \nonumber \\ &+ & \Bigg(\hat{b}^\dagger_{\vk,+}\, v_-(\vk)+\hat{b}^\dagger_{\vk,-} \varepsilon(k)\, v_+(\vk) \Bigg)e^{i(E_f(k)\,t-\vk\cdot\vx)}\Bigg]\label{majoleft}\eea

 From the representation (\ref{repspinors}), it follows that
 \be v^\dagger_h(-\vk)\cdot v_h(\vk) =0~~;~~ h = \pm \,.\label{orto}\ee
 It is straightforward to confirm that the Hamiltonian for the Majorana fields
 \be \frac{1}{2}\int d^3 x \chi^\dagger(\vx,t)\Big[-i\vec{\alpha}\cdot\vec{\nabla} + \beta m_f\Big]\chi(\vx,t) = \sum_{k,h}E_f(k)\,\hat{b}^\dagger_{\vk,h}\,\hat{b}_{\vk,h} \label{hamimaj}\ee where the zero point energy has been subtracted.

\section{Wigner-Weisskopf method for $M\rightarrow l\overline{\nu}$:}\label{app:ww}

The purpose of this appendix is to provide technical details of the Wigner-Weisskopf approximation as applied to the $M \rightarrow l \bar{\nu}$ process. For a more extended discussion see refs.\cite{boya,desiternuestro}.

The total  Hamiltonian is given by $H = H_0 + H_i$, where $H_0$ is the free Hamiltonian and $H_i$ is the interaction part.
The time evolution of a state in the interaction picture is given by
\be
i \frac{d}{dt} | \Psi(t) \rangle_I = \hat{H}_I | \Psi(t) \rangle_I \label{timevol}
\ee where $\hat{H}_I(t) = e^{i\hat{H}_0 t} \hat{H}_i(t) e^{-i\hat{H}_0 t}$. The formal solution of (\ref{timevol}) is given by

\be
| \Psi(t) \rangle_I = \hat{U}(t,t_o) | \Psi(t_o) \rangle_I
\ee
where $\hat{U}(t,t_o) = T(e^{-i\int_{t_o}^{t} \hat{H}_I(t') dt'})$ .
Expanding the state $| \Psi(t) \rangle_I $ in the basis of eigenstates of $H_0$  we have
\be
|\Psi(t) \rangle_I = \sum_{n} C_n(t) | n \rangle
\ee
where $\hat{H}_0 | n \rangle = E_n |n \rangle$.   It is   straightforward   to show that $\sum_n |C_n (t)|^2 = const$ which is a consequence of unitary time evolution.

Now consider the initial state at time $t=0$  to be one  meson state of definite momentum, namely
\be
|\Psi(t=0) \rangle_I = | M \rangle_{\vp} = \sum_{n} C_n(t=0) | n \rangle
\ee
which gives the initial condition $C_n(t=0) = \delta_{n,M_{\vp}}$.

From eq.(\ref{timevol}), upon expanding in basis states, it follows that

\be
\frac{d}{dt} C_{n}(t) = -i \sum_{m} \langle n | H_{I}(t) | i \rangle C_{m}(t)
\ee

\begin{figure}[h!]
\includegraphics[keepaspectratio=true,width=3.2in,height=3.2in]{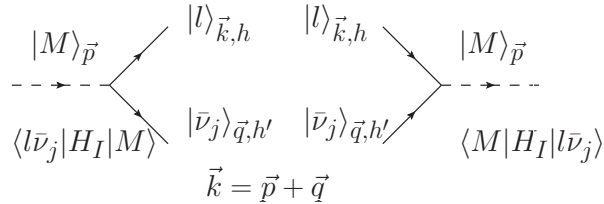}
\caption{Transitions $| M \rangle \rightarrow |l \rangle | \bar{\nu}_j \rangle$}
\label{fig:ww}
\end{figure}

The interaction Hamiltonian (\ref{Hint}) connects the initial meson state, $| M_{\vp} \rangle$   to leptonic/neutrino states, $\{|l\rangle\ \otimes | \bar{\nu} \rangle \}$. These states in turn are coupled back to $| M_{\vp} \rangle$ via $H_I$,  but also to other multiparticle states which describe processes that are higher order in perturbation theory. However, we will only be considering states connected to $|M_{\vp}\rangle$ via first order in perturbation theory. The case that will be of interest to us will be $M \rightarrow l \bar{\nu}$ and is shown in Figure (\ref{fig:ww}).

Considering the set of equations for these states, we obtain
\be
\frac{d}{dt}\, C_{M}(t) = -i \sum_{\kappa} \langle M | H_{I}(t) | \kappa \rangle C_{\kappa}(t)\label{Meq}
\ee

\be
\frac{d}{dt}\, C_{\kappa}(t) = -i \langle \kappa | H_{I}(t) | M \rangle C_{M}(t) \label{kapaeqn}
\ee
where $| \kappa \rangle$ is the intermediate state, $|l_{\alpha}(\vk,h)\rangle |\bar{\nu}_{\alpha}(\vq,h') \rangle$. Using the initial conditions for $t=0$, one obtains

\be
C_{\kappa}(t) = -i \int_{0}^{t} dt' \langle \kappa | H_{I}(t') | M \rangle C_{M}(t')\label{ckaps}
\,, \ee which when inserted into (\ref{Meq}) leads to

\be
\frac{d}{dt} C_{M}(t) = - \int_{0}^{t} dt' \sum_{\kappa} \langle M | H_{I}(t) | \kappa \rangle \langle \kappa | H_{I}(t') | M \rangle C_{M}(t') = - \int_{0}^{t} dt' \Sigma_{M}(t-t') C_M(t')\label{meq2}
\ee
Where the meson self energy has been introduced

\be
\Sigma_{M}(t-t') \equiv \sum_{\kappa} \langle M | H_{I}(t) | \kappa \rangle \langle \kappa | H_{I}(t') | M \rangle = \sum_{\kappa} |\langle M | \hat{H}_I(0)| \kappa \rangle|^{2} e^{i(E_{M}-E_{\kappa})(t-t')}
\ee This self-energy is recognized as the one-loop retarded self energy with the $|l\rangle |\overline{\nu}\rangle$ intermediate state.

Solving eq.(\ref{meq2}) produces a solution for the time evolution of the meson amplitude. We can use the solution for $C_M(t)$ to obtain an expression for the amplitudes $C_{\kappa}(t)$ which allows for computation of the probability of occupying a particular state at any given time. We may solve eq.((\ref{meq2})) either via Laplace transform, or in the case of weak coupling, a derivative expansion which yields the same result at long times ($t\gg 1/m_M$).  Here, we follow the latter method which is the original Wigner-Weisskopf approximation.

We begin by defining the quantity

\be
W_0(t,t') = \int^{t'}_{0} dt'' \Sigma_{M}(t-t'')
\ee

so that

\be
\frac{d}{dt'} W_0(t,t') = \Sigma_{M}(t-t')~~,~~  W_0(t,0)=0
\ee

Integrating eq.(\ref{meq2}) by parts yields

\be
\frac{d}{dt} C_{M}(t) = -\int_{0}^{t} dt' \Sigma_{M}(t-t') C_{M}(t') = -W_0(t,t)C_{M}(t)+\int_{0}^{t} dt'W_0(t,t')\frac{d}{dt'}C_{M}(t')
\ee

The  first is term  second order in $H_I$ whereas the second term is of  fourth order in $H_I$ and will be neglected. This approximation is equivalent to the Dyson resummation of the one-loop self energy diagrams.   Thus to leading order, eq.(\ref{meq2}) becomes

\be
\frac{d}{dt}C_{M}(t) + W_0(t,t) C_M(t) = 0 \,,\label{finmeq}
\ee  where

\be
 W_0(t,t) = \int^{t}_{0} dt' \Sigma_{M}(t-t') = \int^{t}_{0} dt' \sum_{\kappa} |\langle M | \hat{H}_I(0)| \kappa \rangle|^{2} e^{i(E_{M}-E_{\kappa})(t-t')} \label{wott}
\ee

Inserting a convergence factor and taking the   limit $t\rightarrow \infty$  consistently with the Wigner-Weisskopf approximation, we find\footnote{The long time limit in the Wigner-Weisskopf approximation is equivalent to the Breit-Wigner approximation of a resonant propagator\cite{desiternuestro}.}

\be
 W_0(t,t)  = \lim_{\epsilon \to 0^{+}} i \frac{\sum_{\kappa} |\langle M | \hat{H}_I(0)| \kappa \rangle|^{2}}{E_M - E_{\kappa}+i\epsilon} = i \Delta E_M + \frac{\Gamma_M}{2}
\ee where

\be \Delta E_M \equiv \mathcal{P}~ \sum_{\kappa}\frac{  |\langle M | \hat{H}_I(0)| \kappa \rangle|^{2}}{E_M - E_{\kappa}}\,,\label{eshift}\ee is the second order shift in the energy which will be absorbed into a renormalized meson energy and
\be \Gamma_M \equiv 2 \pi \sum_{\kappa} |\langle M | \hat{H}_I(0)| \kappa \rangle|^{2} \delta(E_M - E_{\kappa})\,\label{Pwidth}\ee is the decay width as per Fermi's Golden rule. Therefore  in this approximation, we arrive at

\be
C_M(t) = e^{-i\Delta E_M t} e^{-\frac{\Gamma_M}{2} t} \,.
\ee

Inserting this result  into   eq. (\ref{ckaps})  leads to

\be
\begin{split}
& C_{\kappa}(t) = -i \langle \kappa | H_{I}(0) | M \rangle \int^{t}_{0} dt' e^{-i(E_M + \Delta E_M - E_{\kappa} - i \frac{\Gamma_M}{2}) t'}\\
& = -\langle \kappa | H_{I}(0) | M \rangle \left[ \frac{1 - e^{-i(E_M + \Delta E_M - E_{\kappa} - i \frac{\Gamma_M}{2}) t}}{E_A + \Delta E_M - E_{\kappa} - i \frac{\Gamma_M}{2}} \right]
\end{split}
\ee

Defining the renormalized energy of the single particle meson state  as $E^r_M = E_M + \Delta E_M$ and passing to  the Schroedinger picture   $| M(t) \rangle_S = e^{-i \hat{H}_0 t} | M(t) \rangle_I$, we find that

\be
\begin{split}
& | M^-_{\vp}(t) \rangle_S = e^{-i \hat{H}_0 t}\Bigg[ C_M(t) | M \rangle + \sum_{\kappa} C_{\kappa}(t) | \kappa \rangle\Bigg] \\
& =  e^{-i E^r_M t} e^{-\frac{\Gamma_M}{2} t} |M^-_{\vp}(0)\rangle - \sum_{\kappa} \langle \kappa | H_{I}(0) | M^-_{\vp} \rangle \left[ \frac{1 - e^{-i(E^r_M - E_{\kappa} - i \frac{\Gamma_M}{2}) t}}{E^r_M - E_{\kappa} - i \frac{\Gamma_M}{2}} \right] e^{-i E_{\kappa} t} |\kappa \rangle
\end{split}
\ee

The interaction  Hamiltonian for $M \rightarrow l_{\alpha} \bar{\nu}_{\alpha}$ is given by eqn. (\ref{hamint}) and   the quantization from Appendix A leads to the matrix element

\be
\langle l^-_\alpha \overline{\nu} | H_I(0) | M^-_{\vp} \rangle = \frac{F_M}{\sqrt{V}} \sum_j U_{\alpha j}    \frac{\overline{\mathcal{U}}_{\alpha,h}(\vk) \gamma^\mu \mathbbm{L} \mathcal{V}_{j,h'}(\vq)  p_\mu}{\sqrt{8E_M(p)E_\alpha(k)E_j(q)}} ~~;~~ \vk=\vp+\vq \,\label{mtxelap}
\ee which yields our final result for the   entangled quantum state resulting from meson decay

\be
\begin{split}
& |M^{-}_{\vp}(t))\rangle   =    e^{-iE_M(p)t} e^{-\Gamma_M(p)\frac{t}{2}} |M^{-}_{\vec{p}}(0)\rangle -  F_M \sum_{\vec{q},\alpha j,h,h'} U_{\alpha j} \frac{\overline{\mathcal{U}}_{\alpha,h}(\vk) \gamma^\mu \mathbbm{L} \mathcal{V}_{j,h'}(\vq)  p_\mu}{\sqrt{8VE_M(p)E_\alpha(k)E_j(q)}} \times \\
& \left[ \frac{1 - e^{-i(E^r_M(p) - E_{\alpha}(k) - E_j(q)- i \frac{\Gamma_M}{2}) t}}{E^r_M(p) - E_{\alpha}(k) -E_{j}(q) - i \frac{\Gamma_M}{2}} \right] e^{-i (E_{\alpha}(k) + E_j(q) ) t} |l^-_{\alpha}(h,\vk) \rangle | \bar{\nu}_{j}(h',-\vq)\rangle \\
\end{split} \label{finaresul}
\ee

\section{On the normalization  (\ref{norma}):}\label{app:norma}

The normalization of the disentangled neutrino state (\ref{norma}) has another important interpretation, it is recognized as the number density of charged leptons produced from the decay of the meson. To see this, consider the expansion of the Dirac field for the charged lepton as in eqn. (\ref{psis}) where $\hat{b}^\dagger_{\vk,h_i}$ creates a charged lepton $l^-$ with momentum $\vk$ and helicity $h_i$. The number operator for \emph{particles} is $\hat{n}_{\vk,h_i} = \hat{b}^\dagger_{\vk,h_i}\hat{b}_{\vk,h_i}$ and its expectation value in the \emph{full meson state} (\ref{wwpistate}) is given by
\be n^l_{\vk,h_i}\equiv\langle M^{-}_{\vp}(t))|\hat{n}_{\vk,h_i}|M^{-}_{\vp}(t))\rangle =  ~\,\sum_{j,h'}\frac{\big|U_{\alpha j}\big|^2\,\big|\mathcal{M}^{\produ}_{\alpha j}(\vk,\vq,h_i,h')\big|^2 }{8VE_M(pE_j(q)E_\alpha(k)} \,\Big|\mathcal{F}_{\alpha j}[\vk,\vq;t_c]\Big|^2  (\vq;h_i)\,,\label{numberofl} \ee which is recognized as the normalization (\ref{norma}), namely
\be \mathcal{N}_\nu (\vq;h_i) =  n^l_{\vk,h_i} \,.\label{normuml}\ee

From the definition of the partial width $\Gamma_{M^-\rightarrow l^-_\alpha \overline{\nu}_j}(p,h_i,h') $ of meson decay into a lepton $\alpha$ of helicity $h$ and neutrino eigenstate $\nu_i$ of helicity $h'$
  \be \Gamma_{M^-\rightarrow l^-_\alpha \overline{\nu}_j}(p,h_i,h') = \frac{1}{2E_M(p)} \int \frac{d^3 q}{(2\pi)^3}\, \frac{ \big|\mathcal{M}^{\produ}_{\alpha j}(\vk,\vq,h_i,h')\big|^2 }{2E_j(q)2E_\alpha(k)}~2\pi\,\delta\big( E_M(p)-E_\alpha(|\vp-\vq|)-E_j(q)\big) \label{parwidth}\ee and the total decay width
 \be \Gamma_M(p)= \sum_{j,h'}\big|U_{\alpha j}\big|^2\,\Gamma_{M^-\rightarrow l^-_\alpha \overline{\nu}_j}(p,h_i,h')\,,\label{totwidthres}\ee
  it follows that the total number of charged leptons   produced at time $t_c$ is given by \be  V \sum_{h_i} \int \frac{d^3 q}{(2\pi)^3}\, n^l_{\vk,h_i} =  V \sum_{h_i} \int \frac{d^3 q}{(2\pi)^3}\, \mathcal{N}_\nu (q,h_i) = \Big[1- e^{-\Gamma_M(p)t_c} \Big] \label{intnorma}\ee

\end{document}